\documentclass[a4paper,11pt]{article}
\pdfoutput=1 

\usepackage{jheppub} 
\usepackage{mathtools}
\usepackage{bm}
\usepackage{amsmath,amssymb,amsthm,amscd,graphicx,simpler-wick}
\usepackage{psfrag}
\input epsf.sty
\usepackage{xparse}
\usepackage{tikz}
\usetikzlibrary{tikzmark}
\usepackage{tikz-cd}
\usepackage{needspace}

\usepackage{silence}
\usepackage[missing=]{gitinfo2}

\usepackage{varioref}
\usepackage{microtype}

\newcommand{\tree}{\begin{tikzpicture}[baseline={(current bounding box.center)}, node distance=0.1cm and 0.15cm]
  \coordinate[] (e1);
  \coordinate[left= 0.1 of e1] (c);
  \coordinate[above left= 0.087 and 0.05 of c] (e2);
   \coordinate[below left= 0.087 and 0.05 of c] (e3);
\draw (e1)--(c);
\draw (e2)--(c);
\draw (e3)--(c);
\end{tikzpicture}}

\newcommand{\treet}{\begin{tikzpicture}[baseline={([yshift=-.6ex]current bounding box.center)}, node distance=0.2cm and 0.3cm]
  \coordinate[] (e1); 
  \coordinate[left= 0.2 of e1] (c);
  \coordinate[above left= 0.155 and 0.1 of c] (e2);
   \coordinate[below left= 0.155 and 0.1 of c] (e3);
\draw (e1)--(c);
\draw (e2)--(c);
\draw (e3)--(c);
\end{tikzpicture}}

\def\bbZ{{\mathbb Z}}
\def\bbR{{\mathbb R}}
\def\bbC{{\mathbb C}}
\def\bbQ{{\mathbb Q}}
\def\bbP{{\mathbb P}}

\newcommand{\psiplus}{{\psi_+}}
\newcommand{\psiminus}{{\psi_-}}

\newcommand{\tC}{{\widetilde C}}
\newcommand{\tS}{{\widetilde S}}
\newcommand{\tnabla}{{\widetilde \nabla}}
\newcommand{\tg}{{\tilde g}}
\newcommand{\tmu}{{\widetilde \mu}}
\newcommand{\tcL}{{\widetilde \cL}}
\newcommand{\tsigma}{{\widetilde \sigma}}

\newcommand{\I}{{\mathrm i}}
\newcommand{\e}{{\mathrm e}}
\newcommand{\de}{\mathrm{d}}
\newcommand{\ri}{\text{i}}

\newcommand{\ti}[1]{\textit{#1}}
\newcommand{\abs}[1]{\lvert#1\rvert}

\newcommand{\IP}[1]{\left\langle#1\right\rangle}

\newcommand{\eps}{\epsilon}

\newcommand{\fq}{{\mathfrak q}}

\newcommand{\cL}{{\mathcal L}}
\newcommand{\cV}{{\mathcal V}}
\newcommand{\cG}{{\mathcal G}}
\newcommand{\cO}{{\mathcal O}}

\newcommand{\cM}{{\mathcal M}}
\newcommand{\cT}{{\mathcal T}}
\newcommand{\cI}{{\mathcal I}}
\newcommand{\tJ}{{\widetilde{J}}}
\newcommand{\tT}{{\widetilde{T}}}
\newcommand{\cS}{{\mathcal S}}
\newcommand{\cN}{{\mathcal N}}
\newcommand{\cE}{{\mathcal E}}

\newcommand{\cK}{{\mathcal K}}
\newcommand{\cW}{{\mathcal W}}

\newcommand{\btau}{{\bm{\tau}}}

\newcommand{\fu}{{\mathfrak u}}

\newcommand{\regular}{{\mathrm{reg}}}
\newcommand{\Conf}{{\mathrm{Conf}}}
\newcommand{\Nek}{{\mathrm{Nek}}}
\newcommand{\Heis}{{\mathrm{Heis}}}
\newcommand{\Vir}{{\mathrm{Vir}}}

\newcommand{\ren}{{\mathrm{ren}}}
\newcommand{\Sk}{{\mathrm{Sk}}}
\newcommand{\tot}{{\mathrm{tot}}}

\newcommand{\Liouville}{{\mathrm{Li}}}

\newcommand{\PSL}{{\mathrm{PSL}}}

\newcommand{\GL}{{\mathrm{GL}}}
\newcommand{\SU}{{\mathrm{SU}}}
\newcommand{\Sp}{{\mathrm{Sp}}}

\newcommand{\E}{\text{e}}
\newcommand{\tr}{\text{tr}}
\newcommand{\rd}{\text{d}}

\newcommand{\tPsi}{{\widetilde \Psi}}

\newcommand{\reg}{{\mathrm{reg}}}

\newcommand{\fJ}{{\mathfrak{J}}}

\newcommand{\lo}{{(1)}}
\newcommand{\lt}{{(2)}}

\DeclareMathOperator{\sech}{sech}
\DeclareMathOperator{\End}{End}
\DeclareMathOperator{\Span}{Span}

\DeclareMathOperator{\disc}{disc}
\DeclareMathOperator{\nab}{{\mathcal F}}
\DeclareMathOperator{\Spec}{Spec}

\DeclareMathOperator{\Id}{Id}
\DeclareMathOperator{\Li}{Li}
\DeclareMathOperator{\Fus}{Fus}
\DeclareMathOperator{\Hol}{Hol}
\DeclareMathOperator{\Unfus}{Unfus}
\DeclareMathOperator{\im}{Im}
\DeclareMathOperator{\re}{Re}

\newcommand{\fsl}{{\mathfrak{sl}}}
\newcommand{\fgl}{{\mathfrak{gl}}}

\newcommand{\nop}[1]{{:}\, #1 {:} }

\usetikzlibrary{positioning,arrows,patterns}
\usetikzlibrary{decorations.markings}
\usetikzlibrary{calc}
\usetikzlibrary{shapes}
\usetikzlibrary{topaths}
\usetikzlibrary{decorations}
\usetikzlibrary{decorations.pathmorphing,fit}
\usetikzlibrary{bending}
\usetikzlibrary{calc}

\newcommand*\circled[1]{\tikz[baseline=(char.base)]{
            \node[shape=circle,draw,inner sep=2pt] (char) {#1};}}

\tikzset{
  fermion/.style={draw=black, postaction={decorate},decoration={markings,mark=at position .55 with {\arrow{>}}}},
  vertex/.style={draw,shape=circle,fill=black,minimum size=3pt,inner sep=0pt},
  vertex2/.style={draw,shape=circle,fill=black,minimum size=1pt,inner sep=0pt},
  vertex3/.style={draw,shape=circle,fill=\psiplusColor,minimum size=3pt,inner sep=0pt},
  vertex4/.style={draw,shape=circle,fill=\psiminusColor,minimum size=3pt,inner sep=0pt}  
}

\tikzset{dot/.style={circle, fill, inner sep=1.5pt}}

\newcommand{\branchpointColor}{orange}
\newcommand{\wallColor}{black}
\newcommand{\psiplusColor}{purple!70}
\newcommand{\psiminusColor}{blue!70}
\newcommand{\vdashLineColor}{gray!60} 
\newcommand{\jpColor}{black}      
\newcommand{\tpColor}{black}       
\newcommand{\coveringCurveColor}{gray} 
\newcommand{\baseCurveColor}{gray}  
\newcommand{\leashColor}{green!55!black}

\newlength{\branchpointmarkersize}
\newcommand{\drawbranchpointmarker}[1]{
  \draw [line width=0.5mm, \branchpointColor] ($(#1)-(\branchpointmarkersize/2,\branchpointmarkersize/2)$) -- ($(#1)+(\branchpointmarkersize/2,\branchpointmarkersize/2)$);
  \draw [line width=0.5mm, \branchpointColor] ($(#1)-(\branchpointmarkersize/2,-\branchpointmarkersize/2)$) -- ($(#1)+(\branchpointmarkersize/2,-\branchpointmarkersize/2)$);
}

\tikzset{
    psipluscontour/.style={thick,\psiplusColor},
    psiminuscontour/.style={thick,\psiminusColor},
    leash/.style={dashed,\leashColor}
  }

\title{\boldmath A new construction of $c=1$ Virasoro blocks}
\author[1,*]{Qianyu Hao}
\author[2]{and Andrew Neitzke}
\affiliation[1]{Section de Math\'{e}matiques, Universit\'{e} de Gen\`{e}ve, 1211 Gen\`{e}ve 4, Switzerland}
\affiliation[2]{Department of Mathematics, Yale
University, PO Box 208283, New Haven, CT 06520-8283, United States}
\affiliation[*]{Corresponding author, qianyu.hao@unige.ch}

\emailAdd{qianyu.hao@unige.ch}
\emailAdd{andrew.neitzke@yale.edu}

\makeatletter
\gdef\@fpheader{\bigskip}
\makeatother

\begin{document}

\makeatletter
\renewcommand{\sectionautorefname}{\S\@gobble}
\renewcommand{\subsectionautorefname}{\S\@gobble}
\renewcommand{\subsubsectionautorefname}{\S\@gobble}
\makeatother

\abstract{
We introduce a \ti{nonabelianization} map for conformal blocks, 
which relates $c=1$ Virasoro blocks on a Riemann surface $C$ to 
Heisenberg blocks on a branched double
cover $\tC$ of $C$. 
The nonabelianization map uses the datum of a spectral network on $C$.
It gives new formulas
for Virasoro blocks in terms of fermion correlation functions 
determined by the Heisenberg block on $\tC$.
The nonabelianization map also intertwines with the action of Verlinde loop
operators, and can be used to construct eigenblocks.
This leads to new Kyiv-type formulas and regularized Fredholm determinant formulas for $\tau$-functions. 
}

\maketitle

\flushbottom

\bibliographystyle{JHEP}

\section{Introduction}\label{sec:intro}

This paper concerns a new approach to the construction and study 
of conformal blocks for the Virasoro algebra at central charge $c=1$.

Our motivation comes from recent work on the geometry of 
various problems associated to a Riemann surface $C$ --- topological strings,
exact WKB, and conformal blocks --- and especially the works \cite{Coman:2020qgf,MR4115013},
together with the related \cite{Cheng:2010yw,Coman:2018uwk,MR4567381,Bridgeland:2022ned,Bridgeland:2023eka,Iwaki:2023cek}.
The picture emerging from these works is that in all of these problems
the perturbative partition function admits a nonperturbative extension, but the extension 
depends on the additional datum of a spectral network on $C$.
This leads to the idea that there should be a construction of a partition function
which uses the spectral network directly. In this paper we propose such
a construction for the $c=1$ Virasoro blocks.

\subsection{Virasoro blocks}

In the introduction we work with a compact Riemann surface $C$, 
and we use a condensed notation, suppressing subtleties about coordinate systems
and normal ordering.

We denote the space of Virasoro conformal blocks on $C$ by $\Conf(C, \Vir_{c})$.
A block $\Psi \in \Conf(C, \Vir_{c})$ is a system of chiral correlation functions, written
\begin{equation}
  \IP{T(z_1) \cdots T(z_n)}_\Psi \, ,
\end{equation}
where the $z_i$ denote points of $C$. The correlation functions are required to be compatible with the OPE
and coordinate transformation laws of the Virasoro vertex algebra.
We recall the definitions and some key properties 
in \autoref{sec:vertex-algebras} below.

Virasoro blocks are not easy to calculate; some of the principal methods available are
the recursion relations of \cite{Zamolodchikov:1987avt,Cho:2017oxl}
and the representations provided by the AGT correspondence \cite{Alday:2009aq}.
For more background on Virasoro conformal blocks see e.g. \cite{DiFrancesco:1997nk,MR3971924,Perlmutter:2015iya,Ribault:2014hia}.

\subsection{The free-field construction}

Our approach to computing Virasoro blocks reduces them to simpler objects,
namely conformal blocks for the Heisenberg vertex algebra $\Heis$ (also known as the $\hat\fu(1)$ vertex algebra,
or the chiral free boson vertex algebra).
A block $\Psi \in \Conf(C, \Heis)$ is a recipe for chiral correlation functions, written
\begin{equation}
  \IP{J(z_1) \cdots J(z_n)}_\Psi \, ,
\end{equation}
compatible with the OPE and coordinate transformation laws of the Heisenberg vertex algebra.

There is a well-known way of making $c=1$ Virasoro blocks from Heisenberg blocks, the \ti{free-field} construction:
writing $T = \frac12 J^2$ gives a map
\begin{equation}
 \Conf(C, \Heis) \to \Conf(C, \Vir_{c=1}) \, .
\end{equation}
However, the Virasoro blocks in the image of this map are very special; we are after a more generic construction.

\subsection{The branched free-field construction}

Another construction of $c=1$ 
Virasoro blocks from Heisenberg blocks was given in \cite{MR0897030}.
Here one uses Heisenberg blocks on a branched double cover $\pi: \tC \to C$. 
On $\tC$ we use the letter $\tJ$ for the Heisenberg field. Then 
let $\tJ^{(-)} = \frac{1}{\sqrt 2} (\tJ^{(1)} - \tJ^{(2)})$ denote the anti-invariant combination of insertions on the two sheets of $\tC$.
This gives a well defined operator on $C$ up to the $\bbZ_2$ action 
$\tJ^{(-)} \to -\tJ^{(-)}$. We define $T = \frac12 (\tJ^{(-)})^2$ and substitute this in 
the Heisenberg correlation functions to get the desired Virasoro correlators. 
We call this the \ti{branched free-field} construction,
and review it in \autoref{sec:branched-free-field}.

The branched free-field construction gives Virasoro blocks on $C$, but they turn out to have
additional singularities at the branch points $b_1, \dots, b_k$ 
of the covering. These additional singularities can be interpreted
as insertions of Virasoro primary fields $W_h(b_i)$ with weight $h = \frac{1}{16}$.
Thus altogether we obtain a map
\begin{equation}
   \Conf(\tC,\Heis) \to \Conf\left(C,\Vir_{c=1}; W_{\frac{1}{16}}(b_1) \cdots W_{\frac{1}{16}}(b_k) \right) \, .
\end{equation}
Again this is not quite what we are after:
we want a construction of pure Virasoro blocks on $C$, 
without these extra operator insertions.

\subsection{Adding the spectral network}

The main new idea of this paper, explained in \autoref{sec:nonabelianization}, is to 
use the branched free-field construction with one modification:
we insert an extra operator $E(\cW)$ in the correlation functions on $\tC$. $E(\cW)$
is built from free fermions $\psi_\pm$ (in turn built out of the Heisenberg
field $\tJ$ via fermionization):
\begin{equation} \label{eq:EW-intro}
 E(\cW) = \exp \left[ \frac{1}{2 \pi \I} \int_\cW \psi_+( z^{(+)} ) \psi_-( z^{(-)} ) \, \de z \right] \, .
\end{equation}
Here $z^{(+)}, z^{(-)} \in \tC$ denote the two preimages of a point $z \in C$. The integration 
runs over a contour $\cW$ on $C$, which is a \ti{spectral network}
of type $\fgl_2$, in the sense of \cite{MR3115984}.
In particular $\cW$ is a collection of arcs on $C$, with three ending on each branch point
of the covering $\pi$:
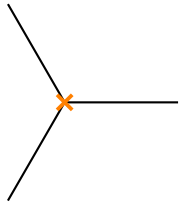
\begin{center}
\begin{tikzpicture}
  \begin{scope}
    \coordinate (center) at (0,0);

    \draw [thick,\wallColor] (center) -- ++(120:1.5cm); 
    \draw [thick,\wallColor] (center) -- ++(240:1.5cm); 
    \draw [thick,\wallColor] (center) -- ++(0:1.5cm);   

    \drawbranchpointmarker{center};

  \end{scope}
\end{tikzpicture}
\end{center}

In \autoref{sec:model-example} we compute the correlators in a simple model example, and
show that with $E(\cW)$ inserted, the normalized correlation functions of $T(z)$
are regular even when $z$ hits a branch point. Thus the insertion of $E(\cW)$
removes the unwanted insertions $W_{\frac{1}{16}}$ at the branch points.

We also show that the $0$-point function with $E(\cW)$ inserted is divergent, 
but can be rendered finite by 
replacing $E(\cW)$ with a renormalized version $E_\ren(\cW)$. This renormalization at first seems like 
a nuisance, but it is important for the consistency 
of the story: it introduces an anomalous dependence on a
local coordinate near each branch point, with weight $-\frac{1}{16}$,
which cancels the insertions of weight $\frac{1}{16}$ there.

Thus we obtain a map between spaces of conformal blocks,
\begin{equation} \label{eq:nonab-map-reduced-intro}
  \Conf(\tC,\Heis) \to \Conf(C,\Vir_{c=1}) \, ,
\end{equation}
as desired.

To be precise, in most of the paper we actually consider a 
slightly different map.
The map \eqref{eq:nonab-map-reduced-intro} uses only the odd part of the
Heisenberg correlators on $\tC$, ignoring
the even part $\tJ^{(+)} = \frac{1}{\sqrt2}(\tJ^{(1)} + \tJ^{(2)})$.
Keeping both the odd and even parts, we obtain an enlarged map,
\begin{equation} \label{eq:nonab-map-intro}
 \nab_\cW: \Conf(\tC,\Heis) \to \Conf(C,\Vir_{c=1} \otimes \Heis) \, .
\end{equation}
We call $\nab_\cW$ the \ti{nonabelianization} map for conformal blocks.

Concretely, for instance, given a block $\tPsi \in \Conf(\tC,\Heis)$,
the $1$-point function of the Virasoro generator in the corresponding block $\nab_\cW(\tPsi)$ is
\eqref{eq:nab-1-point} below, reproduced here:
\begin{equation}
  \IP{T(z)}_{\nab_\cW(\tPsi)} = \frac{1}{4} \IP{ (\tJ(z^{(1)}) - \tJ(z^{(2)}))^2 \, E_\ren(\cW) }_\tPsi \, .
\end{equation}
The correlation functions in the block $\nab_\cW(\tPsi)$ 
can in principle be computed directly using the definition \eqref{eq:EW-intro} of $E(\cW)$: that amounts to 
evaluating an infinite sum of iterated integrals of free fermion $2n$-point functions on $\tC$. These computations look difficult,
but they can actually be carried out in at least one case (this is what we do in \autoref{sec:model-example}).

\subsection{Explicit Heisenberg blocks}

So far the story does not depend on the particular 
Heisenberg block $\tPsi$ we consider.
To get more explicit information, though, we need to fix some specific $\tPsi$.
This is the subject of \autoref{sec:explicit-blocks}:
we fix a choice of $A$ and $B$ cycles
on $\tC$, then construct a collection of linearly independent Heisenberg blocks parameterized
by $\tg$ continuous parameters,
\begin{equation}
 \tPsi_a \in \Conf(\tC,\Heis), \quad a = (a_1, \dots, a_{\tg}) \in \bbC^\tg \, .  
\end{equation}
This part of the story does not involve a spectral network; it lives purely in the theory of Heisenberg blocks on $\tC$. 
The blocks $\tPsi_a$ are characterized by the properties 
\begin{equation} \label{eq:Psi-cycle-properties-intro}
\ell_{A_i} \tPsi_a = a_i \tPsi_a \, , \qquad \ell_{B_i} \tPsi_a = 2 \pi \I \partial_{a_i} \tPsi_a \, , \qquad \IP{1}_{\tPsi_{a=0}} = 1 \, ,
\end{equation}
where we introduce the log-Verlinde operators acting on $\Conf(\tC,\Heis)$,
\begin{equation} \label{eq:log-verlinde-intro}
\ell_\gamma = \oint_\gamma \tJ \, .
\end{equation}
Changing the choice of $A$ and $B$ cycles by an action of $\Sp(2 \tg, \bbZ)$
transforms the blocks $\tPsi_a$ by a generalized Fourier transform.
Indeed, the $\tPsi_a$ can be thought of as delta-function states
in the quantization of a linear symplectic space $\bbR^{2 \tg}$,
with the choice of $A$ and $B$ cycles giving a choice of real polarization.

\subsection{Fenchel-Nielsen blocks, Liouville momenta, and Goncharov-Shen blocks}

Choosing a spectral network $\cW$ and 
applying nonabelianization to the Heisenberg blocks $\tPsi_a$ gives a family of Virasoro
blocks, $\nab_\cW(\tPsi_a)$.
Said otherwise, each type of spectral network $\cW$ gives rise to a corresponding type of Virasoro block.

In particular, there is a class of spectral networks $\cW_P$, called ``Fenchel-Nielsen'' in \cite{Hollands:2013qza},
which correspond to pants decompositions $P$ of the surface $C$. One might call the corresponding conformal blocks
$\nab_{\cW_P}(\tPsi_a)$ ``Fenchel-Nielsen blocks.''
We propose in \autoref{sec:pants-decompositions} that for $a \in \I \bbR^g$ the Fenchel-Nielsen blocks should coincide with the blocks
$\Psi^\Liouville_P(a)$
usually used to describe $c=1$ Liouville theory, 
with the $a_i$ identified as the Liouville momenta through the pant-legs.
Our proposed construction of the Fenchel-Nielsen blocks by nonabelianization 
looks rather different from the usual description of Liouville blocks, and it would be
very interesting to verify directly that they indeed match.

If we have some marked points on $C$, we can consider another class of spectral networks $\cW_T$, called ``Fock-Goncharov'' \cite{MR3115984,Hollands:2013qza}. Fock-Goncharov networks $\cW_T$ correspond 
to ideal triangulations $T$ of the surface $C$, with vertices at the marked points. These networks give rise to another class of blocks $\nab_{\cW_T}(\tPsi_a)$, with primary fields inserted at the marked points; we could call these ``Goncharov-Shen blocks.'' 
The parameters $a_i$ in this case are some analogue of Liouville momenta, associated with the decomposition of $C$ into triangles instead of pants.
It was conjectured in \cite{Goncharov2019} that there should exist Virasoro blocks associated to ideal triangulations of $C$,
building on results of \cite{MR2470108,MR2233852}; as we discuss in \autoref{sec:triangulations}, 
our construction of conformal blocks gives a route to proving this conjecture, but not yet a proof.

\subsection{Verlinde loop operators}

One of the most important structures on the spaces of conformal blocks
which we consider is the action of the \ti{Verlinde loop operators}.
This is the main subject of \autoref{sec:verlinde-operators}.

For each loop $\wp$ on $C$ there is a loop operator
$L_\wp$ acting on $\Conf(C, \Vir_{c=1} \otimes \Heis)$, 
as discussed in e.g. \cite{Alday:2009fs,Drukker:2009id,Coman:2015lna,Gaiotto:2024tpl}.
We also introduce loop operators $L_\gamma$ 
acting on $\Conf(\tC, \Heis)$, labeled by loops $\gamma$ on $\tC$.
The operators $L_\gamma$ are much simpler to describe and study than their
Virasoro counterparts $L_\wp$.

In both cases, the loop operators commute with one another.
The loop operators on $C$ generate the commutative skein algebra
$\Sk_{-1}(C, \GL(2))$, while those on $\tC$ generate the commutative skein algebra
$\Sk_{-1}(\tC, \GL(1))$ (also known as the twisted torus algebra).
Dually, the algebras of loop operators are the algebras
of functions on moduli spaces of (twisted) flat connections,
$\cM(C,\GL(2))$ and $\cM(\tC,\GL(1))$ respectively.

The action of the loop operators gives one way of picking out distinguished conformal blocks:
we can look for simultaneous eigenblocks of all loop operators. Then:

\begin{itemize}
\item
In the case of $\Conf(\tC, \Heis)$, a simultaneous eigenvalue $X$ of the loop operators
means a point of $\cM(\tC, \GL(1))$. Decomposing $X$ according to our basis of $A$ and $B$ cycles
as $X = (\e^x, \e^y)$, we can write an 
eigenblock $\tPsi_{x,y} \in \Conf(\tC, \Heis)$ as a linear combination of the $\tPsi_a$:
\begin{equation}\label{eq:diagonalizing-abelian-verlinde-intro}
  \tPsi_{x,y} = \sum_{n \in \bbZ^\tg} \exp\left(-\frac{(x + 2 \pi \I n) \cdot y}{2 \pi \I}\right) \tPsi_{a = x + 2 \pi \I n} \, .
\end{equation}
This discrete Fourier transform operation, 
passing from eigenblocks of the $\ell_{A_i}$ to eigenblocks of the
$L_\gamma$, corresponds to the Gelfand-Zak transform
in the quantization of $\bbR^{2\tg}$.

\item
In the case of $\Conf(C, \Vir_{c=1} \otimes \Heis)$, a simultaneous
eigenvalue $\lambda$ of the loop operators means a point of $\cM(C, \GL(2))$.
Eigenblocks $\Psi \in \Conf(C, \Vir_{c=1} \otimes \Heis)$ are harder to construct, and 
one of our main points is that nonabelianization gives a systematic approach to this problem.

The tool we use to construct eigenblocks $\Psi$ is covariance of $\nab_\cW$ with respect to the Verlinde loop operators:
we formulate this in \autoref{sec:abelianization-and-verlinde}.
It implies that if $\tPsi$ is an eigenblock of the Verlinde operators on 
$\tC$, then $\Psi = \nab_\cW(\tPsi)$ is an eigenblock of the Verlinde operators on $C$.
This gives a family of eigenblocks $\Psi^\cW_{x,y} = \nab_\cW(\tPsi_{x,y})$.
The corresponding eigenvalues are $\lambda = \nab^\flat_\cW((\e^x,\e^y))$, where $\nab^\flat_\cW$ is the nonabelianization map for flat connections
\cite{MR3115984,Hollands:2013qza}.
\end{itemize}

\subsection{The line bundles of Verlinde eigenblocks}

For each $X \in \cM(\tC, \GL(1))$ the corresponding space of Verlinde eigenblocks $\tPsi \in \Conf(\tC, \Heis)$ is 1-dimensional.
Thus the Verlinde eigenblocks make up a line bundle $\tcL \to \cM(\tC, \GL(1))$, with local trivializations given by 
the blocks $\tPsi_{x,y}$. 
Likewise, for each $\lambda \in \cM(C, \GL(2))$ we can consider the corresponding space of
Verlinde eigenblocks $\Psi \in \Conf(C, \Vir_{c=1} \otimes \Heis)$.
These eigenblocks thus make up a sheaf $\cL \to \cM(C, \GL(2))$, which we conjecture 
is generically a line bundle. Both $\tcL$ and $\cL$ carry interesting holomorphic connections,
whose curvature is a symplectic form.

In \autoref{sec:heisenberg-line-bundle} and \autoref{sec:bundle-geometry} we briefly discuss the geometry of $\tcL$ and $\cL$
respectively. For $\tcL$ we can be completely explicit. For 
$\cL$ the basic point is that 
$\Psi^\cW_{x,y}$ discussed above give local trivializations, and thus
give a description of $\cL$ by patching, with explicit
transition functions.
In this way our picture of $\cL$ is related to previous works 
\cite{MR2233852,Alexandrov:2011ac,Neitzke:2011za,Coman:2020qgf,Bertola:2019nvr,Freed:2022yae}
where essentially the same line bundle was considered, from various perspectives.

\subsection{Verlinde eigenblocks and \texorpdfstring{$\tau$}{tau} functions}

We have just discussed a line bundle $\cL \to \cM(C,\GL(2))$ of eigenblocks for a fixed Riemann surface $C$.
We can also let the surface $C$ vary, and obtain a line bundle $\cL \to \cM(C,\GL(2)) \times \cM_g$.
If $\Psi$ is a section of this bundle, we can consider the 0-point function 
\begin{equation} \label{eq:tau-zero-point-intro}
\btau = \IP{1}_\Psi
\end{equation}
as a function on $\cM(C,\GL(2)) \times \cM_g$.  

One reason to pay attention to $\btau$ was explained in \cite{MR3322384}:
if $C$ is a sphere with four primary field insertions, and $\Psi$ is a certain carefully normalized
section of $\cL$, then $\btau$ is a $\tau$-function for the Painlev\'e VI equation.
This is an interpretation of the celebrated Kyiv formula \cite{Gamayun:2012ma}: the 
particular combination of conformal blocks which was considered there 
has the property that it diagonalizes the Verlinde operators. 

Extending this philosophy, we also think of $\btau$ for more general $C$ as a kind of $\tau$-function. 
We formulate this more precisely, and explain what we mean by ``carefully normalized,'' in \autoref{sec:tau-functions}.
Then we obtain a concrete formula for $\btau$, given in \eqref{eq:tau-formula} below, reproduced here:
\begin{equation} \label{eq:tau-formula-intro}
  \btau = \frac{\Theta\left[\frac{x}{2 \pi \I} \big\vert \frac{-y}{2 \pi \I}\right] (\tau, 0)}{\eta_{\pi^* S}} \times \det_{\reg}(1 + \cI_{x,y}) \, .
\end{equation}
The most nontrivial ingredient in this formula is $\cI_{x,y}$, 
an integral operator acting on sections 
of $K_C^\frac12$ over $\cW$,
whose kernel is a normalized fermion 2-point function
on $\tC$:
\begin{equation} \label{eq:integral-kernel-intro}
  \cK(p,q) = \frac{1}{2\pi \I} \frac{\IP{\psiplus(p^{(+)}) \psiminus(q^{(-)})}_{\tPsi_{x,y}}}{\IP{1}_{\tPsi_{x,y}}} \, .
\end{equation}
(Note that $\cK(p,q)$ 
has no singularity at $p=q$, because the $\psiplus$ and $\psiminus$ insertions are
taken on different sheets of $\tC$.) 

Fredholm determinant representations of $\tau$-functions have appeared
before, e.g. \cite{MR0573370,Widom:1997vx,Tracy:1998vb,Bonelli:2017ptp,Cafasso:2017xgn,Gavrylenko:2016zlf,DelMonte:2020wty,MR4304489,Gavrylenko:2023ewx}.
The determinants in \cite{Cafasso:2017xgn,Gavrylenko:2016zlf,DelMonte:2020wty}
somewhat resemble ours, though they involve 
different contours on $C$ and a different integral operator. It would be desirable to understand whether there is some
procedure which would reduce our determinant to theirs.
This would be especially useful because in
\cite{Cafasso:2017xgn,Gavrylenko:2016zlf,DelMonte:2020wty} 
there is a detailed explanation of how to recover the 
Liouville blocks (in the form coming from \cite{Alday:2009aq})
from the Fredholm determinant, which could help settle
our conjecture in \autoref{sec:pants-decompositions}.

\subsection{Relation to free fermion field theory}

Relations between $\tau$-functions, free fermions and conformal field
theory have been developed extensively from many different points of view,
e.g. \cite{MR0499666,Sato:1979kg,Moore:1990mg,Moore:1990cn,palmer1993tau,korotkin2001,Nekrasov:2003rj,Gavrylenko:2016moe,Coman:2018uwk,Coman:2020qgf}.
In particular, \cite{Moore:1990mg,Moore:1990cn} describes $\tau$-functions 
using operators very similar to our operator $E(\cW)$.

In this paper we use exclusively the abstract language of
conformal blocks, rather than committing ourselves to any
particular field theory.
Still, we can suggest a tentative translation, as follows.
Correlation functions on $\tC$ in the eigenblocks $\tPsi_{x,y}$ 
should be understood as having to do with the theory of a chiral
free fermion on $\tC$, twisted by a background $\GL(1,\bbC)$ gauge field on $\tC$
with holonomies $(\e^x, \e^y)$.
Correlation functions on $C$ in the eigenblocks $\Psi^\cW_{x,y}$
should likewise have to do with the theory of $2$ chiral free 
fermions on $C$,
twisted by a background $\GL(2,\bbC)$ gauge field on $C$
with holonomies $\lambda = \nab^\flat_\cW((\e^x,\e^y))$.
From this point of view, the nonabelianization map $\nab_\cW$ would
become a passage between these two field theories:
 it should say e.g. that
the two fermion determinants are not equal on the nose, but 
that they become equal (up to an overall constant)
after inserting the operator
$E_\ren(\cW)$ in the theory on $\tC$.

\subsection{Open questions and extensions}

In this paper we only discuss the most basic version of the nonabelianization of conformal blocks, and we leave many open questions. Here is a long wish-list of problems to explore:

\begin{itemize}

  \item
  Although we set out our recipe in detail, in this paper we give no explicit computations of blocks using our recipe (apart
  from the case of $C = \bbC\bbP^1$ without primary field insertions, in which case the spaces of blocks are $1$-dimensional.)
  It would be very desirable to make some concrete computations, either analytic or numerical. In particular, it would be good to establish explicitly that the Fenchel-Nielsen blocks indeed agree with the usual basis of 
  Liouville blocks, as we expect.

  \item
  If $\cW$ and $\cW'$ are two spectral networks which differ by a ``flip''  in the sense
  of \cite{MR3115984},
  then the nonabelianization maps $\nab_\cW$, $\nab_{\cW'}$ 
  should differ by a certain operator $\cK_\gamma$
  built from the dilogarithm function. We formulate this statement in
  \autoref{sec:mutations} but do not prove it. It would be desirable to fill this gap.
  In particular, this would be important for proving that our blocks
  indeed coincide with the ones called for in \cite{MR3761995} when $\cW$ is
  a Fock-Goncharov network.

  \item In this paper we focus on constructing Virasoro (or Virasoro-Heisenberg) blocks.
  We expect a closely parallel story for the principal $W$-algebra $W(\fsl_N)$ (or $W(\fgl_N)$) with $c=N-1$.
  Given a branched $N$-fold cover $\pi: \tC \to C$, and a spectral network $\cW$ of type $\fgl_N$ \cite{MR3115984},
  we should obtain a map
  \begin{equation}
    \nab_\cW: \Conf(\tC, \Heis) \to \Conf(C, W(\fgl_N)_{c = N-1}) \, .
  \end{equation}
  The dictionary \eqref{eq:nab-operator-dictionary}
  will be replaced by one coming from the free-field construction of $W(\fgl_N)$ inside $N$ copies of $\Heis$ (see e.g. \cite{Bouwknegt:1992wg,Gavrylenko:2016moe}).
  The spectral network $\cW$ will be used in essentially the same way as it is in this paper (at least in the case of
  simple ramification, which is the generic case).

  \item
  For instance, suppose $C$ is a sphere with $3$ generic primary field insertions.
  For Virasoro, the space of conformal blocks on $C$ is $1$-dimensional, and it is not hard to construct a
  block directly. In contrast, for $W(\fgl_N)$ with $N > 2$, the space of conformal blocks on $C$ is infinite-dimensional, and no construction of a continuous family of independent blocks is known (see however 
  \cite{Coman:2017qgv} which gives a discrete
  family in the case $N=3$ using screening contours). What we are proposing is that, once we fix a spectral
  network $\cW$ on $C$ of type $\fgl_N$, and a choice of $A$ and $B$ cycles on the corresponding spectral cover $\tC$, then $\nab_\cW(\tPsi_a)$ will be the desired continuous family of $W(\fgl_N)$-blocks.

  \item
  It seems likely that there is also an extension of nonabelianization to $c \neq N-1$. Indeed almost all of the ingredients in the story
  have a straightforward deformation to this case. Although the algebras of Verlinde operators are not commutative for general $c$, 
  there is still an intertwining map $\nab_\cW^\Sk$ between them, as discussed in \cite{Neitzke:2020jik,Neitzke:2021gxr} 
  for $N=2$ and $N=3$ (see also closely related \cite{Galakhov:2014xba,Gabella:2016zxu}).
  Thus it makes sense to ask for a map
  \begin{equation}
    \nab_\cW: \Conf(\tC, \Heis) \to \Conf(C, W(\fgl_N)_{c})
  \end{equation}
  which is compatible with the action of Verlinde operators. The key difficulty which needs to be overcome
  is the fact that the free-field construction of $W(\fgl_N)$ is not $S_N$-invariant except at $c = N-1$.

  \item
  In most of this paper we consider conformal blocks on a compact surface, with primary field insertions
  allowed, but not irregular vertex operators in the sense of \cite{Gaiotto:2009ma,Gaiotto:2012sf}.
  We expect that there is an extension of 
  the nonabelianization map to incorporate irregular vertex operators. We discuss one example
  in \autoref{appen:irreg}.
  Some of the most fundamental applications of our construction should involve these irregular vertex operators,
  so it would be useful to develop their theory more systematically.

\item In particular, it should be possible to use our formula \eqref{eq:tau-formula-intro} for $\tau$-functions to produce
a new Fredholm determinant form of the Painlev\'e III$_3$ $\tau$-function, by taking $C = \bbC \bbP^1$ with two irregular singularities. Upon taking an appropriate limit where the kernel $\cK$ simplifies, we would hope that this reproduces a known determinant formula for the $\tau$-function with special initial conditions, studied in \cite{Widom:1997vx,Tracy:1998vb,Bonelli:2017ptp,Gavrylenko:2023ewx}. More generally we might hope that \eqref{eq:tau-formula-intro} can be used to produce new Fredholm determinant formulas for other Painlev\'e equations.

  \item
  It would be very interesting to extend our considerations from vertex algebras to their $q$-analogues.
  It seems likely that this will require replacing spectral networks by exponential networks as introduced in
  \cite{Eager:2016yxd}.
  The conformal blocks we considered in this paper give examples
  of nonperturbative topological string partition functions \cite{Cheng:2010yw,Coman:2020qgf}, in 
  the case where the relevant target space physics is 4-dimensional gauge theory.
  The $q$-Virasoro case would be related instead to 5-dimensional gauge theory compactified on a circle.
  One might hope in this way to re-derive the TS/ST correspondence 
  \cite{Grassi:2014zfa}, which in these 5-dimensional
  examples identifies a nonperturbative version of the topological string partition function as the
  Fredholm determinant of an integral operator.

  \item
  Our considerations in this paper are mostly insensitive to the particular choice of spectral 
  network: any spectral network gives a nonabelianization map for conformal blocks.
  This is parallel to the fact that any spectral network gives a
  nonabelianization map between moduli spaces of flat connections \cite{MR3115984,Hollands:2013qza}.

  In the context of flat connections there is also a deeper story, where the choice of spectral
  network definitely does matter. This is the story of
  exact WKB analysis of one-parameter families of flat connections, of the form 
  $\nabla(\hbar) = \hbar^{-1} \varphi + \cdots$. In that setting the
  Higgs field $\varphi$ determines a spectral network $\cW(\hbar)$ (also called Stokes graph),
  and one gets the sharpest information about $\nabla(\hbar)$ only when one uses
  the network $\cW(\hbar)$.

  We expect an analogous phenomenon for conformal blocks.
  Namely, we can consider a family 
  $\Psi(\hbar) \in \Conf(C, \Vir_{c=1} \otimes \Heis)$ 
  whose $\hbar \to 0$ behavior is controlled
  by a meromorphic quadratic differential on $C$, in an appropriate sense.
  For instance, if the $\Psi(\hbar)$ are Verlinde eigenblocks, their eigenvalues $\lambda(\hbar) \in \cM(C, \GL(2))$ will
  diverge as $\hbar \to 0$, with the usual exponential WKB behavior.
  It is for these families $\Psi(\hbar)$ that we expect to get the sharpest information
  from nonabelianization of conformal blocks: namely,
  we will have a corresponding distinguished network $\cW(\hbar)$,
  and we should get a description of $\Psi(\hbar)$ as 
  $\nab_{\cW(\hbar)}(\tPsi(\hbar))$, obtained by Borel summation
  of a series in $\hbar$.

  \item
The method of Deift-Zhou \cite{dz} in integrable systems
involves a strategy which is quite similar to ours.\footnote{We thank Marco Bertola, Pavlo Gavrylenko, and Dmitry Korotkin for
explaining this method to us.} 
One considers a Riemann-Hilbert problem involving 
jump contours lying along a spectral network, 
with the jumps given by unipotent matrices.
Such a Riemann-Hilbert problem is most effectively solvable 
when the jump matrices are small; for the 
case relevant in \cite{dz}, they are indeed small, 
except near the branch points. To deal with the region near the 
branch points, one cuts out a disc around each branch point and pastes in an exact solution 
of an ODE there (roughly the Airy function). 
It would be interesting to know whether this 
tactic is useful also in the conformal-block context, as an alternative to the renormalization scheme
we use here.

\end{itemize}

\subsection*{Acknowledgements}

We thank Mina Aganagic, David Ben-Zvi, Marco Bertola, Tom Bridgeland, Thomas Creutzig, Fabrizio Del Monte, Gurbir Dhillon, Davide Gaiotto, Pavlo Gavrylenko, Alexander Goncharov, Alba Grassi, Lotte Hollands, Cristoforo Iossa, Kohei Iwaki, Dmitry Korotkin, Oleg Lisovyy, Swarnava Mukhopadhyay, Subrabalan Murugesan, David Poland, Vivek Shende, Maximilian Schwick, Sri Tata, Joerg Teschner, Richard Wentworth, and Yan Zhou for helpful
explanations and discussions. 
We especially thank Sri Tata for explaining an important computation which helped us to write \autoref{sec:model-example}, and Joerg Teschner for many explanations
concerning conformal blocks.
We also thank the anonymous referees for several helpful suggestions and explanations.

This work is partially supported by the Swiss National Science Foundation Grant No. 185723,
by National Science Foundation grant 2005312, and by a Simons Fellowship in Mathematics.
We also thank the Simons Center for Geometry and Physics for its hospitality while
this work was being completed.

\section{Vertex algebras, conformal blocks and fermions} \label{sec:vertex-algebras}

\subsection{The Heisenberg and Virasoro vertex algebras}

We briefly recall the Heisenberg and Virasoro vertex algebras here,
to fix conventions:

\begin{itemize}

 \item Fix a constant $c \in \bbC$.
The Virasoro vertex algebra $\Vir_c$ is generated by one field $T$. 
In a local coordinate $z$ we sometimes write this field as $T^z$;
when the coordinate is clear from context we omit it.
The $T$-$T$ operator product is
\begin{equation} \label{eq:virasoro-ope}
  T(p) T(q) = \frac{c/2}{(z(p)-z(q))^4} + \frac{2 T(q)}{(z(p)-z(q))^2} + \frac{\partial_{z(q)} T(q)}{z(p)-z(q)} + \regular \, .
\end{equation}
Virasoro primary fields $W_h$, for $h \in \bbC$, are defined by the condition\footnote{Here we are considering $q$ to be fixed.
If we consider a family of conformal blocks parameterized by $q$, and require that the family is parallel for the connection on conformal blocks discussed below in \autoref{sec:connections-on-blocks}, then the first-order pole in \eqref{eq:virasoro-primary-ope} determines the variation of the correlation functions with $q$: thus this term is often written as $\frac{\partial_{z(q)} W_h(q)}{z(p)-z(q)}$.}
\begin{equation} \label{eq:virasoro-primary-ope}
  T(p) W_h(q) = \frac{h W_h(q)}{(z(p)-z(q))^2} + \frac{\regular}{z(p) - z(q)} \, .
\end{equation}
The constant $h$ is the \ti{conformal weight} of the primary $W_h$.
Under change of coordinates, $T$ transforms as
\begin{equation} \label{eq:virasoro-coordinate-change}
  T(p)^{z} = \left( \frac{\de w(p)}{\de z(p)} \right)^2 T(p)^{w} + \frac{c}{12} \{w,z\} \, ,
\end{equation}
where $\{\cdot,\cdot\}$ denotes the Schwarzian derivative:
$\{f(z),z\} = \frac{f'''(z)}{f'(z)} - \frac32 \left( \frac{f''(z)}{f'(z)} \right)^2$.

\item The Heisenberg vertex algebra $\Heis$ is generated by one field $J$, with the OPE relation
\begin{equation} \label{eq:heisenberg-ope}
  J(p) J(q) = \frac{1}{(z(p)-z(q))^2} + \regular.
\end{equation}
Its primary fields $V_\alpha$, for $\alpha \in \bbC$, are defined by the condition
\begin{equation} \label{eq:heisenberg-primary-ope}
  J(p) V_\alpha(q) = \frac{\alpha V_\alpha(q)}{z(p) - z(q)} + \regular .
\end{equation}
Under change of coordinates, $J$ transforms as
\begin{equation} \label{eq:heisenberg-coordinate-change}
      J(p)^{z} = \left(\frac{\de w(p)}{\de z(p)}\right) J(p)^{w} \, . 
\end{equation}
The Heisenberg algebra contains a Virasoro algebra of central charge $1$, with generator 
$T^{\Heis}$ given by\footnote{Here and below, the ``normal ordering'' symbol $\nop{\cdots}$ means a specific 
way of regulating a singular OPE: we split points, expand in a local coordinate, 
drop the polar part and then take the limit of
coincident points; for example, here
$\nop{J(p)^2}^z$ means $\lim_{p' \to p} J(p')^z J(p)^z - \frac{1}{(z(p)-z(p'))^2}$.}
\begin{equation} \label{eq:t-heis}
T^{\Heis}(p) = \frac12 \nop{J(p)^2} \, .
\end{equation}
Restricting attention to $T^\Heis$, the primary field $V_\alpha$ for $\Heis$
becomes a $\Vir_{c=1}$ primary $W_{\frac{\alpha^2}{2}}$.

\item We will also consider the combined vertex algebra $\Vir_{c}\otimes\Heis$, generated by fields $J$ and $T$ as above,
with no singularity in the $J$-$T$ operator product.
Primary insertions for $\Vir_{c}\otimes\Heis$ can be written as $V_\alpha W_h$, with $\alpha, h \in \bbC$.

It will be convenient to consider the total Virasoro algebra with central charge $c_\tot = c + 1$,
\begin{equation} \label{eq:virasoro-total}
  T^{\tot}(p) = T(p)+T^{\Heis}(p) \, .
\end{equation}
Restricting attention to $T^\tot$, 
the primary field $V_\alpha W_h$ for $\Vir_{c}\otimes\Heis$ becomes a $\Vir_{c+1}$ primary
$W_{h+\frac{\alpha^2}{2}}$.

\end{itemize}

\subsection{Conformal blocks}

By a \ti{conformal block} we mean a system of correlation functions obeying chiral Ward identities. The space of conformal blocks, written $\Conf(C, \mathcal{V}; \cdots)$, is a canonically defined vector space, depending only on the data of a vertex algebra $\mathcal{V}$ and a Riemann surface $C$, plus the list $\cdots$ 
of primary field insertions at marked points of $C$ (if any).
In this paper, the main players will be $\Conf(C,\Heis)$ and $\Conf(C,\Vir_{c=1} \otimes \Heis)$. We give a quick reminder about them here; more details can be found in \autoref{app:conformal-blocks}.

A conformal block $\Psi\in\Conf(C,\Heis)$ is a system of correlation functions 
\begin{equation} \label{eq:heisenberg-correlator}
\IP{J(p_1)^{z_1}\cdots J(p_n)^{z_n}}_{\Psi}
\end{equation}
defined for all $n \ge 0$. For each $i$, $p_i$ is a point of $C$,
and $z_i$ is a local holomorphic 
coordinate on a chart containing $p_i$.
The correlation functions \eqref{eq:heisenberg-correlator} are required to be 
holomorphic away from the diagonals
$p_i = p_j$, with the singularities at the diagonal governed by \eqref{eq:heisenberg-ope}. The behavior of the correlation functions 
under changes of the local
coordinate systems $z_i$ is controlled by \eqref{eq:heisenberg-coordinate-change}.

Similarly, a conformal block $\Psi\in\Conf(C,\Vir_{c=1} \otimes \Heis)$ consists of correlation functions
\begin{equation} \label{eq:heisenberg-virasoro-correlator}
\IP{T(p_1)^{z_1}\cdots T(p_n)^{z_n} J(q_1)^{w_1} \cdots J(q_m)^{w_m}}_{\Psi} \, ,
\end{equation}
which are holomorphic away from $p_i = p_j$ or $q_i = q_j$,
obey the OPEs \eqref{eq:virasoro-ope}, \eqref{eq:heisenberg-ope}, and obey the coordinate transformation rules
\eqref{eq:virasoro-coordinate-change}, \eqref{eq:heisenberg-coordinate-change}.

We will also need to define spaces of conformal blocks with
primary fields inserted. We define
$\Conf(C,\Heis;V_{\alpha_1}(q_1) \cdots V_{\alpha_k}(q_k))$ to
be the space of systems of correlation functions 
\begin{equation} \label{eq:heisenberg-correlator-with-primaries}
\IP{J(p_1)^{z_1}\cdots J(p_n)^{z_n} V_{\alpha_1}(q_1) \cdots V_{\alpha_k}(q_k)}_{\tPsi}
\end{equation}
with the same OPE and coordinate
transformations for the $J$ insertions as before, but 
now with extra first-order poles when any $p_i$ meets any $q_j$, as dictated 
by \eqref{eq:heisenberg-primary-ope}.\footnote{Note that, if $C$ is compact, 
and $\sum_{i=1}^k \alpha_i \neq 0$, then all correlation functions must vanish: this is the law of charge conservation, which one can show concretely using the fact that
the sum of residues of a meromorphic $1$-form on $C$ is always zero. Thus we will only be interested in the
case when $\sum_{i=1}^k \alpha_i = 0$.}
Similarly we use \eqref{eq:heisenberg-primary-ope} 
and \eqref{eq:virasoro-primary-ope} to define
$\Conf(C,\Vir_{c=1} \otimes \Heis;V_{\alpha_1}W_{h_1}(q_1) \cdots V_{\alpha_k}W_{h_k}(q_k))$.

In these definitions the primary insertions are held fixed,
and we do not fix coordinate systems around them; 
thus we are treating the primary insertions
unsymmetrically from the vertex algebra generators $J$ and $T$.
We will discuss a more symmetrical version in \autoref{sec:connections-on-blocks}
below.

\subsection{Connections on conformal block spaces} \label{sec:connections-on-blocks}

The space $\Conf(C, \cV)$ depends on the Riemann surface $C$. 
As $C$ varies, these spaces make up a bundle
$\Conf(\cdot, \cV)$ over the moduli space $\cM_g$ of Riemann surfaces.
As we now recall, choosing a Virasoro generator $T$ inside $\cV$ equips $\Conf(\cdot, \cV)$ with a 
twisted connection $\nabla$ (see e.g. \cite{MR0869564,MR2082709}.)

Suppose given a family $\Psi$ of conformal blocks over $\cM_g$.
A tangent vector to $\cM_g$ at $C$ can be represented by an infinitesimal
Beltrami differential, $\mu \in \Omega^{0,1}(TC)$.
We write $\mu^z$ for the local coordinate expression of $\mu$, i.e. $\mu = \mu^z \partial_z \de \overline{z}$.
The covariant derivative of $\Psi$ along $\mu$ is given by
\begin{equation} \label{eq:T-connection-rough}
  \IP{\cdots}_{\nabla_\mu \Psi} = \partial_\mu \left( \IP{\cdots}_{\Psi} \right) - \frac{1}{2 \pi \I} \int_C \mu(p)^z \IP{T(p)^z \cdots}_\Psi \de z \de \overline{z} 
\end{equation}

There is a subtlety to address here.
The product $\mu^z T^z \de z \de \overline{z}$ is not coordinate-invariant,
because of the Schwarzian derivative term in \eqref{eq:virasoro-coordinate-change}. It is invariant only
under M\"obius transformations.
Thus, the right side in \eqref{eq:T-connection-rough} depends on the choice of an atlas of holomorphic 
charts on $C$ related by M\"obius transformations. Such an atlas is also known as a complex projective 
structure on $C$. 
So the bundle $\Conf(\cdot, \cV)$ 
acquires a connection $\nabla$ only after choosing a section $S$ 
of a bundle over $\cM_g$, whose fiber over $C$ is the space of 
complex projective structures on $C$.
Any two complex projective structures differ by a holomorphic quadratic differential
$\phi = \{w,z\} \de z^2$, where $w$ and $z$ are coordinates
in the two atlases. Changing $S \to S + \phi$
shifts $T \to T + \frac{c}{12} \phi$, and thus shifts $\nabla_\mu$ by
a multiple of the identity operator,
\begin{equation}
  \nabla^{S+\phi}_\mu = \nabla^S_\mu - \left( \frac{c}{24 \pi \I} \int_C \mu^z(p) \phi^z(p) \, \de z \de \overline{z} \right) \Id  \, .
\end{equation}
This is what we mean by saying that $\nabla$ is a twisted connection in $\Conf(\cdot, \cV)$ over $\cM_g$.

When we have primary fields $P_i$ inserted, the space $\Conf(C,\cV;P_1(p_1) \cdots P_n(p_n))$
depends also on the points $p_i$, so now we have a bundle 
$\Conf(C,\cV;P_1(\cdot) \cdots P_n(\cdot))$
over $C^n \setminus \Delta$, where $\Delta$ 
is the locus where some insertions collide.
This bundle has a twisted connection given by
\begin{equation} \label{eq:point-connection-rough}
  \IP{\cdots P(p)}_{\nabla_{z(p)} \Psi} = \partial_{z(p)} \left( \IP{\cdots P(p)}_{\Psi} \right) - \frac{1}{2 \pi \I} \oint_p \IP{\cdots T(q)^z P(p)}_\Psi \de z(q) \, .
\end{equation}
A short calculation using \eqref{eq:virasoro-primary-ope}
and \eqref{eq:virasoro-coordinate-change} shows that
changing the choice of local coordinate around $p$, from $z$ to $w(z)$, changes the connection $\nabla$ by
\begin{equation} \label{eq:connection-shift}
\nabla^z = \nabla^w - h_P \, \de \log (\de w(p) / \de z(p)) \cdot \Id
\end{equation}
where $h_P$ is the conformal weight of the insertion $P(p)$.
This is what we mean by saying that $\nabla$ is a twisted connection in $\Conf(C,\cV;P_1(\cdot) \cdots P_n(\cdot))$
over $C^n \setminus \Delta$.

Here is a variant, which treats the primary insertions
more symmetrically with the vertex algebra generators, at the cost of depending on 
more auxiliary data. Suppose 
\begin{equation}
h_P = a/b \in \bbQ  
\end{equation}
and we have a holomorphic line bundle $\cL$ over $C$ with an isomorphism
\begin{equation}
 \cL^b \simeq K_C^a \, .  
\end{equation}
Then we can consider systems of correlation functions
where the dependence on the primary $P(p)$ is $\cL(p)$-valued, i.e. consider elements of 
$\Conf(C,\cV;P(p)\cdots) \otimes \cL(p)$.\footnote{A useful way of thinking of this is to say
that we consider correlation functions which
depend on a choice of local coordinate $z$ around the point $p$
where $P(p)$ is inserted, 
changing by a factor $(\de w(p) / \de z(p))^{h_P}$
when we change coordinates from $w$ to $z$; this is parallel to the
coordinate dependence we have for the vertex algebra insertions, but with
the extra complication that $h_P$ is not an integer.
From that point of view, the line bundle $\cL$ is
being used to choose a branch of the fractional exponent.}
An important virtue of these line-bundle-valued conformal blocks is that
$\Conf(C,\cV;P(\cdot)) \otimes \cL$ 
has an actual connection over $C$, not a twisted connection.
The explicit expression of this connection is again given by \eqref{eq:point-connection-rough},
now with the understanding that the correlation functions are written relative
to a trivialization of $\cL(p)$ by some choice of $(\de z(p))^{h_P}$;
then, when we change coordinates, the shift \eqref{eq:connection-shift}
is compensated by the change of trivialization of $\cL(p)$.

We can do similarly for multiple 
insertions, obtaining a bundle 
$\Conf(C,\cV;P_1(\cdot) \cdots P_n(\cdot)) \otimes \boxtimes_{i=1}^n \cL_i$
with connection over $C^n \setminus \Delta$.\footnote{We use the symbol $\boxtimes$ for the outer tensor product of vector bundles: 
if $V$ and $W$ are vector bundles over $X$ and $Y$ respectively, then $V \boxtimes W$ is the vector bundle over $X \times Y$ whose fibers are $V_x \otimes W_y$. Here each $\cL_i$ is a vector bundle over $C$, and $\boxtimes_{i=1}^n \cL_i$ is a vector bundle
over $C^n$.}

\subsection{Free fermion correlators} \label{sec:free-fermions}

Now consider $\cV = \Heis$, and consider
primary insertions $V_{\pm 1}$, 
which we also denote as $\psi_\pm$ (free fermions).
As we will now explain, a conformal block for $\Heis$
canonically determines conformal blocks 
with these primaries inserted.
We briefly summarize the key properties of these blocks 
here, deferring the verifications to
\autoref{app:unfusion-heisenberg}.

Fix points $p_i$, $q_i$ in $C$.
Choose a homotopy class of path $\ell_i$
from $q_i$ to $p_i$, for each $i$.
Also fix a spin
structure $K^{\frac12}$ on a neighborhood of each $\ell_i$.
(One way to do this would be to fix a global spin structure
on $C$, but it will be convenient to be a little more flexible.)
We call the path equipped with the chosen 
spin structure a \ti{leash}, and just denote it
$\ell_i$.
Finally, fix a block $\Psi \in \Conf(C,\Heis)$.
Then there is a canonical induced block 
\begin{equation}
\Psi_{\ell_1,\dots,\ell_n} \in \Conf(C,\Heis;\psi_+(p_1) \cdots \psi_+(p_n) \psi_-(q_1) \cdots \psi_-(q_n)) \otimes \prod_{i=1}^n \left( K^{\frac12}(p_i) \otimes K^{\frac12}(q_i) \right)  \, .
\end{equation}

\begin{center}
\begin{tikzpicture}[
    point/.style={circle, fill, inner sep=1.5pt},
    arrow/.style={->,>={Latex[bend]},shorten >=1pt,thick},
    auto
]

\draw[smooth,thick,\baseCurveColor] (0,1) to[out=30,in=150] (2,1) to[out=-30,in=210] (3,1) to[out=30,in=150] (5,1) to[out=-30,in=30] (5,-1) to[out=210,in=-30] (3,-1) to[out=150,in=30] (2,-1) to[out=210,in=-30] (0,-1) to[out=150,in=-150] (0,1);
\draw[smooth,thick,\baseCurveColor] (0.4,0.1) .. controls (0.8,-0.25) and (1.2,-0.25) .. (1.6,0.1);
\draw[smooth,thick,\baseCurveColor] (0.5,0) .. controls (0.8,0.2) and (1.2,0.2) .. (1.5,0);
\draw[smooth,thick,\baseCurveColor] (3.4,0.1) .. controls (3.8,-0.25) and (4.2,-0.25) .. (4.6,0.1);
\draw[smooth,thick,\baseCurveColor] (3.5,0) .. controls (3.8,0.2) and (4.2,0.2) .. (4.5,0);

\def\pOne{(2.8,0.2)}
\def\pTwo{(1.45,0.8)}
\def\pThree{(0.5,-0.9)}
\def\pFour{(2.85,-0.25)}
\def\pFive{(4.5,-1)}
\def\pSix{(4.5,1)}

\node[point,\psiplusColor,label=above:$\psi_+(p_1)$] (p1) at \pOne {};
\node[point,\psiplusColor,label=left:$\psi_+(p_2)$] (p2) at \pTwo {};
\node[point,\psiplusColor,label=right:$\psi_+(p_3)$] (p3) at \pThree {};
\node[point,\psiminusColor,label=below:$\psi_-(q_2)$] (q2) at \pFour {};
\node[point,\psiminusColor,label=above:$\psi_-(q_3)$] (q3) at \pFive {};
\node[point,\psiminusColor,label=below:$\psi_-(q_1)$] (q1) at \pSix {};

\draw[leash] (p1) to[bend right=5] (q1);
\draw[leash] (p2) to[bend left=10] (q2);
\draw[leash] (p3) to[bend left=50] (q3);

\end{tikzpicture}
\end{center}
Said otherwise, given the block $\Psi$ there is a
canonical definition of correlation functions
\begin{equation} \label{eq:fermion-correlator}
\left\langle
J(r_1)^{z_1} \cdots J(r_k)^{z_k} \,
\tikz[remember picture, baseline]{
  \node[anchor=base, inner sep=0pt] (psiplus1) {{$\psi_+(p_1)^{w_1}$}};
}
\,
\tikz[remember picture, baseline]{
  \node[anchor=base, inner sep=0pt] (psiminus1) {{$\psi_-(q_1)^{y_1}$}};
}
\, \raisebox{0.0ex}{$\cdots$} \,
\tikz[remember picture, baseline]{
  \node[anchor=base, inner sep=0pt] (psiplusN) {{$\psi_+(p_n)^{w_n}$}};
}
\,
\tikz[remember picture, baseline]{
  \node[anchor=base, inner sep=0pt] (psiminusN) {{$\psi_-(q_n)^{y_n}$}};
}
\right\rangle_\Psi \, .
\end{equation}
\begin{tikzpicture}[overlay, remember picture]
  \coordinate (start1) at ($(psiplus1.south)!0.5!(psiplus1.north)$);
  \coordinate (end1) at ($(psiminus1.south)!0.5!(psiminus1.north)$);
  \coordinate (start1Shifted) at ([yshift=-1.8ex]start1); 
  \coordinate (end1Shifted) at ([yshift=-1.8ex]end1);
  \draw[thick] (start1Shifted) -- (end1Shifted);
  \draw[thick] (start1Shifted) -- ++(0,0.5ex);
  \draw[thick] (end1Shifted) -- ++(0,0.5ex);

  \coordinate (mid1) at ($(start1Shifted)!0.5!(end1Shifted)$);
  \node at ($(mid1)+(0,-1.8ex)$) {$\ell_1$}; 

  \coordinate (startN) at ($(psiplusN.south)!0.5!(psiplusN.north)$);
  \coordinate (endN) at ($(psiminusN.south)!0.5!(psiminusN.north)$);
  \coordinate (startNShifted) at ([yshift=-1.8ex]startN); 
  \coordinate (endNShifted) at ([yshift=-1.8ex]endN);
  \draw[thick] (startNShifted) -- (endNShifted);
  \draw[thick] (startNShifted) -- ++(0,0.5ex);
  \draw[thick] (endNShifted) -- ++(0,0.5ex);

  \coordinate (midN) at ($(startNShifted)!0.5!(endNShifted)$);
  \node at ($(midN)+(0,-1.8ex)$) {$\ell_n$}; 
\end{tikzpicture}

\noindent The correlation functions \eqref{eq:fermion-correlator}
depend on local coordinate patches $z_i$, $w_i$, $y_i$
around the insertions, and on leashes $\ell_i$\footnote{There is no difference whether the leashes are drawn above or below the text.}, 
as indicated. They also depend on a discrete choice of
square roots of $\de w_i$ and $\de y_i$ for each $i$,
not indicated explicitly in the notation.
Under a change of local coordinate and square root
around a fermion, the correlators transform according to the rule
\begin{equation} \label{eq:fermion-coordinate-transformation}
  \psi_\pm(q)^w = \frac{\sqrt{\de w'(q)}}{\sqrt{\de w(q)}} \psi_\pm(q)^{w'} \, .
\end{equation}
They can be given by an explicit construction (\ti{fermionization}): when $p$, $q$, and the leash $\ell$
are all contained in a single patch with local coordinate $z$
and a choice of $\sqrt{\de z}$, we write
\Needspace{4\baselineskip}
\begin{equation}\label{eq:heisenberg-unfusion}
  \tikz[remember picture, baseline]{
    \node[anchor=base, inner sep=0pt] (psiplus) {$\psi_+(p)^z$};
  }
  \,
  \tikz[remember picture, baseline]{
    \node[anchor=base, inner sep=0pt] (psiminus) {$\psi_-(q)^z$};
  }
  =
  \frac{1}{z(p)-z(q)} \ \nop{\exp \int_\ell J \, } 
\end{equation}
\begin{tikzpicture}[overlay, remember picture]
  \coordinate (start) at (psiplus.north);
  \coordinate (end) at (psiminus.north);
  \coordinate (startShifted) at ([yshift=1ex]start);
  \coordinate (endShifted) at ([yshift=1ex]end);
  \draw[thick] (startShifted) -- (endShifted);
  \draw[thick] (startShifted) -- ++(0,-0.5ex);
  \draw[thick] (endShifted) -- ++(0,-0.5ex);
  \coordinate (mid) at ($(startShifted)!0.5!(endShifted)$);
  \node at ($(mid)+(0,1.5ex)$) {$\ell$};
\end{tikzpicture}\noindent\unskip
where we define the normal-ordered exponential by
\begin{equation}
  \nop{\exp \int_\ell J \, } =
  1 + \int_\ell \de z(r) J(r)^z  + \frac{1}{2} \int_\ell \int_\ell \de z(r_1) \de z(r_2) \left(J(r_1)^z J(r_2)^z - \frac{1}{(z(r_1)-z(r_2))^2} \right) + \cdots \, .
\end{equation}
This formula expresses the desired correlators
\eqref{eq:fermion-correlator} in terms of
correlators involving only $J$.
Direct computation 
shows that the resulting correlators have the requisite
analytic properties; in particular, they have a first-order
pole when a $J$ insertion meets one of the $\psi_\pm$
insertions.

Now, consider the Virasoro 
generator $T^\Heis$ from \eqref{eq:t-heis}.
With respect to $T^\Heis$, 
the primary insertions $\psi_\pm$ have 
conformal weight $h = \frac12$.
Thus, following the general discussion in 
\autoref{sec:connections-on-blocks}, 
there is a connection in the bundle over 
$C^{2n} \setminus \Delta$ where
the blocks $\Psi_{\ell_1, \dots, \ell_n}$ lie.
In fact, these blocks are covariantly constant for this
connection. This condition amounts (using 
\eqref{eq:point-connection-rough} and \eqref{eq:t-heis}) to the explicit equation
\begin{equation} \label{eq:fermion-cc-condition}
  \partial_z \psi_\pm(q) = \pm \nop{J \psi_\pm(q)} \, .
\end{equation}

Finally we consider the OPE between fermions $\psi_+$
and $\psi_-$.
When $\ell$ is a short path from $p$ to $q$, 
\eqref{eq:heisenberg-unfusion} immediately gives the 
$p \to q$ expansion
\begin{equation} \label{eq:fermion-OPE}
  \tikz[remember picture, baseline]{
    \node[anchor=base, inner sep=0pt] (psiplus) {\raisebox{0.15ex}{$\psi_+(p)^z$}};
  }
  \,
  \tikz[remember picture, baseline]{
    \node[anchor=base, inner sep=0pt] (psiminus) {\raisebox{0.15ex}{$\psi_-(q)^z$}};
  }
  = \frac{1}{z(p)-z(q)} + J(q) + \cdots \, .
\end{equation}
\begin{tikzpicture}[overlay, remember picture]
  \coordinate (start) at ($(psiplus.south)!0.5!(psiplus.north)$);
  \coordinate (end) at ($(psiminus.south)!0.5!(psiminus.north)$);
  \coordinate (startShifted) at ([yshift=-1.8ex]start);
  \coordinate (endShifted) at ([yshift=-1.8ex]end);
  \draw[thick] (startShifted) -- (endShifted);
  \draw[thick] (startShifted) -- ++(0,0.5ex);
  \draw[thick] (endShifted) -- ++(0,0.5ex);

  \coordinate (mid) at ($(startShifted)!0.5!(endShifted)$);
  \node at ($(mid)+(0,-1.5ex)$) {$\ell$};
\end{tikzpicture}

\noindent When we bring together
two fermions which are not connected by a leash,
we have as $p_2 \to q_1$ the relation
\bigskip
\Needspace{4\baselineskip} 

\begin{equation} \label{eq:fermion-OPE-general}
  \tikz[remember picture, baseline]{
    \node[anchor=base, inner sep=0pt] (psi3plus1) {$\psi_+(p_1)$};
  }
  \,
  \tikz[remember picture, baseline]{
    \node[anchor=base, inner sep=0pt] (psi3minus1) {$\psi_-(q_1)^z$};
  }
  \,\,
  \tikz[remember picture, baseline]{
    \node[anchor=base, inner sep=0pt] (psi3plus2) {$\psi_+(p_2)^z$};
  }
  \,
  \tikz[remember picture, baseline]{
    \node[anchor=base, inner sep=0pt] (psi3minus2) {$\psi_-(q_2)$};
  }
  =
  - \frac{
    \tikz[remember picture, baseline]{
      \node[anchor=base, inner sep=0pt] (psi3plus1rhs) {$\psi_+(p_1)$};
    }
    \,
    \tikz[remember picture, baseline]{
      \node[anchor=base, inner sep=0pt] (psi3minus2rhs) {$\psi_-(q_2)$};
    }
  }{z(p_2) - z(q_1)}
  + \regular \, ,
\end{equation}
\begin{tikzpicture}[overlay, remember picture]
  \coordinate (start31) at ($(psi3plus1.south)!0.5!(psi3plus1.north)$);
  \coordinate (end31) at ($(psi3minus1.south)!0.5!(psi3minus1.north)$);
  \coordinate (start31Shifted) at ([yshift=-2ex]start31);
  \coordinate (end31Shifted) at ([yshift=-2ex]end31);
  \draw[thick] (start31Shifted) -- (end31Shifted);
  \draw[thick] (start31Shifted) -- ++(0,0.5ex);
  \draw[thick] (end31Shifted) -- ++(0,0.5ex);
  \coordinate (mid31) at ($(start31Shifted)!0.5!(end31Shifted)$);
  \node at ($(mid31)+(0,-1.5ex)$) {$\ell_1$};

  \coordinate (start32) at ($(psi3plus2.south)!0.5!(psi3plus2.north)$);
  \coordinate (end32) at ($(psi3minus2.south)!0.5!(psi3minus2.north)$);
  \coordinate (start32Shifted) at ([yshift=-2ex]start32);
  \coordinate (end32Shifted) at ([yshift=-2ex]end32);
  \draw[thick] (start32Shifted) -- (end32Shifted);
  \draw[thick] (start32Shifted) -- ++(0,0.5ex);
  \draw[thick] (end32Shifted) -- ++(0,0.5ex);
  \coordinate (mid32) at ($(start32Shifted)!0.5!(end32Shifted)$);
  \node at ($(mid32)+(0,-1.5ex)$) {$\ell_2$};

  \coordinate (start3rhs) at ($(psi3plus1rhs.north)!0.5!(psi3plus1rhs.south)$);
  \coordinate (end3rhs) at ($(psi3minus2rhs.north)!0.5!(psi3minus2rhs.south)$);
  \coordinate (start3rhsShifted) at ([yshift=2ex]start3rhs);
  \coordinate (end3rhsShifted) at ([yshift=2ex]end3rhs);
  \draw[thick] (start3rhsShifted) -- (end3rhsShifted);
  \draw[thick] (start3rhsShifted) -- ++(0,-0.5ex);
  \draw[thick] (end3rhsShifted) -- ++(0,-0.5ex);
  \coordinate (mid3rhs) at ($(start3rhsShifted)!0.5!(end3rhsShifted)$);
  \node at ($(mid3rhs)+(0,1.5ex)$) {$\ell_1 + \ell_2$};
\end{tikzpicture}

\noindent where to build the spin structure on the leash
$\ell_1 + \ell_2$ we use the isomorphism
between the spin structures on $\ell_1$ and $\ell_2$
determined by the chosen square roots of $\de z$.

\section{The branched free-field construction} \label{sec:branched-free-field}

Suppose that we have a surface $C$ and a smooth branched double cover
\begin{equation}
  \pi: \tC \to C \, .
\end{equation}
It is known (e.g. \cite{MR0897030,MR0869937,Gavrylenko:2015cea})
that, beginning with a conformal block $\tPsi \in \Conf(\tC,\Heis)$,
one can produce a conformal block for $\Vir_{c=1}$ on $C$,
with an insertion of a 
primary $W_{\frac{1}{16}}$ at each of the branch points $b_1$, \dots, $b_k$
of the covering $\pi$.
The method is a 
$\bbZ_2$-twisted version of the usual free-field
construction of $\Vir_{c=1}$.
We will call it the \ti{branched free-field} construction.
In the rest of this section we review how it works.

In the version of the story which we will discuss,
the two sheets of $\tC$ give us locally
two free Heisenberg fields on $C$ rather than one; 
related to this, 
we will get blocks for $\Vir_{c=1} \otimes \Heis$ rather than 
$\Vir_{c=1}$.
Thus, altogether, we will describe a linear map
\begin{equation} \label{eq:branched-free-field-map}
 \nab_0: \quad \Conf(\tC,\Heis) \quad \to \quad \Conf\left(C,\Vir_{c=1} \otimes \Heis; W_{\frac{1}{16}}(b_1) \cdots W_{\frac{1}{16}}(b_k) \right) \, .
\end{equation}
Note that this involves Heisenberg fields both on $\tC$ and on $C$. From now on,
in an effort to reduce confusion, we 
use $\tJ$ for the Heisenberg generator on $\tC$,
and $J$ for the one on $C$.

\subsection{The basic dictionary}

Practically speaking, 
giving a map \eqref{eq:branched-free-field-map} 
means giving a recipe for correlation functions
\eqref{eq:heisenberg-virasoro-correlator} on $C$,
in terms of correlation functions \eqref{eq:heisenberg-correlator}
on $\tC$.

The main ingredient in this recipe is a dictionary at the level of the local operators. 
Consider a point $p \in C$, which is not a branch point 
of $\pi$, and let $p^{(1)}$, $p^{(2)}$ be its two preimages
in $\tC$.
The dictionary is:
\begin{equation} \label{eq:nab-operator-dictionary}
  J(p) \rightsquigarrow \frac{1}{\sqrt2} \left( \tJ(p^{(1)}) + \tJ(p^{(2)}) \right) \, , \qquad T(p) \rightsquigarrow \frac14 \nop{(\tJ(p^{(1)}) - \tJ(p^{(2)}))^2} \, .
\end{equation}
This dictionary is to be understood as holding in correlation functions.
We emphasize that these formulas are invariant under the interchange
$p^{(1)} \leftrightarrow p^{(2)}$, and thus they do not depend on our
local choice of how to label the two sheets of $\tC$.

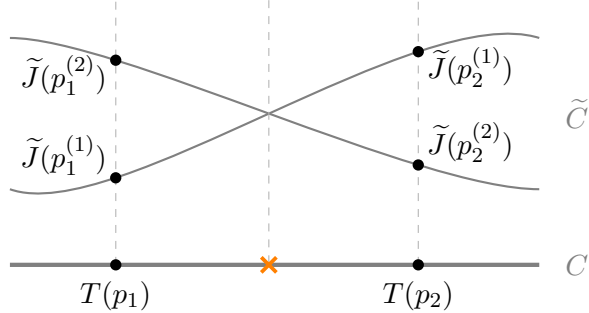
\begin{center}
\begin{tikzpicture}
\usetikzlibrary{intersections}

\newcommand{\firstCol}{1.9}
\newcommand{\secondCol}{5.9}
\newcommand{\curveLowerY}{1.0}
\newcommand{\curveUpperY}{3.0}
\newcommand{\startX}{0.5}
\newcommand{\finalX}{7.5}
\newcommand{\curveHeight}{2.0}
\newcommand{\dashedHeight}{3.5}

\draw[ultra thick,\baseCurveColor] (\startX,0) -- (\finalX,0) node[right] {};

\draw[dashed,\vdashLineColor,name path=vertical1] (\firstCol,0) -- (\firstCol,\dashedHeight);
\draw[dashed,\vdashLineColor,name path=vertical2] (\secondCol,0) -- (\secondCol,\dashedHeight);

\draw[name path=curve1,thick,\coveringCurveColor] (\startX,\curveLowerY) .. controls (\firstCol,\curveHeight-1.5) and (\secondCol,\curveHeight+1.5) .. (\finalX,\curveUpperY);
\draw[name path=curve2,thick,\coveringCurveColor] (\startX,\curveUpperY) .. controls (\firstCol,\curveHeight+1.0) and (\secondCol,\curveHeight-1.0) .. (\finalX,\curveLowerY);

\path [name intersections={of=curve1 and curve2, by=branch}];
\coordinate (branchb) at (branch |- 0,0);

\draw[dashed,\vdashLineColor] (branch |- 0,\dashedHeight) -- (branchb);

\node[dot,label=below:{\color{\tpColor}\( T(p_1) \)}] (Tp1) at (\firstCol,0) {};
\node[dot,label=below:{\color{\tpColor}\( T(p_2) \)}] (Tp2) at (\secondCol,0) {};

\path [name intersections={of=vertical1 and curve1, by=Jp11}];
\path [name intersections={of=vertical1 and curve2, by=Jp12}];
\path [name intersections={of=vertical2 and curve1, by=Jp21}];
\path [name intersections={of=vertical2 and curve2, by=Jp22}];

\node[dot,label={[xshift=1mm, yshift=3mm]left:{\color{\jpColor}\(  \tJ(p_1^{(1)}) \)}}] at (Jp11) {};
\node[dot,label={[xshift=1mm, yshift=-2mm]left:{\color{\jpColor}\(  \tJ(p_1^{(2)}) \)}}] at (Jp12) {};
\node[dot,label={[xshift=-1mm, yshift=-2mm]right:{\color{\jpColor}\( \tJ(p_2^{(1)}) \)}}] at (Jp21) {};
\node[dot,label={[xshift=-1mm, yshift=3mm]right:{\color{\jpColor}\( \tJ(p_2^{(2)}) \)}}] at (Jp22) {};

\drawbranchpointmarker{branchb};


\node at (\finalX+0.5,0) {\color{\baseCurveColor}\( C \)};
\node at (\finalX+0.5,\curveHeight) {\color{\coveringCurveColor}\( \tC \)};

\end{tikzpicture}
\end{center}

More precisely: take a coordinate
disc $z: U \to \bbC$ around $p$, small enough that
$\pi^{-1}(U)$ is the union of two discs $U^{(1)}$, $U^{(2)}$ 
in $\tC$, containing the preimages $p^{(1)}$, $p^{(2)}$ of $p$. 
Each of these discs
inherits a local coordinate $z^{(i)}: U^{(i)} \to \bbC$,
given by 
\begin{equation}
z^{(i)} = z \circ \pi \, .
\end{equation}
Then, the correlation functions of the block $\nab_0(\tPsi)$
in the coordinate $z$ 
are defined to be the correlation functions of
the block $\tPsi$ in the coordinates
$z^{(i)}$,
 using the dictionary \eqref{eq:nab-operator-dictionary} to match up the operators.

So, for example, the 0-point function of $\nab_0(\tPsi)$ is the same as
that of $\tPsi$,
\begin{equation}
  \IP{1}_{\nab_0(\tPsi)} = \IP{1}_\tPsi,
\end{equation}
and the 2-point function of Virasoro generators $T$ 
in the block $\nab_0(\tPsi)$ on $C$ is
a combination of 4-point functions of Heisenberg generators $\tJ$
in the block $\tPsi$ on $\tC$,
\begin{equation}
  \IP{T(p)^z T(q)^w}_{\nab_0(\tPsi)} = \frac{1}{16} \IP{ \nop{(\tJ(p^{(1)})^{z^{(1)}} - \tJ(p^{(2)})^{z^{(2)}})^2} \, \nop{(\tJ(q^{(1)})^{w^{(1)}} - \tJ(q^{(2)})^{w^{(2)}})^2}  }_\tPsi \, .
\end{equation}

The correlation functions $\IP{\cdots}_{\nab_0(\tPsi)}$ 
have the desired
short-distance singularities, as long as all insertions 
are away from the branch points of $\pi$.
Indeed, as far as the local singularities are concerned, 
$\tJ(p^{(1)})$ and $\tJ(p^{(2)})$
are decoupled from one another: we could equally well think of them as
two fields $\tJ^{(1)}(p)$ and $\tJ^{(2)}(p)$ on $C$.
Changing basis to $\tJ^{(\pm)} = \frac{1}{\sqrt2} (\tJ^{(1)} \pm \tJ^{(2)})$,
each of $\tJ^{(\pm)}$ again has the OPE \eqref{eq:heisenberg-ope} and thus 
generates a copy of $\Heis$, and there is no singularity in the OPE between $\tJ^{(+)}$ and $\tJ^{(-)}$.
Our dictionary \eqref{eq:nab-operator-dictionary} then becomes
\begin{equation} \label{eq:nab-operator-dictionary-decoupled}
  J \rightsquigarrow \tJ^{(+)} \, , \qquad T \rightsquigarrow \frac12 \nop{(\tJ^{(-)})^2} \, .
\end{equation}
By a short calculation
it follows that $J$ and $T$ obey the OPEs \eqref{eq:heisenberg-ope} and \eqref{eq:virasoro-ope} 
of $\Heis$ and $\Vir_{c=1}$ respectively, that 
there is no singularity in the OPE between $J$ and $T$, and 
that $J$ and $T$ also obey the coordinate transformation laws
\eqref{eq:heisenberg-coordinate-change} and \eqref{eq:virasoro-coordinate-change} (with $c=1$).
This is essentially the same calculation one makes in the standard
free-field construction of $\Vir_{c=1}$ from $\Heis$.

\subsection{Singularities at branch points} \label{sec:branch-point-singularities}

We are ready to consider what happens at the branch points. Our computation will 
be similar to one in \cite{Gavrylenko:2015cea,Bershtein:2017lmg}.

It is convenient to calculate for the total Virasoro generator $T^\tot$ instead of $T$,
and then deduce the behavior of $T$ afterward.
We need to be careful about local coordinate systems.
Let $w$ be a local coordinate on $C$
which vanishes at a branch point $b$.
Then choose $p$ near $b$, and an open
$U \subset C$ containing $p$, such that $w \vert_U$ has a single-valued root $z = \sqrt{w}$,
and $b$ is in the closure of $U$. ($b$ cannot be in $U$, since $\sqrt{w}$ exists on $U$;
having $b$ in the closure of $U$ is the next best thing.)
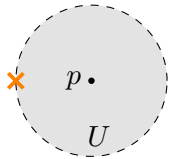
\begin{center}
\begin{tikzpicture}   
    \draw[fill=gray!22,dashed](0,0) circle (1.0) node [black,yshift=-1.5cm]{};
    \drawbranchpointmarker{-1,0};
    \draw[fill=black](0,0) circle (1 pt) node [left] {$p$};
    \draw[fill=white](0.1,-0.73)  node [] {$U$};
\end{tikzpicture}
\end{center}
Now we apply the rule, following from \eqref{eq:nab-operator-dictionary} and \eqref{eq:virasoro-total},
\begin{equation}
  T^\tot(p)^z \rightsquigarrow \frac12 \nop{(\tJ(p^{(1)})^{z^{(1)}})^2 + (\tJ(p^{(2)})^{z^{(2)}})^2} \, .
\end{equation}
The right side is a sum of two terms, each of which is finite because of the normal ordering.\footnote{If we studied $T$ instead of $T^\tot$, our life would be slightly harder at this moment, because there would be a cross-term which would have a singularity as $p \to b$; this is why we compute for $T^\tot$ instead.}
The local coordinates $z^{(1)}$ and $z^{(2)}$ both extend to good coordinate systems in $\tC$ 
including the point $\pi^{-1}(b)$. Finally,
there are no operators inserted at $\pi^{-1}(b)$.
We conclude that the right side is bounded 
as $p \to b$, and thus
$T^\tot(p)^z$ is bounded as $p \to b$.

This does not yet tell us what we really want to know, because
the coordinate $z$ on $U \subset C$ does not extend to a
coordinate on a neighborhood of $b$; for that we need to return
to the coordinate $w = z^2$. Using
the rule \eqref{eq:virasoro-coordinate-change}
at $c=2$ and the relation $\{w,z\} = -\frac32 \frac{1}{z^2}$ gives
\begin{equation}
  T^\tot(p)^{z} = (2z(p))^2 T^\tot(p)^w + \frac{1}{6} \left(-\frac32 \frac{1}{z(p)^2} \right) \, ,
\end{equation}
so finally
\begin{equation} \label{eq:T-branch-point-contributions-add}
  T^\tot(p)^w = \frac{1}{(2z(p))^2} \left(  \frac{1}{4 z(p)^2} + \regular \right) = \frac{1}{16 z(p)^4} + \frac{\regular}{z(p)^2} = \frac{1}{16 w(p)^2} + \frac{\regular}{w(p)} \, ,
\end{equation}
so $T^\tot(p)^w$ 
has a second-order pole at the branch point $b$,
with coefficient $\frac{1}{16}$.

We should also consider the behavior of $J(p)$ near the branch point.
We switch to the common coordinate system $z^{(1)}$ for both insertions
on $\tC$, again using the fact that this coordinate extends
over a neighborhood of $\pi^{-1}(b)$. 
Then we have
\begin{equation}
  J(p)^z \rightsquigarrow \tJ(p^{(1)})^{z^{(1)}} - \tJ(p^{(2)})^{z^{(1)}} \, ,
\end{equation}
with the relative minus sign coming
from the fact that $z^{(2)} = -z^{(1)}$, so $\de z^{(2)} / \de z^{(1)} = -1$.
It follows that $J(p)^z$ vanishes (to first order in $z$) as 
$z \to 0$. Changing coordinates to $w = z^2$ using
$J(p)^z = (2z(p)) J(p)^w$, we find that $J(p)^w$ is regular at $b$.

Finally, having the behavior of both $T^\tot$ and $J$ at the branch point,
we can deduce the behavior of $T = T^\tot - T^\Heis$: it has
\begin{equation} \label{eq:T-singularity-branch-point}
  T(p)^w = \frac{1}{16 w(p)^2} + \frac{\regular}{w(p)} \, .
\end{equation}
The interpretation of \eqref{eq:T-singularity-branch-point} 
is that $\nab_0(\tPsi)$ is a conformal block 
with the primary field $W_{\frac{1}{16}}$ inserted at each
branch point, as we claimed at the beginning of this 
section.

\subsection{Inserting primaries}

Now we briefly discuss the extension of the branched free-field construction to 
include insertions of additional primary fields. This is relatively straightforward: a $\Vir_{c=1} \otimes \Heis$
primary inserted at $p \in C$ corresponds to a 
combination of $\Heis$ primaries inserted at $p^{(1)}, p^{(2)} \in \tC$. 
This leads to generalizations of \eqref{eq:branched-free-field-map}
with additional primary fields inserted on both sides of the map.

First suppose we fix $\beta \in \bbC$, and consider 
an insertion of the $\Vir_{c=1}$ primary $W_{\beta^2}(p)$ on $C$. 
We require that $p$ is not a branch point of $\pi$. 
Then we introduce a dictionary extending \eqref{eq:nab-operator-dictionary}: 
\begin{equation} \label{eq:W-dictionary}
  W_{\beta^2}(p) \ \rightsquigarrow \ V_\beta(p^{(i)}) V_{-\beta}(p^{(j)})
\end{equation}
with either choice of the sheet ordering $(i,j)$.
This dictionary is engineered to produce the expected
analytic properties as the insertions $J(q)$ or $T(q)$ approach
$W_{\beta^2}(p)$. For instance, 
using \eqref{eq:nab-operator-dictionary} and \eqref{eq:W-dictionary}
we have
\begin{equation}
  J(q) W_{\beta^2}(p) \ \rightsquigarrow \ \frac{1}{\sqrt2}\left(\tJ(q^{(1)}) + \tJ(q^{(2)})\right) \, V_\beta(p^{(i)}) V_{-\beta}(p^{(j)})
\end{equation}
and note using 
\eqref{eq:heisenberg-primary-ope}
that the right side is regular as $q \to p$
(the singular contributions from $\beta$ and $-\beta$ cancel),
matching the expectation that $J(q) W_{\beta^2}(p)$ is regular
as $q \to p$.
Similarly, we have
\begin{equation}
  T(q) W_{\beta^2}(p) \ \rightsquigarrow \ \frac14 \, \nop{\left(\tJ(q^{(1)}) - \tJ(q^{(2)})\right)^2} \, V_\beta(p^{(i)}) V_{-\beta}(p^{(j)})
\end{equation}
and in a local coordinate $z$ 
the right side has a singularity with leading term
$\frac{\beta^2}{(z(p)-z(q))^2}$, as expected.
This is more or less the standard free-field construction
of primaries for $\Vir_{c=1}$.

The dictionary
\eqref{eq:W-dictionary} involves correlated 
insertions on both sheets of $\tC$. We can also 
work with insertions on only one sheet of $\tC$:
these correspond to primaries on $C$ which are charged 
under both factors of $\Vir_{c=1} \otimes \Heis$.
Namely, consider the primary field
\begin{equation} \label{eq:chi-def}
  \chi_\beta = W_{\beta^2} V_{\sqrt{2} \beta}
\end{equation}
on $C$.
By similar computations to the above, we can check that
this insertion can be obtained using the dictionary
\begin{equation} \label{eq:chi-dictionary}
  \chi_\beta(p) \ \rightsquigarrow \ V_{2 \beta}(p^{(i)})
\end{equation}
for either choice of $i$.

One important case
is $\beta = \pm \frac12$. This corresponds to the simplest
degenerate primary for $\Vir_{c=1}$, with weight $h = \frac14$. 
The realization \eqref{eq:W-dictionary} of $W_{\frac14}$ involves insertions of 
primaries $V_{\frac12}$ and $V_{-\frac12}$ on the two sheets of $\tC$.
The realization \eqref{eq:chi-dictionary} of $\chi_{\pm \frac12} = W_{\frac14} V_{\pm 1/\sqrt2}$,
with both $\Vir_{c=1}$ and $\Heis$ charge,
involves a single insertion of $V_{\pm 1} = \psi_{\pm}$ 
on one of the two sheets.

In the case of a degenerate insertion, we can ask whether the conformal blocks
which we obtain by this dictionary are really degenerate in the sense that they satisfy 
the null-vector constraint. It turns
out that they do. Let us check this explicitly in the case of $\beta = \pm \frac12$,
using the dictionary \eqref{eq:W-dictionary} (so that we are just using the Virasoro
algebra, with no Heisenberg part on the base).
The null-vector constraint in this case is that
$(L_{-2} - L_{-1}^2) W_{\frac14}$ should be zero in correlation functions.\footnote{Here and below we use a coordinate patch
on $C$ in which the insertion is placed at $z = 0$, and the usual mode expansions 
$T(p)^z = \sum\limits_{n \in \bbZ} L_n z(p)^{-n-2}$, $J(p)^z = \sum\limits_{n \in \bbZ} J_n z(p)^{-n-1}$.}
We check this as follows: our dictionary gives
\begin{align}
L_{-2} W_{\beta^2} & \rightsquigarrow \left[ \frac12 \tJ^{(-)}_{-1} \tJ^{(-)}_{-1} + \tJ^{(-)}_{-2} \tJ^{(-)}_{0} \right] V^{(i)}_\beta V^{(j)}_{-\beta} \\
&= \left[ \frac12 (\tJ^{(-)}_{-1})^2 + \sqrt2 \beta \tJ^{(-)}_{-2} \right] V^{(i)}_\beta V^{(j)}_{-\beta}
\end{align}
and
\begin{align}
L_{-1}^2 W_{\beta^2} & \rightsquigarrow \left[ \tJ^{(-)}_{-2}\tJ^{(-)}_{1} + \tJ^{(-)}_{-1}\tJ^{(-)}_{0} \right] \left[ \tJ^{(-)}_{-1}\tJ^{(-)}_{0} \right]  V^{(i)}_\beta V^{(j)}_{-\beta}  \\
&= \left[ 2\beta^2 (\tJ^{(-)}_{-1})^2  + \sqrt2 \beta \tJ^{(-)}_{-2} \right]  V^{(i)}_\beta V^{(j)}_{-\beta} \, .
\end{align}
Subtracting these two and choosing
$\beta = \pm \frac12$ we obtain
\begin{equation}
  (L_{-2} - L_{-1}^2) W_{\frac14} \rightsquigarrow 0 \, .
\end{equation}

\subsection{Walls as branched screening contours} \label{sec:walls}

As we reviewed in \autoref{sec:free-fermions},
given a block
$\tPsi \in \Conf(\tC, \Heis)$, we can define 
correlation functions on $\tC$ with 
free-fermion insertions.
We are now going to define a specific sort of free-fermion
insertion which is essentially topological in nature.

We consider a contour $\cG$ on $C$, 
with a bit of extra discrete data: 
\begin{itemize}
\item An orientation of $\cG$.
\item A spin structure $K_C^{\frac12}$
in a neighborhood of $\cG$.
\item
A labeling of the two sheets of $\tC$ over $\cG$ 
by $\pm$.
\item For each $q \in \cG$, 
a choice of a leash $\ell_\cG(q)$ on $\tC$
running from $q^{(-)}$ to $q^{(+)}$, not passing
through any branch points of $\pi: \tC \to C$, 
equipped with the spin structure $\pi^* K_C^{\frac12}$.
The leash $\ell_\cG(q)$ must depend continuously on $q$.
\end{itemize}
The contour $\cG$ 
equipped with this extra data is called a \ti{wall}.
Given a wall $\cG$ we define an extended operator, built from free fermions on $\tC$ lying over $\cG$:\footnote{The definition \eqref{eq:bilocal-wall-operator} involves a
local coordinate $z$ around $\cG$ and a choice of 
square-root $\sqrt{\de z}$, which then induces choices of
$\sqrt{\de z^{(+)}}$ and $\sqrt{\de z^{(-)}}$. Happily, 
using \eqref{eq:fermion-coordinate-transformation} we see
that this dependence cancels out in correlation functions
involving $W(\cG)$. Indeed the coordinate $z$ does not even
need to exist globally around $\cG$; we could use different
coordinates on different parts of $\cG$ if that is more
convenient.}
\begin{equation} \label{eq:bilocal-wall-operator}
W(\cG) = \int_\cG 
\tikz[remember picture, baseline]{
  \node[anchor=base, inner sep=0pt] (psiplus) {$\psi_+(q^{(+)})^{z^{(+)}}$};
}
\, 
\tikz[remember picture, baseline]{
  \node[anchor=base, inner sep=0pt] (psiminus) {$\psi_-(q^{(-)})^{z^{(-)}}$};
}
\, \de z(q) \, .
\end{equation}
\begin{tikzpicture}[overlay, remember picture]
  \coordinate (start) at ($(psiplus.south)!0.5!(psiplus.north)$);
  \coordinate (end) at ($(psiminus.south)!0.5!(psiminus.north)$);
  \coordinate (startShifted) at ([yshift=-2ex]start);
  \coordinate (endShifted) at ([yshift=-2ex]end);
  \draw[thick] (startShifted) -- (endShifted);
  \draw[thick] (startShifted) -- ++(0,0.5ex);
  \draw[thick] (endShifted) -- ++(0,0.5ex);
  \coordinate (mid) at ($(startShifted)!0.5!(endShifted)$);
  \node at ($(mid)+(0,-1.5ex)$) {$\ell_\cG(q)$};
\end{tikzpicture}

We emphasize that, although the wall 
$\cG$ lies on the base $C$,
the insertion points $q^{(+)}$, $q^{(-)}$ lie on $\tC$.
From the point of view of $C$, $W(\cG)$ appears like an ordinary
line defect; from the point of view of $\tC$, it is a bit exotic,
in the sense that it is the integral of a bilocal expression rather than a local one. 

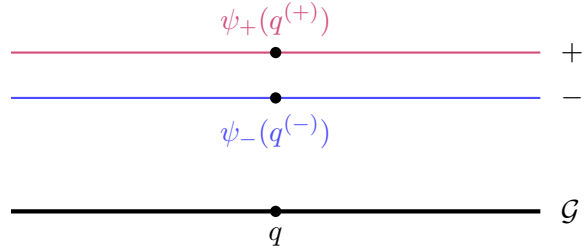
\begin{center}
\begin{tikzpicture}

\newcommand{\leng}{7}

\coordinate (bl) at (0,0);
\coordinate (br) at (\leng,0);
\coordinate (bc) at (0.5*\leng,0);
\coordinate (c2l) at (0,1.5);
\coordinate (c2r) at (\leng,1.5);
\coordinate (c2c) at (0.5*\leng,1.5);
\coordinate (c1l) at (0,2.1);
\coordinate (c1r) at (\leng,2.1);
\coordinate (c1c) at (0.5*\leng,2.1);

\node [label=right:$\cG$] at (br) {};
\node [label=right:$+$] at (c1r) {};
\node [label=right:$-$] at (c2r) {};

\draw [ultra thick,\wallColor] (bl) -- (br);  
\draw [psipluscontour] (c1l) -- (c1r);
\draw [psiminuscontour] (c2l) -- (c2r);

\node[dot,label=below:{\color{black}\( q \)}] (q) at (bc) {};
\node[dot,label=above:{\color{\psiplusColor}\( \psiplus(q^{(+)}) \)}] (q1) at (c1c) {};
\node[dot,label=below:{\color{\psiminusColor}\( \psiminus(q^{(-)}) \)}] (q2) at (c2c) {};

\end{tikzpicture}
\end{center}

From holomorphy of the
correlation functions it follows that $W(\cG)$ is
topological, in the sense that it is invariant under 
deformations of $\cG$ which do not cross any other insertions
(with fixed endpoints if $\cG$ is open).
In fact, more is true: $\cG$ can be moved freely across insertions 
of $J$ or $T$ on $C$. Indeed, when we bring the contour $\cG$ close
to an insertion $J(p)$, we have
\begin{equation}
  J(p) \psiplus(q^{(+)}) \psiminus(q^{(-)}) \rightsquigarrow \frac{1}{\sqrt2} \left(\tJ(p^{(1)}) + \tJ(p^{(2)})\right) \psiplus(q^{(+)}) \psiminus(q^{(-)}) \, ,
\end{equation}
and the right side is actually regular at $q = p$ (the two first-order poles cancel one another).
For a $T$ insertion the story is a bit more interesting: we have
\begin{equation} \label{eq:T-wall-total-deriv}
\begin{split} 
  T(p) & \psiplus(q^{(+)}) \psiminus(q^{(-)}) \rightsquigarrow \frac{1}{4} \nop{\left(\tJ(p^{(1)}) - \tJ(p^{(2)})\right)^2} \, \psiplus(q^{(+)}) \psiminus(q^{(-)}) \\
  &= \frac{\psiplus(q^{(+)}) \psiminus(q^{(-)})}{(z(p)-z(q))^2} + \frac{\nop{\tJ \psiplus(q^{(+)})} \psiminus(q^{(-)})}{z(p)-z(q)} - \frac{ \psiplus(q^{(+)}) \nop{\tJ \psiminus(q^{(-)})}}{z(p)-z(q)} + \regular \\
  &= \partial_{z(q)} \left( \frac{\psiplus(q^{(+)}) \psiminus(q^{(-)})}{z(p)-z(q)} \right) + \regular \, .
\end{split}
\end{equation}
This implies that, although the integrand in $W(\cG) T(p)$
can have a pole at $q = p$, this pole has zero residue.
Thus the contour $\cG$ 
can be freely deformed across $p$.

Altogether, then,
correlation functions involving $J(p) W(\cG)$ or $T(p) W(\cG)$ do
not have singularities when $p$ meets the interior of 
$\cG$.
If $\cG$ has no endpoints, this means the singularity structure of correlation
functions of $J$'s and $T$'s is not disturbed by the insertion of $W(\cG)$.
It follows that we can modify the branched free-field construction, by
inserting $W(\cG)$ for any wall $\cG$ without boundary.
This leads to a new map
\begin{equation} \label{eq:branched-free-field-map-with-walls}
 \nab_\cG: \quad \Conf(\tC,\Heis) \quad \to \quad \Conf\left(C,\Vir_{c=1} \otimes \Heis; W_{\frac{1}{16}}(b_1) \cdots W_{\frac{1}{16}}(b_k) \right) \, .
\end{equation}
The map $\nab_\cG$ depends only on the homotopy class of the contour $\cG$.

We remark that $W(\cG)$ is similar to the \ti{screening contours} which appear in the 
free-field construction of Virasoro blocks \cite{DOTSENKO1985691} (see e.g. \cite{DiFrancesco:1997nk} for an account). 
Indeed, in the special case where $\tC$ is actually
a trivial cover $\tC = C \sqcup C$, the branched free-field construction would reduce
to the ordinary free-field construction, and $W(\cG)$ would reduce essentially to a screening contour.
In that case the insertion of $W(\cG)$ makes a particularly drastic difference: the $\psi_\pm$ insertions 
are on different connected components of $\tC$, and so they shift the total Heisenberg charge on
each component by $\pm 1$.
This leads to the familiar fact that the free-field construction can only 
produce Virasoro blocks for which the conformal weights take certain discrete values, such that
the total Heisenberg charge on each component is an integer. Moreover, that integer then determines
how many screening contours $W(\cG)$ need to be inserted if we want the correlators to be nonzero.
In contrast, when $\tC$ is a smooth cover with nontrivial branching ---
the case we will usually consider --- the insertion of $W(\cG)$ does not change the total
Heisenberg charge, and so the number of $W(\cG)$ insertions is not fixed. Indeed the main construction of this paper involves inserting an \ti{exponential} of
$W(\cG)$, and all terms in the expansion of this 
exponential generally contribute
to the correlation functions.

In the presence of primary field insertions we can also 
take a wall $\cG$ with both ends on primary field insertions,
instead of a closed loop.
To see whether this makes sense, we should ask whether
the integral \eqref{eq:bilocal-wall-operator} defining $W(\cG)$ is convergent.
The covariant-constancy equation $\partial_{z(p)} \psi_\pm(p)^z = \pm \nop{J \psi_\pm(p)}^z$,
combined with the singular behavior of $J$ near an insertion $V_\alpha(q)$,
implies the power-law behavior
\begin{equation}
 \psi_\pm(p)^z V_\alpha(q) \sim (z(p)-z(q))^{\pm\alpha} V_\alpha(q)
\end{equation}
as $p \to q$.
It follows that \eqref{eq:bilocal-wall-operator} is indeed convergent, 
provided that any wall which ends on an insertion with $\re \alpha > 0$ is labeled
$+$, and any wall which ends on an insertion with $\re \alpha < 0$ is labeled $-$.
Moreover, using \eqref{eq:T-wall-total-deriv} we can compute the contribution from the wall to the
singular part of $T(p)$ as $p \to q$: it is 
$\frac{\psi_+(q^{(+)}) \psi_-(q^{(-)})}{z(p)-z(q)}$,
which vanishes under the same assumption on the insertions.
Thus we conclude that walls ending on primary
insertions do not alter the singularity
of $T$ at the insertion, so our dictionary for primary insertions 
is not affected by the insertion of $W(\cG)$.

The properties of the modified branched free-field maps
$\nab_\cG$ might be interesting
to investigate, but that is not our main purpose here.
In the next section we will instead insert
$W(\cG)$ for walls $\cG$ which end on the branch points.

\section{The nonabelianization map} \label{sec:nonabelianization}

The main new idea of this paper is that one can modify
the branched free-field construction in a way which eliminates the $W_{\frac{1}{16}}$
insertions at branch points, without creating any extra 
singularities anywhere else.
Thus we will obtain a linear map
\begin{equation} \label{eq:nab-map}
 \nab_\cW: \quad \Conf(\tC,\Heis) \quad \to \quad \Conf(C,\Vir_{c=1} \otimes \Heis) \, ,
\end{equation}
or with primary fields inserted,
\begin{equation} \label{eq:nab-map-with-primaries}
 \nab_\cW: \quad \Conf(\tC,\Heis;V_{\beta_1}(p^{(i_1)}_1) \cdots V_{\beta_k}(p^{(i_k)}_k)) \quad \to \quad \Conf(C,\Vir_{c=1} \otimes \Heis ; \chi_{\beta_1}(p_1) \cdots \chi_{\beta_k}(p_k)) \, .
\end{equation}
We call $\nab_\cW$ the nonabelianization map for conformal blocks.

\subsection{Spectral networks} \label{sec:nonabelianization-map}

We recall from \cite{MR3115984,Freed:2022yae} the
notion of spectral network (for $\fgl(2)$).

As in the previous section, we consider a smooth branched double cover $\pi: \tC \to C$.
A spectral network $\cW$
subordinate to $\pi$ is a collection of walls on $C$.
We consider the generic situation: each branch point is an endpoint of 
exactly $3$ walls, meeting
at an angle $\frac{2 \pi}{3}$, with the sheet labelings $+$, $-$ over the walls alternating as indicated in the figure.
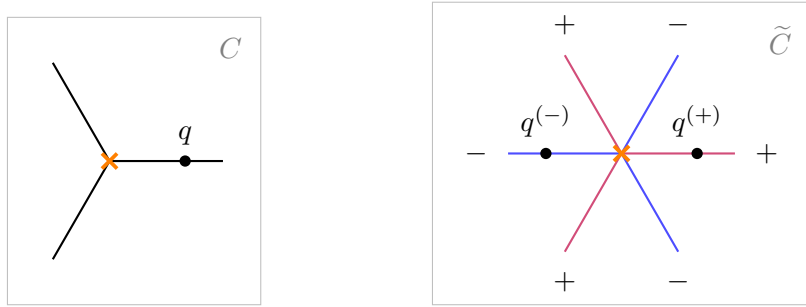
\begin{center}
\begin{tikzpicture}
  \begin{scope}
    \coordinate (center) at (0,0);

    \draw [thick,\wallColor] (center) -- ++(120:1.5cm); 
    \draw [thick,\wallColor] (center) -- ++(240:1.5cm); 
    \draw [thick,\wallColor] (center) -- ++(0:1.5cm);   

    \drawbranchpointmarker{center};

    \node[dot,label=above:{\color{black}\( q \)}] (q) at (1,0) {};

    \node at (1.6,1.5) {\color{\baseCurveColor}\( C \)};
  \end{scope}
  \draw [lightgray, thin] (-1.35,-1.9) rectangle (2,1.9);
\end{tikzpicture}
\hspace{2cm}
\begin{tikzpicture}
  \begin{scope}
    \coordinate (center) at (0,0);

    \draw [psipluscontour] (center) -- ++(120:1.5cm);   
    \draw [psipluscontour] (center) -- ++(240:1.5cm); 
    \draw [psipluscontour] (center) -- ++(0:1.5cm);
    \draw [psiminuscontour] (center) -- ++(60:1.5cm);   
    \draw [psiminuscontour] (center) -- ++(180:1.5cm); 
    \draw [psiminuscontour] (center) -- ++(300:1.5cm);

    \node [label=above:$+$] at ++(120:1.5cm) {};
    \node [label=below:$+$] at ++(240:1.5cm) {};
    \node [label=right:$+$] at ++(0:1.5cm) {};
    \node [label=above:$-$] at ++(60:1.5cm) {};
    \node [label=left:$-$] at ++(180:1.5cm) {};
    \node [label=below:$-$] at ++(300:1.5cm) {};

    \drawbranchpointmarker{center};

    \node[dot,label=above:{\color{black}\( q^{(+)} \)}] at (1,0) {};
    \node[dot,label=above:{\color{black}\( q^{(-)} \)}] at (-1,0) {};

    \node at (2.1,1.5) {\color{\coveringCurveColor}\( \tC \)};
  \end{scope}
  \draw [lightgray, thin] (-2.5,-2) rectangle (2.5,2);
\end{tikzpicture}
\end{center}

The walls may be half-infinite, running around the surface $C$ forever, or they may end on insertions of
primaries. If they end on primaries, then we require that the preimages 
labeled $+$ end on $V_\alpha$ with $\re \alpha > 0$, and preimages labeled $-$ 
end on $V_\alpha$ with $\re \alpha < 0$, as we discussed in \autoref{sec:walls}.

We choose a spin structure $K_C^{\frac12}$
over a neighborhood of $\cW$.
We also choose a leash $\ell_\cG(q)$ 
for each point $q$ of a 
wall $\cG$; $\ell_\cG(q)$ runs from $q^{(-)}$
along the wall to the branch point, 
follows a semicircle around the branch point,
then goes back out along the wall to $q^{(+)}$.
The semicircle we pick is dictated by the orientation of the wall:
we turn 90 degrees right starting from the positive direction
on the wall.
The figure below shows the leash we take if the wall is oriented outward
from the branch point.\footnote{We now have two ingredients in $W(\cW)$ 
which use an orientation of the wall: we use the orientation in defining the integral along the wall, and also in determining the leash. These two dependences cancel one another, so in the end $W(\cW)$ does not depend on the orientation we choose for the wall.}
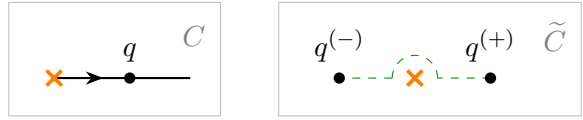
\begin{center}
\begin{tikzpicture}
  \begin{scope}
    \coordinate (center) at (0,0);

    \node[dot,label=above:{\color{black}\( q \)}] at (1,0) {};

    \draw [thick,\wallColor,-Stealth] (center) -- (0.65,0);
    \draw [thick,\wallColor] (0.5,0) -- (1.8,0);

    \drawbranchpointmarker{center};

    \node at (1.85,0.55) {\color{\coveringCurveColor}\( C \)};
  \end{scope}
  \draw [lightgray, thin] (-0.6,-0.5) rectangle (2.2,1.0);
\end{tikzpicture}
\hspace{0.5cm}
\begin{tikzpicture}
  \begin{scope}
    \coordinate (center) at (0,0);

    \draw[leash] (1,0) -- (0.3,0);

    \draw[leash] (0.3,0) arc (0:180:0.3);

    \draw[leash] (-0.3,0) -- (-1,0);

    \drawbranchpointmarker{center};

    \node[dot,label=above:{\color{black}\( q^{(+)} \)}] at (1,0) {};
    \node[dot,label=above:{\color{black}\( q^{(-)} \)}] at (-1,0) {};

    \node at (1.85,0.55) {\color{\coveringCurveColor}\( \tC \)};

  \end{scope}
  \draw [lightgray, thin] (-1.8,-0.5) rectangle (2.2,1.0);
\end{tikzpicture}
\end{center}
One natural way to get a double cover $\pi: \tC \to C$ and a
subordinate spectral network would be to use a meromorphic
quadratic differential on $C$; this is how they arose in \cite{MR3115984}.
In this paper we do not require that $\tC$ and $\cW$ should arise in this way,
but our construction will be particularly well behaved if they do:
see \autoref{sec:image} below.

\subsection{Defining the nonabelianization map} \label{sec:defining-nab-map}

Fix a choice of a 
spectral network $\cW$ subordinate to $\pi$.

Now we can describe the nonabelianization map.
We apply the branched free-field dictionary 
\eqref{eq:nab-operator-dictionary} 
as before, but in addition we
insert in every correlation function the operator 
\begin{equation}
E(\cW) = \exp\left(\frac{1}{2 \pi \I} W(\cW)\right) \, ,
\end{equation}
where $W(\cW)$ was defined in \eqref{eq:bilocal-wall-operator}. 

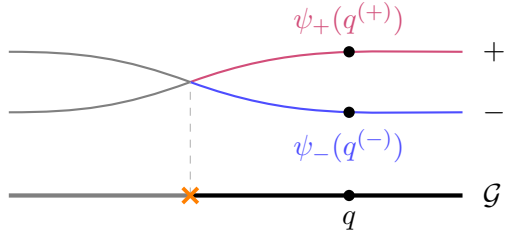
\begin{center}
\begin{tikzpicture}

\newcommand{\leng}{6}

\coordinate (bl) at (0,0);
\coordinate (br) at (\leng,0);
\coordinate (brh) at (0.75*\leng,0);
\coordinate (c1l) at (0,1.1);
\coordinate (c2r) at (\leng,1.1);
\coordinate (c2rh) at (0.75*\leng,1.1);
\coordinate (c2l) at (0,1.9);
\coordinate (c1r) at (\leng,1.9);
\coordinate (c1rh) at (0.75*\leng,1.9);
\coordinate (branch) at (0.4*\leng,1.5);
\coordinate (branchb) at (0.4*\leng,0);

\draw [ultra thick,gray] (bl) -- (branchb);
\draw [ultra thick,\wallColor] (branchb) -- (br);  
\draw [thick,gray] (c1l) to[out=0,in=200 ] (branch);
\draw [psipluscontour] (branch) to[out=20,in=180 ] (c1r);  
\draw [psiminuscontour] (c2r) to[out=180,in=-20] (branch); 
\draw [thick,gray] (branch) to[out=160,in=0 ] (c2l);
\draw [dashed,\vdashLineColor] (branchb) -- (branch);

\drawbranchpointmarker{branchb};

\node [label=right:$\cG$] at (br) {};
\node [label=right:$+$] at (c1r) {};
\node [label=right:$-$] at (c2r) {};

\node[dot,label=below:{\color{black}\( q \)}] (q) at (brh) {};
\node[dot,label=above:{\color{\psiplusColor}\( \psiplus(q^{(+)}) \)}] (q1) at (c1rh) {};
\node[dot,label=below:{\color{\psiminusColor}\( \psiminus(q^{(-)}) \)}] (q2) at (c2rh) {};

\end{tikzpicture}
\end{center}

There are various issues which have to be understood. 
The most urgent question is whether
the insertion of $E(\cW)$ really makes sense. \ti{A priori},
computing correlation functions with this insertion means doing an infinite sum of iterated integrals, and one could worry about convergence, both for the individual integrals and for their sum.
Indeed, there is a clear possibility of trouble, 
because the points $q^{(+)}$ and $q^{(-)}$ in \eqref{eq:bilocal-wall-operator}
come together as $q$ approaches the branch point at the beginning of each wall. This gives a logarithmic 
divergence in every term of the sum.

To understand this issue we choose a local coordinate function 
$y$ around each branch point on $C$ and a
parameter $\eps > 0$, and cut off
the integrals \eqref{eq:bilocal-wall-operator} at a distance $\abs{y} = \eps^2$; call the resulting cutoff wall operators
$W_\eps(\cW)$.
Correlation functions with $W_\eps(\cW)$ inserted instead
of $W(\cW)$ are no longer divergent, but they have unwanted
singularities, both at the branch points and at the endpoints
of the cutoff contours.
Our interest is in taking the limit
$\eps \to 0$.

Now comes a key point: we claim that correlation functions 
involving $\exp \left( \frac{1}{2 \pi \I} W_\eps(\cW)\right)$ vanish
like $\eps^{\frac{k}{8}}$ as $\eps \to 0$, where $k$ is the number of branch points.
Since this assertion only concerns 
what happens in the neighborhood of a branch point, we can prove it
by studying a simple model example. 
We do this in \autoref{sec:model-example} below.
With this behavior in mind we 
define renormalized spectral network operators by
\begin{equation} \label{eq:renormalized-correlator}
 E_\ren(\cW) = \lim_{\eps \to 0} \eps^{-\frac{k}{8}} \exp\left(\frac{1}{2 \pi \I} W_\eps(\cW)\right) \, . 
\end{equation}
Correlation functions involving $E_\ren(\cW)$
are well defined and (generically) nonzero.
The renormalized operator $E_\ren(\cW)$ 
is topological away from the
branch points, but not at the branch points: 
using \eqref{eq:renormalized-correlator} we see that it depends on the choice of 
local coordinate $y$ around each branch point $b$, with scaling dimension $-\frac{1}{16}$, i.e.
\begin{equation}
  E^y_\ren(\cW) = \left\lvert \frac{\de y'(b)}{\de y(b)} \right\rvert^{- \frac{1}{16}} E^{y'}_\ren(\cW)  \, .
\end{equation}
This is the first encouraging sign that our construction
may work: indeed, an insertion at each branch point with 
dimension $-\frac{1}{16}$ is
just what is needed to cancel the $\frac{1}{16}$ 
we had in the branched free-field construction.

Next we consider the analytic properties of the 
correlators, as functions of the insertion points.
As we noted above, when the operator $W(\cW)$ is inserted, 
the correlation functions of $T$ and $J$ do not develop
any extra singularities in the interior of the 
contours $\cW$; it follows that the same is true 
when $E_\ren(\cW)$ is inserted.
What remains is to see what happens at the branch points.
We claim that after the insertion
of $E_\ren(\cW)$ there are no 
singularities in the correlation functions 
of $T$ and $J$ at the 
branch points.
Again we prove this in \autoref{sec:model-example} by studying a simple model example.

Let us summarize. We have given a definition 
of a Virasoro-Heisenberg block $\nab_\cW(\tPsi)$ on $C$,
beginning from a Heisenberg block $\tPsi$ on $\tC$,
using the extra data of a spectral network $\cW$ on $C$ and local coordinates
around branch points.
In the block $\nab_\cW(\tPsi)$, the first few Virasoro correlation functions are
\begin{align} \label{eq:nab-0-point}
  \IP{1}_{\nab_\cW(\tPsi)} &= \IP{E_\ren(\cW)}_{\tPsi}, \\
\label{eq:nab-1-point}
  \IP{T(p)^z}_{\nab_\cW(\tPsi)} &= \frac{1}{4} \IP{ \nop{(\tJ(p^{(1)})^{z^{(1)}} - \tJ(p^{(2)})^{z^{(2)}})^2} \, E_\ren(\cW) }_\tPsi \, , \\
\label{eq:nab-2-point}
  \IP{T(p)^z T(q)^z}_{\nab_\cW(\tPsi)} &= \frac{1}{16} \IP{ \nop{(\tJ(p^{(1)})^{z^{(1)}} - \tJ(p^{(2)})^{z^{(2)}})^2} \, \nop{(\tJ(q^{(1)})^{z^{(1)}} - \tJ(q^{(2)})^{z^{(2)}})^2} \, E_\ren(\cW) }_\tPsi \, .
\end{align}
The $n$-point functions are defined similarly,
using the dictionary \eqref{eq:nab-operator-dictionary} for each operator inserted on $C$,
and inserting the extra operator $E_\ren(\cW)$ in each correlator.
With these definitions, the correlation functions $\IP{\cdots}_{\nab_\cW(\tPsi)}$ have all the expected properties for a conformal block on $C$. We conclude that we have indeed built
a nonabelianization map \eqref{eq:nab-map} 
for conformal blocks, as desired.

\subsection{Compatibility with connections} \label{sec:compatibility}

The dictionary \eqref{eq:nab-operator-dictionary}
takes
\begin{equation}
  T^\tot(p) \ \rightsquigarrow \ \tT^\Heis(p^{(1)}) + \tT^\Heis(p^{(2)}) \, .
\end{equation}
It follows that the nonabelianization map $\nab_\cW$
is compatible with the connections on conformal blocks induced
by $T^\tot$ (on $C$) and $\tT^\Heis$ (on $\tC$), in the following sense.

We will consider variations of $(C, \tC, \pi)$ which are ``even'' under the deck transformation, as follows.
Suppose we have a Beltrami differential $\mu$ on $C$, giving an infinitesimal variation of $C$,
and assume that $\mu$ vanishes around the branch points of the covering $\pi: \tC \to C$ (we can always achieve this 
by a shift $\mu \to \mu + \bar\partial X$ for some $(1,0)$ vector field $X$).
Then we have the pullback Beltrami differential $\pi^* \mu$
on $\tC$, which gives an infinitesimal variation of $\tC$.
The map $\pi: \tC \to C$ remains holomorphic as we simultaneously vary the complex structures 
of $C$ and $\tC$.

Now choose a complex projective structure
$S$ on $C$. Then $\pi^*S$ is a complex projective
structure on $\tC$ (singular at the branch points,
but this will be irrelevant for us since $\pi^* \mu$ vanishes there).
Using these projective structures we can define
covariant derivative operators $\nabla_\mu$
and $\tnabla_{\pi^* \mu}$ on conformal blocks, by the recipe
described in \autoref{sec:connections-on-blocks}, using $T^\tot$ on $C$ and $\tT^\Heis$ on $\tC$.

Chasing through the definitions, we see that the map $\nab_\cW$ intertwines these two covariant derivatives.
In particular, if we have a family of Heisenberg blocks $\tPsi$
which is $\tnabla$-covariantly constant, 
and $(C, \tC, \pi)$ vary by an even variation in the above sense,
then $\Psi = \nab_\cW(\tPsi)$ will be
$\nabla$-covariantly constant.

To get a more complete picture, it
would be useful to consider more general variations of the tuple $(C, \tC, \pi)$.
In this paper, though, we will stick to even variations.

\subsection{Degenerate primaries and nonabelianization of flat connections} \label{sec:degenerate-primaries}

Next we look at how nonabelianization acts on conformal blocks with insertions of degenerate primaries.
In this way we will make a connection to the way spectral networks appeared in \cite{MR3115984,Hollands:2013qza},
and the notion of nonabelianization of flat connections.

Fix a block $\tPsi \in \Conf(\tC, \Heis)$. Also fix $p,q \in C$, with lifts $p^{(i)}$ and $q^{(j)}$ to $\tC$, a leash $\ell$ from $p^{(i)}$ to $q^{(j)}$ on $\tC$, and a spin structure $K_C^{\frac12}$ on $C$. 
Equip $\ell$ with the spin structure $\pi^* K_C^{\frac12}$.
Let 
\begin{equation}
\tPsi_\ell \in \Conf(\tC,\Heis; \psi_+(p^{(i)}) \psi_-(q^{(j)})) \otimes K_C^{\frac12}(p) \otimes K_C^{\frac12}(q)
\end{equation}
denote the free fermion block determined by these data,
as discussed in \autoref{sec:free-fermions}.
Then, consider its nonabelianization:
this is a block on $C$ with degenerate insertions $\chi_{\frac12}(p)
 \chi_{-\frac12}(q)$,
\begin{equation}
\nab_\cW(\tPsi_\ell) \in \Conf(C,\Vir_{c=1} \otimes \Heis; \chi_{\frac12}(p) \chi_{-\frac12}(q)) \otimes K_C^{\frac12}(p) \otimes K_C^{\frac12}(q) \, .  
\end{equation}
The block $\tPsi_\ell$ depends continuously
on the endpoints $p$, $q$, and indeed it is covariantly
constant.
The same is not true for $\nab_\cW(\tPsi_\ell)$: the latter
is covariantly constant away from the walls of $\cW$, but 
discontinuous at the walls. To see this, 
note that expanding out the definition \eqref{eq:bilocal-wall-operator} we
have
\begin{equation}
\tikz[remember picture, baseline]{
  \node[anchor=base, inner sep=0pt] (psiLeftMinus) {$\psi_-(q^{(j)})$};
}
\, 
\tikz[remember picture, baseline]{
  \node[anchor=base, inner sep=0pt] (psiLeftPlus) {$\psi_+(p^{(i)})$};
}
\, \cW(\cG) = \int_{\cG}
\tikz[remember picture, baseline]{
  \node[anchor=base, inner sep=0pt] (psiRightMinus1) {$\psi_-(q^{(j)})$};
}
\,
\tikz[remember picture, baseline]{
  \node[anchor=base, inner sep=0pt] (psiRightPlus1) {$\psi_+(p^{(i)})$};
}
\,
\tikz[remember picture, baseline]{
  \node[anchor=base, inner sep=0pt] (psiRightMinus2) {$\psi_-(r^{(-)})^{z^{(-)}}$};
}
\,
\tikz[remember picture, baseline]{
  \node[anchor=base, inner sep=0pt] (psiRightPlus2) {$\psi_+(r^{(+)})^{z^{(+)}}$};
}
\, \de z(r) \, .
\end{equation}

\begin{tikzpicture}[overlay, remember picture]
  \coordinate (startLeft) at ($(psiLeftMinus.south)!0.5!(psiLeftMinus.north)$);
  \coordinate (endLeft) at ($(psiLeftPlus.south)!0.5!(psiLeftPlus.north)$);
  \coordinate (startLeftShifted) at ([yshift=-2ex]startLeft);
  \coordinate (endLeftShifted) at ([yshift=-2ex]endLeft);
  \draw[thick] (startLeftShifted) -- (endLeftShifted);
  \draw[thick] (startLeftShifted) -- ++(0,0.5ex);
  \draw[thick] (endLeftShifted) -- ++(0,0.5ex);
  \coordinate (midLeft) at ($(startLeftShifted)!0.5!(endLeftShifted)$);
  \node at ($(midLeft)+(0,-1.5ex)$) {$\ell$};

  \coordinate (startRight1) at ($(psiRightMinus1.south)!0.5!(psiRightMinus1.north)$);
  \coordinate (endRight1) at ($(psiRightPlus1.south)!0.5!(psiRightPlus1.north)$);
  \coordinate (startRight1Shifted) at ([yshift=-2ex]startRight1);
  \coordinate (endRight1Shifted) at ([yshift=-2ex]endRight1);
  \draw[thick] (startRight1Shifted) -- (endRight1Shifted);
  \draw[thick] (startRight1Shifted) -- ++(0,0.5ex);
  \draw[thick] (endRight1Shifted) -- ++(0,0.5ex);
  \coordinate (midRight1) at ($(startRight1Shifted)!0.5!(endRight1Shifted)$);
  \node at ($(midRight1)+(0,-1.5ex)$) {$\ell$};

  \coordinate (startRight2) at ($(psiRightPlus2.south)!0.5!(psiRightPlus2.north)$);
  \coordinate (endRight2) at ($(psiRightMinus2.south)!0.5!(psiRightMinus2.north)$);
  \coordinate (startRight2Shifted) at ([yshift=-2ex]startRight2);
  \coordinate (endRight2Shifted) at ([yshift=-2ex]endRight2);
  \draw[thick] (startRight2Shifted) -- (endRight2Shifted);
  \draw[thick] (startRight2Shifted) -- ++(0,0.5ex);
  \draw[thick] (endRight2Shifted) -- ++(0,0.5ex);
  \coordinate (midRight2) at ($(startRight2Shifted)!0.5!(endRight2Shifted)$);
  \node at ($(midRight2)+(0,-1.5ex)$) {$\ell_\cG(r)$};
\end{tikzpicture}

\noindent Now suppose $p$ lies near the contour $\cG$.
For $i=+$ the integrand is regular, but for
$i=-$ it has a first-order pole at $r = p$, 
arising from the singular
OPE between $\psiplus(p^{(-)})$ and $\psiminus(r^{(-)})$.
This pole leads to a discontinuity of the integral 
when $p$ crosses $\cG$,
given by the residue of the integrand at $r = p$.
Using \eqref{eq:fermion-OPE-general} we can compute 
this residue. The result is that when $p$ crosses
$\cG$ from right to left (with respect to the
orientation of $\cG$) we have the additive discontinuity
\begin{equation}
  \disc_{p \in \cG} \nab_\cW(\tPsi_\ell) = \begin{cases} 
  0 & \text{ for } i = +, \\
  \nab_\cW(\tPsi_{\ell+\ell_\cG(p)}) & \text{ for } i = -,
  \end{cases}
\end{equation}

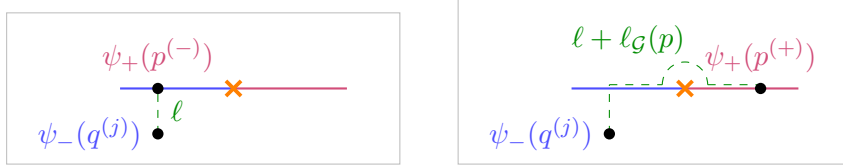
\begin{center}
\begin{tikzpicture}
  \begin{scope}
    \coordinate (center) at (0,0);

    \draw [psipluscontour] (center) -- ++(0:1.5cm);
    \draw [psiminuscontour] (center) -- ++(180:1.5cm);

    \drawbranchpointmarker{center};


    \draw[leash] (-1,-0.6) -- (-1,0);
    \node at (-0.75, -0.3) {\color{\leashColor}\( \ell \)};

    \node[dot,label=above:{\color{\psiplusColor}\( \psiplus(p^{(-)}) \)}] at (-1,0) {};
    \node[dot,label=left:{\color{\psiminusColor}\( \psiminus(q^{(j)}) \)}] at (-1,-0.6) {};

  \end{scope}
  \draw [lightgray, thin] (-3,-1) rectangle (2.2,1);
\end{tikzpicture}
\hspace{0.5cm}
\begin{tikzpicture}
  \begin{scope}
    \coordinate (center) at (0,0);

    \draw [psipluscontour] (center) -- ++(0:1.5cm);
    \draw [psiminuscontour] (center) -- ++(180:1.5cm);

    \drawbranchpointmarker{center};


    \draw[leash] (-1,-0.6) -- (-1,0.05);

    \draw[leash] (-1,0.05) -- (-0.3,0.05);

    \draw[leash] (-0.3,0.05) arc (180:0:0.3);

    \draw[leash] (0.3,0.05) -- (1,0.05);

    \node at (-0.75, 0.65) {\color{\leashColor}\( \ell + \ell_\cG(p) \)};

    \node[dot,label=left:{\color{\psiminusColor}\( \psiminus(q^{(j)}) \)}] at (-1,-0.6) {};
    \node[dot,label=above:{\color{\psiplusColor}\( \psiplus(p^{(+)}) \)}] at (1,0) {};

  \end{scope}
  \draw [lightgray, thin] (-3,-1) rectangle (2.2,1.2);
\end{tikzpicture}

\end{center}

Similarly there is a discontinuity when $q$ meets $\cG$,
\begin{equation}
  \disc_{q \in \cG} \nab_\cW(\tPsi_\ell) = \begin{cases} 
  \nab_\cW(\tPsi_{\ell + \ell_\cG(p)}) & \text{ for } j = +, \\
  0 & \text{ for } j = - \, .
  \end{cases}
\end{equation}

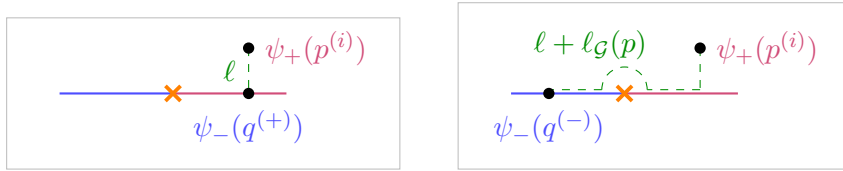
\begin{center}
\begin{tikzpicture}
  \begin{scope}
    \coordinate (center) at (0,0);

    \draw [psipluscontour] (center) -- ++(0:1.5cm);
    \draw [psiminuscontour] (center) -- ++(180:1.5cm);

    \drawbranchpointmarker{center};


    \draw[leash] (1,0.6) -- (1,0);
    \node at (0.75, 0.3) {\color{\leashColor}\( \ell \)};

    \node[dot,label=below:{\color{\psiminusColor}\( \psiminus(q^{(+)}) \)}] at (1,0) {};
    \node[dot,label=right:{\color{\psiplusColor}\( \psiplus(p^{(i)}) \)}] at (1,0.6) {};

  \end{scope}
  \draw [lightgray, thin] (-2.2,-1) rectangle (3,1);
\end{tikzpicture}
\hspace{0.5cm}
\begin{tikzpicture}
  \begin{scope}
    \coordinate (center) at (0,0);

    \draw [psipluscontour] (center) -- ++(0:1.5cm);
    \draw [psiminuscontour] (center) -- ++(180:1.5cm);

    \drawbranchpointmarker{center};


    \draw[leash] (1,0.6) -- (1,0.05);

    \draw[leash] (1,0.05) -- (0.3,0.05);

    \draw[leash] (0.3,0.05) arc (0:180:0.3);

    \draw[leash] (-0.3,0.05) -- (-1,0.05);

    \node at (-0.45, 0.65) {\color{\leashColor}\( \ell + \ell_\cG(p) \)};

    \node[dot,label=right:{\color{\psiplusColor}\( \psiplus(p^{(i)}) \)}] at (1,0.6) {};
    \node[dot,label=below:{\color{\psiminusColor}\( \psiminus(q^{(-)}) \)}] at (-1,0) {};

  \end{scope}
  \draw [lightgray, thin] (-2.2,-1) rectangle (3,1.2);
\end{tikzpicture}

\end{center}

Here is a useful perspective on these discontinuities.
As we have discussed, 
when the degenerate primary insertions are away from $\cW$,
the connections on conformal blocks intertwine
under our dictionary: given free fermion blocks 
$\tPsi$ on $\tC$ which are covariantly constant for $\tnabla$, 
the corresponding blocks $\nab_\cW(\tPsi)$ on $C$
are covariantly constant for $\nabla$.
When the degenerate insertions lie on $\cW$, 
however, the map $\nab_\cW$ is 
not defined. Thus, if we consider a path 
where one of the degenerate insertions crosses $\cW$, 
there is no reason why the parallel transports
of the two connections need to intertwine. 
Rather, what we have just seen is that the 
$\nabla$-parallel transport along a path $\wp$ on $C$ corresponds to a certain
\ti{linear combination} of $\tnabla$-parallel 
transports along paths on $\tC$.
This relation has appeared before: it is the nonabelianization map of \cite{MR3115984,Hollands:2013qza},
which expresses the parallel transport of a connection $\nabla$ on $C$ in terms
of the parallel transport of a corresponding connection $\tnabla$ on $\tC$.
In the context of \cite{MR3115984,Hollands:2013qza}, 
$\nabla$ is a connection of rank $2$ and $\tnabla$ of rank $1$. In that case nonabelianization induces 
a map between moduli spaces of (twisted) local systems,\footnote{
More precisely: the version of abelianization in \cite{MR3115984} applies to $K^{-\frac12}$-twisted connections on both
$C$ and $\tC$; the version in \cite{Hollands:2013qza} applies
to ordinary flat connections on $C$, and to ``almost-flat'' connections on $\tC$, i.e. flat connections except for 
holonomy $-1$ around branch points. 
The two versions of the story can be identified after choosing a spin structure on $C$.}
\begin{equation} \label{eq:nab-classical}
 \nab_\cW^\flat: \cM(\tC, \GL(1)) \to \cM(C, \GL(2)) \, .
\end{equation}
In our present context, $\nabla$ and $\tnabla$ are connections in the infinite-dimensional bundles of conformal blocks, but the 
relation between them is exactly as in \cite{MR3115984,Hollands:2013qza}.

In \autoref{sec:eigenblocks-and-connections} below, we will see that the infinite-dimensional bundles of conformal blocks admit finite-dimensional
subbundles preserved by the connections, and after restricting to those subbundles we recover exactly the story of \cite{MR3115984,Hollands:2013qza}.

\subsection{The image of nonabelianization} \label{sec:image}

We have described a map $\nab_\cW: \Conf(\tC, \Heis) \to \Conf(C, \Vir_{c=1} \otimes \Heis)$ 
for any covering $\tC$ and spectral network $\cW$, without regard for how $\tC$ and $\cW$ are 
constructed. In general, though, 
one cannot expect that $\nab_\cW$ will have any good properties; first, it may not be 1-1; 
second, its image may not contain the conformal blocks one most wants to study.

To get oriented, let us recall what happens for the classical nonabelianization map \eqref{eq:nab-classical}.
This map is defined for any spectral network $\cW$, but
for arbitrary $\cW$, the map has no particularly good properties.
For $\cW$ arising from generic holomorphic or meromorphic quadratic differentials as
in \cite{Gaiotto:2009hg}, 
the situation is much better: then $\dim \cM(\tC, \GL(1)) = \dim \cM(C, \GL(2))$,
$\nab_\cW^\flat$ is finite-to-one (one-to-one if we include appropriate decoration data 
in the definition of $\cM(C,\GL(2))$), and the image of $\nab_\cW^\flat$ is an open subset
of $\cM(C,\GL(2))$, which always contains the \ti{Teichm\"uller
component} $\cT \subset \cM(C,\GL(2))$.

We expect a similar picture for the map $\nab_\cW$ on conformal blocks: if $\cW$ arises from a
holomorphic or meromorphic quadratic differential, then $\nab_\cW$ should be 
injective, and its image should contain all $c=1$ Liouville conformal blocks on $C$.

\section{A model example} \label{sec:model-example}

In this section, we discuss the simplest nontrivial example of our setup. We take
the Riemann surface $C = \bbC\bbP^1$, and the double cover $\tC = \bbC\bbP^1$, with the double cover map
\begin{align}
\pi: \tC & \to C \\
z & \mapsto x = z^2
\end{align}
This map has two branch points, at $z = 0$ and $z = \infty$.
We choose the walls of the spectral network $\cW$
to be $\arg x = \frac{2 \pi n}{3}$ for $n=0,1,2$, and
specify the sheet labels $+$, $-$ over each wall 
as indicated below.

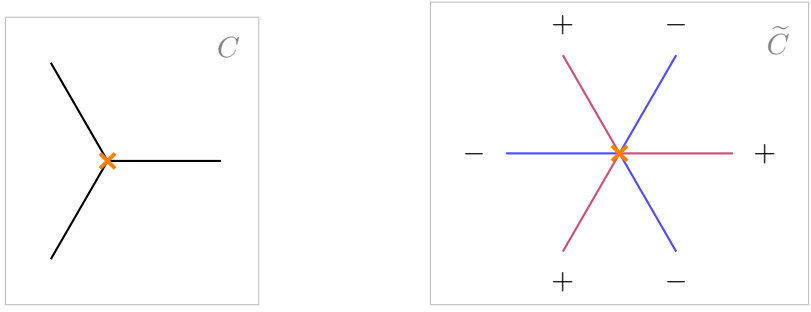
\begin{center}
\begin{tikzpicture}
  \begin{scope}
    \coordinate (center) at (0,0);

    \draw [thick,\wallColor] (center) -- ++(120:1.5cm); 
    \draw [thick,\wallColor] (center) -- ++(240:1.5cm); 
    \draw [thick,\wallColor] (center) -- ++(0:1.5cm);   

    \drawbranchpointmarker{center};

    \node at (1.6,1.5) {\color{\baseCurveColor}\( C \)};
  \end{scope}
  \draw [lightgray, thin] (-1.35,-1.9) rectangle (2,1.9);
\end{tikzpicture}
\hspace{2cm}
\begin{tikzpicture}
  \begin{scope}
    \coordinate (center) at (0,0);

    \draw [psipluscontour] (center) -- ++(120:1.5cm);   
    \draw [psipluscontour] (center) -- ++(240:1.5cm); 
    \draw [psipluscontour] (center) -- ++(0:1.5cm);
    \draw [psiminuscontour] (center) -- ++(60:1.5cm);   
    \draw [psiminuscontour] (center) -- ++(180:1.5cm); 
    \draw [psiminuscontour] (center) -- ++(300:1.5cm);

    \node [label=above:$+$] at ++(120:1.5cm) {};
    \node [label=below:$+$] at ++(240:1.5cm) {};
    \node [label=right:$+$] at ++(0:1.5cm) {};
    \node [label=above:$-$] at ++(60:1.5cm) {};
    \node [label=left:$-$] at ++(180:1.5cm) {};
    \node [label=below:$-$] at ++(300:1.5cm) {};

    \drawbranchpointmarker{center};

    \node at (2.1,1.5) {\color{\coveringCurveColor}\( \tC \)};
  \end{scope}
  \draw [lightgray, thin] (-2.5,-2) rectangle (2.5,2);
\end{tikzpicture}
\end{center}

We emphasize that in this example the branch points at $z=0$ and $z=\infty$ play symmetric roles. (One 
important point for consistency is that the leash around $z=0$ is homologous
to the one around $z=\infty$; this uses the fact that $\tC$ has genus zero and we have not inserted any primary fields.)
This spectral network may look unfamiliar to readers
familiar with e.g. \cite{MR3115984}; it cannot arise from a meromorphic quadratic differential on $C$.\footnote{It 
has however arisen in the context of 3-dimensional spectral networks
\cite{Freed:2022yae}; one can get
it by starting with a conventional 2-d spectral network on a disc with one branch point, crossing with $\bbR$ to get 
a translation-invariant 3-d spectral network, then intersecting
that 3-d network with a sphere around a point of the branch locus.}
In particular, we should not think of it as
corresponding to the meromorphic quadratic differential $x \, \de x^2$ on $\bbC \bbP^1$; that one involves an irregular singularity
at $x = \infty$ instead of a simple branch point.

At any rate, since $\tC = \bbC\bbP^1$ there is a unique 
conformal block $\tPsi \in \Conf(\tC,\Heis)$ normalized by $\IP{1}_\tPsi = 1$.
We want to apply the map $\nab_\cW$ to this block.

\subsection{Fermion correlators on \texorpdfstring{$\bbC\bbP^1$}{CP1}} \label{eq:simple-fermions-cp1}

We first recall the fermion correlation functions in the block $\tPsi$.
First note that on $\tC = \bbC \bbP^1$ we have a standard inhomogeneous coordinate $z$,
a standard global spin structure $K_{\tC}^{\frac12}$, and a standard
section $\sqrt{\de z}$. 
Moreover, given $z, w \in \tC$ there is a unique leash $\ell$ from $w$ to $z$, 
up to homotopy.

If we use these choices,
the $2$-point function is simple:
\begin{equation} \label{eq:cp1-two-point-fermion}
\IP{
\tikz[remember picture, baseline]{
  \node[anchor=base, inner sep=0pt] (psiplus) {$\psi_+(z)$};
}
\,
\tikz[remember picture, baseline]{
  \node[anchor=base, inner sep=0pt] (psiminus) {$\psi_-(w)$};
}
}_\tPsi = \frac{1}{z-w} \, .  
\end{equation}
\begin{tikzpicture}[overlay, remember picture]
  \coordinate (start) at (psiplus.north);
  \coordinate (end) at (psiminus.north);
  \coordinate (startShifted) at ([yshift=1ex]start);
  \coordinate (endShifted) at ([yshift=1ex]end);
  \draw[thick] (startShifted) -- (endShifted);
  \draw[thick] (startShifted) -- ++(0,-0.5ex);
  \draw[thick] (endShifted) -- ++(0,-0.5ex);
\end{tikzpicture}\noindent\unskip
More generally the $2n$-point function is
\begin{equation} \label{eq:free-fermion-determinant-formula-cp1}
    \left\langle \prod_{i=1}^n 
    \tikz[remember picture, baseline]{
        \node[anchor=base, inner sep=0pt] (psiplus_num) {$\psi_+(z_i)$};
    }
    \,
    \tikz[remember picture, baseline]{
        \node[anchor=base, inner sep=0pt] (psiminus_num) {$\psi_-(w_i)$};
    }
    \right\rangle_{\tPsi} = \det \left[ \frac{1}{z_i - w_j} \right]_{i,j=1}^{n} \, .
\end{equation}
\begin{tikzpicture}[overlay, remember picture]
    \coordinate (start1) at (psiplus_num.north);
    \coordinate (end1) at (psiminus_num.north);
    \coordinate (start1Shifted) at ([yshift=1ex]start1);
    \coordinate (end1Shifted) at ([yshift=1ex]end1);
    \draw[thick] (start1Shifted) -- (end1Shifted);
    \draw[thick] (start1Shifted) -- ++(0,-0.5ex);
    \draw[thick] (end1Shifted) -- ++(0,-0.5ex);
\end{tikzpicture}\noindent\unskip
and the $2n$-point function with other operator insertions
is similarly
\begin{equation} \label{eq:free-fermion-determinant-formula-with-insertions-cp1}
    \left\langle \cdots \prod_{i=1}^n 
    \tikz[remember picture, baseline]{
        \node[anchor=base, inner sep=0pt] (psiplus_num) {$\psi_+(z_i)$};
    }
    \,
    \tikz[remember picture, baseline]{
        \node[anchor=base, inner sep=0pt] (psiminus_num) {$\psi_-(w_i)$};
    }
    \right\rangle_{\tPsi} = \det \left[ 
    \frac{
    \IP{ \cdots
    \tikz[remember picture, baseline]{
        \node[anchor=base, inner sep=0pt] (psiplus_rhs) {$\psiplus(z_i)$};
    }
    \, 
    \tikz[remember picture, baseline]{
        \node[anchor=base, inner sep=0pt] (psiminus_rhs) {$\psi_-(w_j)$};
    }
    }_\tPsi}{\IP{\cdots}_\tPsi}
    \right]_{i,j=1}^{n} \, .
\end{equation}
\begin{tikzpicture}[overlay, remember picture]
    \coordinate (start1) at (psiplus_num.north);
    \coordinate (end1) at (psiminus_num.north);
    \coordinate (start1Shifted) at ([yshift=1ex]start1);
    \coordinate (end1Shifted) at ([yshift=1ex]end1);
    \draw[thick] (start1Shifted) -- (end1Shifted);
    \draw[thick] (start1Shifted) -- ++(0,-0.5ex);
    \draw[thick] (end1Shifted) -- ++(0,-0.5ex);

    \coordinate (start2) at (psiplus_rhs.north);
    \coordinate (end2) at (psiminus_rhs.north);
    \coordinate (start2Shifted) at ([yshift=1ex]start2);
    \coordinate (end2Shifted) at ([yshift=1ex]end2);
    \draw[thick] (start2Shifted) -- (end2Shifted);
    \draw[thick] (start2Shifted) -- ++(0,-0.5ex);
    \draw[thick] (end2Shifted) -- ++(0,-0.5ex);
\end{tikzpicture}\noindent\unskip
We will not prove these formulas here; 
they are special cases of more general ones
which we discuss in \autoref{sec:diagonalizing-abelian-verlinde}.

\subsection{A tricky sign}

To compute $\nab_\cW(\tPsi)$ requires us to evaluate correlation functions involving insertions of the form
\begin{equation} \label{eq:wall-basic-example}
\int_\cG \de x(q) \,
\tikz[remember picture, baseline]{
  \node[anchor=base, inner sep=0pt] (psiplus) {$\psi_+(q^{(+)})^{x^{(+)}}$};
}
\,
\tikz[remember picture, baseline]{
  \node[anchor=base, inner sep=0pt] (psiminus) {$\psi_-(q^{(-)})^{x^{(-)}}$};
}
\, .
\end{equation}
\begin{tikzpicture}[overlay, remember picture]
  \coordinate (start) at ($(psiplus.south)!0.5!(psiplus.north)$);
  \coordinate (end) at ($(psiminus.south)!0.5!(psiminus.north)$);
  \coordinate (startShifted) at ([yshift=-2ex]start);
  \coordinate (endShifted) at ([yshift=-2ex]end);
  \draw[thick] (startShifted) -- (endShifted);
  \draw[thick] (startShifted) -- ++(0,0.5ex);
  \draw[thick] (endShifted) -- ++(0,0.5ex);
  \coordinate (mid) at ($(startShifted)!0.5!(endShifted)$);
  \node at ($(mid)+(0,-1.5ex)$) {$\ell_\cG(q)$};
\end{tikzpicture}\noindent\unskip
It is technically inconvenient that this involves two 
different local coordinate systems
$x^{(\pm)}$ on $\tC$, and also that $\ell_\cG(q)$ carries 
the spin structure $\pi^* K_C^{\frac12}$, which is not globally
defined on $\tC$.
We want to replace these objects with the simpler ones discussed in \autoref{eq:simple-fermions-cp1},
in order to be able to use the concrete formula \eqref{eq:free-fermion-determinant-formula-cp1}.
Thus let $\ell'_\cG(q)$ be the same path but now with the
standard spin structure $K_\tC^{\frac12}$ on $\tC = \bbC\bbP^1$, 
and fix an 
isomorphism $\iota: \pi^* K_C^{\frac12} \to K^{\frac12}_\tC$ along the path $\ell_\cG(q)$.
Using \eqref{eq:fermion-coordinate-transformation}, we
rewrite \eqref{eq:wall-basic-example} as
\Needspace{5\baselineskip}
\begin{equation} \label{eq:wall-basic-rewritten}
  \int_\cG \de x(q) \, \left(
    \tikz[remember picture, baseline]{
      \node[anchor=base, inner sep=0pt] (psiplus) {$\psi_+(q^{(+)})^z$};
    }
    \frac{\sqrt{\de z(q^{(+)})}}{\iota(\pi^*\sqrt{\de x(q)})} \right)
    \left(
    \tikz[remember picture, baseline]{
      \node[anchor=base, inner sep=0pt] (psiminus) {$\psi_-(q^{(-)})^z$};
    }
    \frac{\sqrt{\de z(q^{(-)})}}{\iota(\pi^*\sqrt{\de x(q)})} \right) \, .
\end{equation}
\begin{tikzpicture}[overlay, remember picture]
  \coordinate (start) at ($(psiplus.south)!0.5!(psiplus.north)$);
  \coordinate (end) at ($(psiminus.south)!0.5!(psiminus.north)$);
  \coordinate (startShifted) at ([yshift=-4.5ex]start);
  \coordinate (endShifted) at ([yshift=-4.5ex]end);
  \draw[thick] (startShifted) -- (endShifted);
  \draw[thick] (startShifted) -- ++(0,2.5ex);
  \draw[thick] (endShifted) -- ++(0,2.5ex);
  \coordinate (mid) at ($(startShifted)!0.5!(endShifted)$);
  \node at ($(mid)+(0,-1.7ex)$) {$\ell'_\cG(q)$};
\end{tikzpicture}

\medskip
\noindent To simplify this, note that 
$x(q) = z(q^{(+)})^2$, so $\de x(q) = 2z(q^{(+)}) \, \de z(q^{(+)})$. It follows that
$\iota(\pi^* \sqrt{\de x(q)}) = \sqrt{2 z(q^{(\pm)})} \sqrt{\de z(q^{(\pm)})}$,
for a branch of $\sqrt{2z}$ continuous along the path $\ell_\cG(q)$.
Following our rule from \autoref{sec:nonabelianization-map},
$\ell_\cG(q)$ goes around clockwise from $q^{(-)}$ to $q^{(+)}$, which implies that this branch has $\sqrt{2z(q^{(+)})} \sqrt{2z(q^{(-)})} = 2 \I z(q^{(+)})$.
Thus the insertion can be simplified to
\begin{equation}
  \int_\cG \frac{2z(q^{(+)}) \, \de z(q^{(+)})}{2 \I z(q^{(+)})} \,
  \tikz[remember picture, baseline]{
    \node[anchor=base, inner sep=0pt] (psiplus1) {$\psi_+(q^{(+)})^z$};
  }
  \,
  \tikz[remember picture, baseline]{
    \node[anchor=base, inner sep=0pt] (psiminus1) {$\psi_-(q^{(-)})^z$};
  }
  = - \I \int_\cG \de z(q^{(+)})
  \tikz[remember picture, baseline]{
    \node[anchor=base, inner sep=0pt] (psiplus2) {$\psi_+(q^{(+)})^z$};
  }
  \,
  \tikz[remember picture, baseline]{
    \node[anchor=base, inner sep=0pt] (psiminus2) {$\psi_-(q^{(-)})^z$};
  }
  \, .
\end{equation}
\begin{tikzpicture}[overlay, remember picture]
  \coordinate (start1) at ($(psiplus1.south)!0.5!(psiplus1.north)$);
  \coordinate (end1) at ($(psiminus1.south)!0.5!(psiminus1.north)$);
  \coordinate (start1Shifted) at ([yshift=-2.3ex]start1);
  \coordinate (end1Shifted) at ([yshift=-2.3ex]end1);
  \draw[thick] (start1Shifted) -- (end1Shifted);
  \draw[thick] (start1Shifted) -- ++(0,0.5ex);
  \draw[thick] (end1Shifted) -- ++(0,0.5ex);
  \coordinate (mid1) at ($(start1Shifted)!0.5!(end1Shifted)$);
  \node at ($(mid1)+(0,-1.7ex)$) {$\ell'_\cG(q)$};

  \coordinate (start2) at ($(psiplus2.south)!0.5!(psiplus2.north)$);
  \coordinate (end2) at ($(psiminus2.south)!0.5!(psiminus2.north)$);
  \coordinate (start2Shifted) at ([yshift=-2.3ex]start2);
  \coordinate (end2Shifted) at ([yshift=-2.3ex]end2);
  \draw[thick] (start2Shifted) -- (end2Shifted);
  \draw[thick] (start2Shifted) -- ++(0,0.5ex);
  \draw[thick] (end2Shifted) -- ++(0,0.5ex);
  \coordinate (mid2) at ($(start2Shifted)!0.5!(end2Shifted)$);
  \node at ($(mid2)+(0,-1.7ex)$) {$\ell'_\cG(q)$};
\end{tikzpicture}

\noindent Finally we simplify our notation as follows. We 
write just $z$ for $z(q^{(+)})$, and use always the standard
coordinate $z$, the standard spin structure $K_\tC^{\frac12}$,
and the standard $\sqrt{\de z}$ on $\tC = \bbC\bbP^1$. 
Since $\tC$ is simply 
connected and $K_\tC^{\frac12}$ is defined everywhere,
the leash is uniquely determined, so we can drop the name
$\ell_\cG'(q)$ from the notation too. Then the insertion is 
\begin{equation} \label{eq:fermion-insertion-simplified}
 -\I \int_{\cG^{(+)}} \de z \,
 \tikz[remember picture, baseline]{
   \node[anchor=base, inner sep=0pt] (psiplus) {$\psi_+(z)$};
 }
 \,
 \tikz[remember picture, baseline]{
   \node[anchor=base, inner sep=0pt] (psiminus) {$\psi_-(-z)$};
 }
\end{equation}
\begin{tikzpicture}[overlay, remember picture]
  \coordinate (start) at ($(psiplus.south)!0.5!(psiplus.north)$);
  \coordinate (end) at ($(psiminus.south)!0.5!(psiminus.north)$);
  \coordinate (startShifted) at ([yshift=-2ex]start);
  \coordinate (endShifted) at ([yshift=-2ex]end);
  \draw[thick] (startShifted) -- (endShifted);
  \draw[thick] (startShifted) -- ++(0,0.5ex);
  \draw[thick] (endShifted) -- ++(0,0.5ex);
\end{tikzpicture}\noindent\unskip
and we can rewrite the operator $E(\cW)$ as
\begin{equation} \label{eq:EW-rewritten}
 E(\cW) = \exp \left( \frac{-1}{2 \pi} \int_{\cW^{(+)}} \de z \,
 \tikz[remember picture, baseline]{
   \node[anchor=base, inner sep=0pt] (psiplus) {$\psi_+(z)$};
 }
 \,
 \tikz[remember picture, baseline]{
   \node[anchor=base, inner sep=0pt] (psiminus) {$\psi_-(-z)$};
 }
 \right) \, .
\end{equation}
\begin{tikzpicture}[overlay, remember picture]
  \coordinate (start) at ($(psiplus.south)!0.5!(psiplus.north)$);
  \coordinate (end) at ($(psiminus.south)!0.5!(psiminus.north)$);
  \coordinate (startShifted) at ([yshift=-2ex]start);
  \coordinate (endShifted) at ([yshift=-2ex]end);
  \draw[thick] (startShifted) -- (endShifted);
  \draw[thick] (startShifted) -- ++(0,0.5ex);
  \draw[thick] (endShifted) -- ++(0,0.5ex);
\end{tikzpicture}\noindent\unskip
The main point of this careful treatment was
to get the correct sign in \eqref{eq:EW-rewritten};
up to that sign, one could have guessed the form of
\eqref{eq:EW-rewritten} by
naively applying \eqref{eq:fermion-coordinate-transformation},
without being careful about branches of square roots.

\subsection{The normalized 2-fermion correlator}

We begin by considering the normalized fermion 2-point function on $\tC$ with the spectral network inserted:
\begin{equation}\label{eq:deffzw}
F(z,w) = \frac{ \langle
\tikz[remember picture, baseline]{
  \node[anchor=base, inner sep=0pt] (psiplus) {$\psi_+(z)$};
}
\,
\tikz[remember picture, baseline]{
  \node[anchor=base, inner sep=0pt] (psiminus) {$\psi_-(w)$};
}
\, E_\ren(\cW) \rangle_\tPsi }{ \langle E_\ren(\cW) \rangle_\tPsi} \, .
\end{equation}
\begin{tikzpicture}[overlay, remember picture]
  \coordinate (start) at (psiplus.north);
  \coordinate (end) at (psiminus.north);
  \coordinate (startShifted) at ([yshift=1ex]start);
  \coordinate (endShifted) at ([yshift=1ex]end);
  \draw[thick] (startShifted) -- (endShifted);
  \draw[thick] (startShifted) -- ++(0,-0.5ex);
  \draw[thick] (endShifted) -- ++(0,-0.5ex);
\end{tikzpicture}
Here are some properties of $F$ which follow from \eqref{eq:deffzw}:
\begin{itemize}
  \item $F(z,w)$ is a single-valued function of $z$ and $w$.
  \item $F(z,w)$ has the standard free fermion OPE as $z \to w$,
  \begin{equation}
    F(z,w) = \frac{1}{z-w} + \regular \, .
  \end{equation}
  \item $F(z,w)$ is piecewise analytic: it 
  jumps by addition of $- \I F(-z,w)$ when $z$ crosses $\cW^{(-)}$ in the counterclockwise direction,
  and jumps by addition of $-\I F(z,-w)$ when $w$ crosses $\cW^{(+)}$ in the counterclockwise direction.
  \item $F(z,w)$ has the symmetry
  \begin{equation}
    F(1/z, 1/w) = \left( -zw \right) F(z,w) \, .
  \end{equation}
\end{itemize}
We will prove that $F(z,w)$ is given by the piecewise analytic function
\begin{equation} \label{eq:two-point-with-walls}
   F(z,w) =\begin{cases} 0 & \text{if $z\in \circled{$n_+$}$ and $w\in \circled{$n_-$}$,}\\
   \frac{2 \sqrt{wz}}{z^2 - w^2} & \text{otherwise},
   \end{cases} 
\end{equation}
where \circled{$n_\pm$} are regions on $\tC$, separated by $\mathcal{W}^{(\mp)}$, 
as shown below.
\begin{figure}[h]
\label{fzwsector}
\centering
\begin{tikzpicture}
  \tikzset{
    psiregionmarker/.style={anchor=center, black}
  }
  \def\sep{0.0}
  \def\wlen{1.5}
  \foreach \angle in {60, 180, 300} {
    \draw[psiminuscontour] (\angle:\sep) -- (\angle:\wlen);
  }
  \node[psiregionmarker] at (0.85*\wlen, 0) {$\circled{$1_+$}$};
  \node[psiregionmarker] at (-0.42*\wlen, 0.72*\wlen) {$\circled{$2_+$}$};
  \node[psiregionmarker] at (-0.42*\wlen, -0.72*\wlen) {$\circled{$3_+$}$};
\end{tikzpicture}
\hspace{0.35in}
\begin{tikzpicture}
  \tikzset{
    tpsiregionmarker/.style={anchor=center, black}
  }
  \def\sep{0.0}
  \def\wlen{1.5}
  \foreach \angle in {0, 120, 240} {
    \draw[psipluscontour] (\angle:\sep) -- (\angle:\wlen);
  }
  \node[tpsiregionmarker] at (-0.85*\wlen, 0) {$\circled{$1_-$}$};
  \node[tpsiregionmarker] at (0.42*\wlen, -0.72*\wlen) {$\circled{$2_-$}$};
  \node[tpsiregionmarker] at (0.42*\wlen, 0.72*\wlen) {$\circled{$3_-$}$};
\end{tikzpicture}
\end{figure}
We need to explain which branch of $\sqrt{wz}$ we take:
when $z = w$ we take $\sqrt{wz} = z = w$, 
and more generally we choose the branch
by continuation from the region where $z$ is close to $w$.
We can check directly that this $F$ has all the expected properties we listed above.

In the rest of this section we give the direct computation of $F$. 
We use the determinant formula
\eqref{eq:free-fermion-determinant-formula-cp1} to expand both the numerator
and denominator of \eqref{eq:deffzw} as a sum of terms. 
This is very much like using Wick's theorem to compute in free fermion field theory, and we borrow
a convenient organizational scheme from that setting. Indeed, each term in the numerator 
can be conveniently represented by a Feynman diagram.
This diagram has two fixed colored vertices representing the fermions $\psiplus(z)$ and $\psiminus(w)$, 
an arbitrary number of black vertices representing fermion pairs $- \frac{1}{2 \pi} \int_{\cW^{(+)}} \psiplus(x_i) \psiminus(-x_i) \, \de x_i$, 
and edges connecting each $\psiplus$ to a $\psiminus$. Thus each black vertex is $2$-valent while each fixed vertex is 
$1$-valent, from which it follows that each diagram consists of a single linear chain with endpoints 
the two fixed vertices, plus some number of bubbles (loops) made of black vertices.
The value of the diagram is obtained by writing a factor \eqref{eq:cp1-two-point-fermion} for each edge, integrating
over the $x_i$, and dividing by a ``symmetry factor'' counting the automorphisms of the diagram.
In the denominator, we have a similar sum over diagrams, except now there are 
only black vertices, and thus each diagram involves only bubbles.

The bubbles cancel between numerator and denominator, and we are left with the sum over 
connected diagrams only. 
(In particular, the divergence of the integrals near the branch points appears only 
in the bubbles, and thus cancels, so we do not need to worry about regularizing it.)
Explicitly, the contribution from a connected diagram is
\begin{equation}\label{eq:defchain}
\begin{tikzpicture}[node distance=1cm and 1.5cm]
\def\radius{0.6}
  \coordinate[vertex3] (e1); 
  \coordinate[vertex, right = 0.2 of e1] (v1);
  \coordinate[ right = 0.1 of v1] (v15);
  \coordinate[vertex2, right = 0.2 of v1] (v2);
  \coordinate[vertex2, right = 0.1 of v2] (v3);
  \coordinate[vertex2, right = 0.1 of v3] (v4);
  \coordinate[ right = 0.1 of v4] (v45);
  \coordinate[vertex, right = 0.2 of v4] (v5);
  \coordinate[vertex4, right = 0.2 of v5] (v6);
\draw (e1) -- (v1);
\draw (v1) -- (v15);
\draw (v45) -- (v5);
\draw (v5) -- (v6);
\end{tikzpicture} \, = \, \left(\frac{-1}{2\pi}\right)^n \int_{\tree}\cdots\int_{\tree} \frac{1}{z+x_1}\left(\prod_{i=1}^{n-1}\frac{1}{x_i+x_{i+1}}\right)\frac{1}{x_n-w} \, \de x_1\cdots\de x_n,
\end{equation}
where $n$ is the number of black vertices, and
we have renamed the three-pronged contour of integration from $\cW^{(+)}$ to $\treet$.
Summing up the connected diagrams gives the desired $F(z,w)$:
\begin{align} 
\label{eq:chain2fermion}
 F(z,w) =& \ \,
    \begin{tikzpicture}[ node distance=1cm and 1.5cm]
\def\radius{0.6}
  \coordinate[vertex3] (e1); 
  \coordinate[vertex4, right = 0.3 of e1] (v6);
\draw (e1) -- (v6);
\end{tikzpicture}
\,+\,
  \begin{tikzpicture}[node distance=1cm and 1.5cm]
\def\radius{0.6}
  \coordinate[vertex3] (e1); 
  \coordinate[vertex, right = 0.3 of e1] (v1);
  \coordinate[vertex4, right = 0.3 of v1] (v6);
\draw (e1) -- (v6);
\end{tikzpicture}
\,+\,
\begin{tikzpicture}[node distance=1cm and 1.5cm]
\def\radius{0.6}
  \coordinate[vertex3] (e1); 
  \coordinate[vertex, right = 0.3 of e1] (v1);
   \coordinate[vertex, right = 0.3 of v1] (v5);
  \coordinate[vertex4, right = 0.3 of v5] (v6);
\draw (e1) --  (v6);
\end{tikzpicture}
\,+\,
\begin{tikzpicture}[ node distance=1cm and 1.5cm]
\def\radius{0.6}
  \coordinate[vertex3] (e1); 
  \coordinate[vertex, right = 0.3 of e1] (v1);
   \coordinate[vertex, right = 0.3 of v1] (v2);
      \coordinate[vertex, right = 0.3 of v2] (v5);
  \coordinate[vertex4, right = 0.3 of v5] (v6);
\draw (e1) --  (v6);
\end{tikzpicture}
\,+\,\cdots\\
=& \sum_{n=0}^\infty \left(\frac{-1}{2\pi}\right)^n \int_{\tree}\cdots\int_{\tree} \frac{1}{z+x_1}\left(\prod_{i=1}^{n-1}\frac{1}{x_i+x_{i+1}}\right)\frac{1}{x_n-w} \, \de x_1\cdots\de x_n\\
=& \sum_{n=0}^\infty \left(\frac{-1}{2\pi}\right)^n \int_0^\infty \cdots \int_0^\infty \left( \prod_{i=1}^n 3 x_i^2 \de x_i \right) (w^2 + (-1)^n wz + z^2) \\
&\quad\quad\times \frac{1}{z^3 + x_1^3} \frac{1}{x_1^3 + x_2^3} \cdots \frac{1}{x_n^3 - w^3} \\
=& \sum_{n=0}^\infty \left(\frac{-1}{2\pi}\right)^n (w^2 + (-1)^n wz + z^2) \int_0^\infty \cdots \int_0^\infty \de t_1 \cdots \de t_n \, \frac{1}{z^3 + t_1} \frac{1}{t_1 + t_2} \cdots \frac{1}{t_n - w^3} \, , \label{eq:iterated-int-model-explicit}
\end{align}
where we used the change of variable $t_i=x_i^3.$

We can compute the first few of these integrals directly.
The result is conveniently expressed in terms of the variable $X = \frac{1}{2\pi}( \log(-w^3)-\log(z^3) )$; it begins 
\begin{equation}
 \frac{1}{z-w} + \frac{X}{z+w} + \frac{X^2 + \frac14}{2 (z-w)} + \frac{X(X^2 + 1)}{6 (z+w)} + \frac{(X^2 + \frac14)(X^2 + \frac94)}{24 (z-w)} + \cdots.
\end{equation}
The $n$-th term is
\begin{equation}\label{eq:defIn}
  I_n(z,w) = \begin{cases} \frac{\prod_{k=1}^{n/2} (X^2 + (k-\frac12)^2)}{n! (z-w)} & \text{ for $n$ even}, \\ 
  \frac{X \prod_{k=1}^{(n-1)/2} (X^2 + k^2)}{n! (z+w)} & \text{ for $n$ odd} \, .
  \end{cases}
\end{equation}
We prove \eqref{eq:defIn} as follows.\footnote{We thank Sri Tata for showing us
how to evaluate these integrals by diagonalizing the semi-infinite Hilbert transform (cf. \cite{Koppelman:1959ud}). The integrals \eqref{eq:iterated-int-model-explicit} are similar to ones studied in \cite{tata20222d}.}
The problem is to calculate $f_n(z^3,-w^3)$, where we define
\begin{equation}\label{eq:fnxy} f_n(x,y) = \left(\frac{1}{2\pi}\right)^n\int_0^\infty \cdots \int_0^\infty \de t_1 \cdots \de t_n \, \frac{1}{x+ t_1} \frac{1}{t_1 + t_2} \cdots \frac{1}{t_n + y} \, .
\end{equation}
In $f_n(x,y)$ we view $y$ as a fixed parameter, and suppress
it from the notation for now.
Then, defining an integral operator $\cS$ by
\begin{equation}
(\mathcal{S}f)(x)=\frac{1}{2\pi}\int_{0}^\infty \frac{f(x')}{x+x'} \, \de x' \, ,
\end{equation}
the desired $f_n$ can be written as
\begin{equation}
  f_n(x) = (\cS^n f_0)(x) \, , \qquad f_0(x) = \frac{1}{x+y} \, .
\end{equation}
To compute it, we diagonalize the operator $\cS$. Indeed,
$\cS$ is a bounded linear operator on $L^2(\bbR^+)$, 
with eigenfunctions parameterized by $\alpha \in \bbR$,
\begin{equation}
\phi_\alpha(x) = x^{-\frac12 - \I \alpha} = \e^{(-\frac12 - \I \alpha)\log x}, \quad \cS \phi_\alpha = \frac12 \sech(\pi\alpha) \phi_\alpha \, .
\end{equation}
Any $f \in L^2(\bbR^+)$ can be expanded in the $\phi_{\alpha}$:
\begin{equation}
f(x) = \int_{-\infty}^\infty \hat f(\alpha)\phi_{\alpha}(x) \, \de \alpha \, ,\qquad 
\hat f(\alpha) = \frac{1}{2\pi}\int_{0}^\infty \phi_{-\alpha}(x) f(x) \, \de x \, .
\end{equation}
In particular, we can expand our input function $f_0$ in this basis: we have
\begin{equation}
  \hat f_0(\alpha) = \frac{1}{2\pi}\int_{0}^\infty \phi_{-\alpha}(x) f_0(x) \, \de x = \frac12 \text{sech}(\pi\alpha) \phi_{-\alpha}(y) \, ,
\end{equation}
and so (after a change of variable $\alpha \to -\alpha$) we have
\begin{equation}
\label{eq:1overww'}
f_0(x) = \frac12 \int_{-\infty}^\infty\text{sech}(\pi\alpha) \phi_{-\alpha}(x)\phi_{\alpha}(y) \, \de\alpha \, .
\end{equation}
Now we are ready to compute:
\begin{align} f_n(x) = (\cS^n f_0)(x)
&= \int_{-\infty}^\infty\de\alpha \, \phi_{-\alpha}(x)\phi_{\alpha}(y)\left(\frac1 2 \,\text{sech}(\pi\alpha)\right)^{n+1}\\
&= \e^{-\frac1 2 (\log(x)+\log(y))}\int_{-\infty}^\infty\de\alpha\,\e^{\ri\alpha(\log(x)-\log(y))}\left(\frac1 2 \,\text{sech}(\pi\alpha)\right)^{n+1}\\
&= \e^{-\frac1 2 (\log(x)+\log(y))}\int_{0}^\infty\de s\,\frac{1}{\pi s}s^{\frac{\ri}{\pi}(\log(x)-\log(y))}\left(\frac{s}{1+s^2}\right)^{n+1}\\
\nonumber &= \e^{-\frac1 2 (\log(x)+\log(y))} \frac{\Gamma(\frac{1}{2}+\frac{n}{2}-\ri \frac{\log(y)-\log(x)}{2\pi})\Gamma(\frac{1}{2}+\frac{n}{2}+\ri \frac{\log(y)-\log(x)}{2\pi})}{2\pi\Gamma(1+n)}\\
&= \begin{cases} \frac{\prod_{k=1}^{n/2} \left(\left(\frac{\log(y)-\log(x)}{2\pi}\right)^2 + (k-\frac12)^2\right)}{n! (x+y)} & \text{ for $n$ even}, \\ 
  -\frac{\left(\frac{\log(y)-\log(x)}{2\pi}\right) \prod_{k=1}^{(n-1)/2} \left(\left(\frac{\log(y)-\log(x)}{2\pi}\right)^2 + k^2\right)}{n! (x-y)} & \text{ for $n$ odd} \, ,
  \end{cases} \label{eq:fn-result}
\end{align}
where we have used the change of coordinate $\alpha=\frac{\log(s)}{\pi}$. 
On the other hand we have
\begin{equation} I_n(z,w) = \begin{cases} (w^2 + wz + z^2)f_n(z^3,-w^3) & \text{ for $n$ even}, \\ 
  -(w^2 - wz + z^2)f_n(z^3,-w^3) & \text{ for $n$ odd} \, .
  \end{cases}
\end{equation}
Combining this with \eqref{eq:fn-result} gives the desired proof of \eqref{eq:defIn}.
Once we have obtained \eqref{eq:defIn}, summing over $n$ gives
\begin{equation}\label{eq:coshsinhIn}
  \sum_{n=0}^\infty I_n(z,w) = \frac{2}{\sqrt3} \left( \frac{\cosh \frac{\pi X}{3}}{z-w} + \frac{\sinh \frac{\pi X}{3}}{z+w} \right)
\end{equation}
which matches the desired \eqref{eq:two-point-with-walls}.

Alternatively, we can give a more direct proof of \eqref{eq:coshsinhIn}
by summing over $n$ first before integrating over $\alpha$. 
Because of the factor $(w^2 + (-1)^n wz + z^2)$ in $I_n$, we sum the even and odd terms separately, getting
\begin{align}
\sum_{n\text{ even}}f_n(x,y) &= \e^{-\frac12 (\log(x)+\log(y))} \int_{-\infty}^\infty \de\alpha\,\e^{\I\alpha(\log(x)-\log(y))}\left(\frac{2\text{sech}(\pi\alpha)}{4-\text{sech}^2(\pi\alpha)}\right) \, , \\
\sum_{n\text{ odd}}f_n(x,y) &= \e^{-\frac1 2 (\log(x)+\log(y))}\int_{-\infty}^\infty\de\alpha\,\e^{\I\alpha(\log(x)-\log(y))}\left(\frac{\text{sech}^2(\pi\alpha)}{4-\text{sech}^2(\pi\alpha)}\right) \, .
\end{align}
Depending on the sign of ${\mathrm{Re}}(\log(x)-\log(y))$, we can close the contours in either the upper or the lower half-plane. Assume ${\mathrm{Re}}(\log(x)-\log(y)) > 0$ and close the contours above (the other case is similar).
Then
\begin{align}
&\sum_{n\text{ even}} I_n(z,w) = (w^2 + wz + z^2) \sum_{n\text{ even}}f_n(z^3,-w^3) \\
&= (w^2 + wz + z^2) \, 2\pi\ri\left(\sum_{k\geq 0}\text{Res}_{\alpha=\frac{\ri}{3}+k\ri}+\sum_{k\geq 0}\text{Res}_{\alpha=\frac{2\ri}{3}+k\ri}\right) \\
&= (w^2 + wz + z^2) \, 2\pi\I \, \e^{-\frac12(\log(z^3)+\log(-w^3))}\left(-\frac{\ri(-w^3)^{2/3}}{2\sqrt{3}\pi (z^3)^{2/3}}-\frac{\ri(-w^3)^{1/3}}{2\sqrt{3}\pi (z^3)^{1/3}}\right)\left(\sum_{k\geq 0}\frac{(-1)^k(-w^3)^k}{z^k}\right) \\
&= \frac{2}{\sqrt3} \frac{\cosh \frac{\pi X}{3}}{z-w} \, ,
\end{align}
and similarly
\begin{align}
&\sum_{n\text{ odd}}I_n(z,w)=-(w^2 - wz + z^2)\sum_{n\text{ odd}}f_n(z^3,-w^3) = \frac{2}{\sqrt3}\frac{\sinh \frac{\pi X}{3}}{z+w} \, .
\end{align}
Thus the sum over all $n$ indeed matches \eqref{eq:coshsinhIn}, as desired.

\subsection{Heisenberg and Virasoro one-point functions}

Recall that in \autoref{sec:branch-point-singularities} we found that, if we use the branched free-field construction
$\nab_0$ (without the spectral network), the branch points naturally come with insertions of $W_{\frac{1}{16}}$.
Now we will show that when we use the full nonabelianization map $\nab_\cW$ in our model example
these insertions are removed.

We first compute the behavior of $T^\tot$ near a branch 
point. Rather than directly computing the iterated integrals in the definition of $\nab_\cW$, we
leverage the fact that we have already computed the fermion 2-point function.
As we did in \autoref{sec:branch-point-singularities} we use 
a local coordinate $z = \sqrt{x}$, and we also simplify our notation a bit,
writing $\tJ(z)$ for $\tJ(p^{(1)})^{z^{(1)}}$ or $\tJ(p^{(2)})^{z^{(2)}}$,
and similarly for $\psi_\pm$. Then we compute:
\begin{align}
\IP{\tT^\Heis(z) E_\ren(\cW) \cdots}_{\tPsi}&= \frac12 \IP{\nop{\tJ(z)\tJ(z)} E_\ren(\cW) \cdots}_\tPsi \\
&=\frac12 \lim_{z\rightarrow w} \left( \IP{\tJ(z)\tJ(w) E_\ren(\cW) \cdots}_\tPsi-\frac{\IP{E_\ren(\cW) \cdots}_\tPsi}{(z-w)^2} \right) \, .
\end{align}
Using $\tJ = \nop{\psi_+ \psi_-}$ this becomes
\begin{align}
&= \frac12 \lim_{z\rightarrow w} \left( \IP{\nop{\psiplus(z)\psiminus(z)}\nop{\psiplus(w)\psiminus(w)}E_\ren(\cW)\cdots}_\tPsi - \frac{\IP{E_\ren(\cW)\cdots}_\tPsi}{(z-w)^2} \right) \, .
\end{align}
Using \eqref{eq:free-fermion-determinant-formula-with-insertions-cp1} we can re-express this in terms of the
fermion 2-point function with operator insertions,
\begin{equation}
 K(z,w) = \frac{\IP{\,
\tikz[remember picture, baseline]{
  \node[anchor=base, inner sep=0pt] (psiplus1) {{$\psiplus(z)$}};
}
\,
\tikz[remember picture, baseline]{
  \node[anchor=base, inner sep=0pt] (psiminus1) {{$\psiminus(w)$}};
}  E_\ren(\cW) \cdots}_\tPsi}{\IP{E_\ren(\cW) \cdots}_\tPsi} \, ,
\end{equation}
\begin{tikzpicture}[overlay, remember picture]
  \coordinate (start) at (psiplus1.north);
  \coordinate (end) at (psiminus1.north);
  \coordinate (startShifted) at ([yshift=1ex]start);
  \coordinate (endShifted) at ([yshift=1ex]end);
  \draw[thick] (startShifted) -- (endShifted);
  \draw[thick] (startShifted) -- ++(0,-0.5ex);
  \draw[thick] (endShifted) -- ++(0,-0.5ex);
  \coordinate (mid) at ($(startShifted)!0.5!(endShifted)$);
  \node at ($(mid)+(0,1.5ex)$) {};
\end{tikzpicture}

obtaining
\begin{multline} \label{eq:interm-3}
  \frac{\IP{\tT^\Heis(z) E_\ren(\cW) \cdots}_{\tPsi}}{\IP{E_\ren(\cW)\cdots}_{\tPsi}} = \\ \frac12 \lim_{z\rightarrow w} \lim_{z'\to z} \lim_{w'\to w} \left( \left( K(z',z)-\frac{1}{z'-z} \right)\left(  K(w',w)-\frac{1}{w'-w} \right) - K(z',w)K(w',z) - \frac{1}{(z-w)^2}  \right) \, .
\end{multline}
As long as the insertions 
$\cdots$ are away from $0$, for the purposes of studying the behavior 
near $z = 0$, we can replace $K$ by $F$ given by \eqref{eq:two-point-with-walls}.
To be more precise: fix some open disc 
$U$ in $\bbC$, whose closure contains $z = 0$;
then thicken it to a small neighborhood of $U \times U$ in $\bbC^2$;
on this neighborhood we have
\begin{equation} \label{eq:K-approx}
   K(z,w) = 2 \sqrt{zw} \left( \frac{1}{z^2-w^2} + a(z,w) \right)
\end{equation}
for a bounded function $a(z,w)$. Indeed, this follows
from \eqref{eq:two-point-with-walls} and \eqref{eq:free-fermion-determinant-formula-cp1}.
Substituting this in \eqref{eq:interm-3} gives
\begin{equation} \label{eq:interm-4}
\frac{\IP{\tT^\Heis(z) E_\ren(\cW) \cdots}_{\tPsi}}{\IP{E_\ren(\cW)\cdots}_{\tPsi}} = - \frac{1}{8z^2} + O(z) \, .
\end{equation}
Now we are ready to compute what we really want, the 
expectation value of $T^\tot$ in the block
$\nab_\cW(\tPsi)$ on $C$. 
Recalling our dictionary $T^\tot(p)^z \rightsquigarrow \tT^\Heis(p^{(1)})^{z^{(1)}} + \tT^\Heis(p^{(2)})^{z^{(2)}}$, we see that we need to sum \eqref{eq:interm-4} over $z$ and $-z$, giving
\begin{equation}
  \frac{\IP{T^\tot(p)^z \cdots}_{\nab_\cW(\tPsi)}}{\IP{\cdots}_{\nab_\cW(\tPsi)}} = - \frac{1}{4z^2} + O(z^2) \, .
\end{equation}
This $- \frac{1}{4 z^2}$ is just what we need to cancel the singularity coming from the change of coordinates. Indeed,
using 
the change-of-coordinate rule \eqref{eq:virasoro-coordinate-change}
at $c=2$ and the relation $\{x,z\} = -\frac32 \frac{1}{z^2}$ gives
\begin{equation}
  T^\tot(p)^{z} = (2z(p))^2 T^\tot(p)^x + \frac{1}{6} \left(-\frac32 \frac{1}{z(p)^2} \right) \, ,
\end{equation}
so
\begin{equation}
  T^\tot(p)^x = \frac{1}{(2z(p))^2} \left( -\frac{1}{4 z(p)^2} + \frac{1}{4 z(p)^2} + O(z(p)^2) \right) = \regular
\end{equation}
as desired.

We should also check that there is no singularity in the Heisenberg correlators $\IP{J(p)^x \cdots}$ as $p$
approaches a branch point.
We directly compute:
\begin{align}
\IP{\tJ(z) E_\ren(\cW) \cdots}_\tPsi
&= \IP{\nop{\psiplus(z) \psiminus(z)} E_\ren(\cW) \cdots}_\tPsi \\
&=  \lim_{z' \to z} \left( \IP{\psiplus(z') \psiminus(z) E_\ren(\cW) \cdots} - \frac{\IP{E_\ren(\cW) \cdots}}{z'-z} \right)_\tPsi
\end{align}
and thus
\begin{equation}
  \frac{\IP{\tJ(z) E_\ren(\cW) \cdots}_\tPsi}{\IP{E_\ren(\cW) \cdots}_\tPsi} = \lim_{z' \to z} K(z',z) - \frac{1}{z'-z}
\end{equation}
Substituting \eqref{eq:K-approx} here gives
\begin{equation}
  \frac{\IP{\tJ(z) E_\ren(\cW) \cdots}_\tPsi}{\IP{E_\ren(\cW) \cdots}_\tPsi} = O(z) \, .
\end{equation}
Then using our dictionary $J(p)^z \rightsquigarrow \tJ(p^{(1)})^{z^{(1)}} + \tJ(p^{(2)})^{z^{(2)}}$ we get
\begin{equation}
  \frac{\IP{J(p)^z \cdots}_{\nab_\cW(\tPsi)}}{\IP{\cdots}_{\nab_\cW(\tPsi)}} = O(z) \, ,
\end{equation}
and using \eqref{eq:heisenberg-coordinate-change} to change coordinates this implies
\begin{equation}
  J(p)^x = J(p)^z / 2z = \regular \, ,
\end{equation}
as desired.

Finally, from the fact that both $T^\tot$ and $J$ are nonsingular at the branch point,
it follows that $T = T^\tot - T^\Heis$ 
is also nonsingular there.

\subsection{The 0-point function and its regularization}

In this section we give the computation of the 0-point function $\IP{E(\cW)}_{\tPsi}$.
As we did above, we expand in Feynman diagrams.
We introduce the notation
\begin{equation}
\begin{tikzpicture}[baseline={(current bounding box.center)}, node distance=1cm and 1.5cm]
\def\radius{0.6}
  \coordinate[] (e1); 
  \coordinate[vertex, above = 0.54 of e1] (v1);
  \coordinate[vertex, below left= 0.27 and  0.467654 of e1] (v2);
  \coordinate[vertex2, below right= 0.27 and  0.467654 of e1] (v3);
  \coordinate[vertex2, below left= 0.02 and  0.01 of v3] (v4);
  \coordinate[vertex2, above right= 0.02 and  0.01 of v3] (v4);
\draw (0.6,0) arc (0:280:\radius);
\end{tikzpicture}\equiv(-1)^{n}\frac{1}{(2\pi)^n}\int_{\tree}\cdots\int_{\tree} \left(\prod_{i=1}^{n-1}\frac{1}{x_i+x_{i+1}}\right)\frac{1}{x_n+x_1}\prod_{i=1}^{n}\de x_i \, ,
\end{equation}
where $n$ is the number of black vertices on the circle.
As usual, keeping track of the combinatorial
factors, we find that the partition function is
the exponential of the sum of connected bubble diagrams. 
Ignoring for now the need to regulate the integrals, this
means that
\begingroup
\allowdisplaybreaks
\begin{align}
\log \IP{E(\cW)}_{\tPsi} &= \,
\begin{tikzpicture}[baseline={(current bounding box.center)}, node distance=1cm and 1.5cm]
\def\radius{0.6}
  \draw (0,0) circle (\radius);
  \coordinate[] (e1); 
  \coordinate[vertex, above = 0.54 of e1] (v1);
\end{tikzpicture}
+ \frac12 \, \begin{tikzpicture}[baseline={(current bounding box.center)}, node distance=1cm and 1.5cm]
\def\radius{0.6}
  \draw (0,0) circle (\radius);
  \coordinate[] (e1); 
  \coordinate[vertex, above = 0.54 of e1] (v1);
  \coordinate[vertex, below= 0.54 of e1] (v2);
\end{tikzpicture}
+ \frac13 \, \begin{tikzpicture}[baseline={(current bounding box.center)}, node distance=1cm and 1.5cm]
\def\radius{0.6}
  \draw (0,0) circle (\radius);
  \coordinate[] (e1); 
  \coordinate[vertex, above = 0.54 of e1] (v1);
  \coordinate[vertex, below left= 0.27 and  0.467654 of e1] (v2);
  \coordinate[vertex, below right= 0.27 and  0.467654 of e1] (v3);
\end{tikzpicture}
+ \frac14 \,\begin{tikzpicture}[baseline={(current bounding box.center)}, node distance=1cm and 1.5cm]
\def\radius{0.6}
  \draw (0,0) circle (\radius);
  \coordinate[] (e1); 
  \coordinate[vertex, above = 0.54 of e1] (v1);
  \coordinate[vertex, below = 0.54 of e1] (v2);
  \coordinate[vertex, left = 0.54 of e1] (v3);
  \coordinate[vertex, right = 0.54 of e1] (v4);
\end{tikzpicture}
+\cdots\\
&= \sum_{n=1}^\infty\frac{1}{n}\left(\frac{-1}{2\pi}\right)^n\int_{\tree}\cdots\int_{\tree} \frac{1}{x_1+x_2}\frac{1}{x_2+x_3}\cdots\frac{1}{x_n+x_1} \,  \de x_1\cdots\de x_n \label{eq:fredholm-determinant-longhand} \\
&=3\sum_{n\text{ even}}\frac{1}{n}\frac{1}{(2\pi)^n}\int_{0}^\infty\cdots\int_{0}^\infty\de t_1\cdots\de t_n \, \frac{1}{t_1+t_2}\cdots\frac{1}{t_n+t_1}\\
&\quad-\sum_{n\text{ odd}}\frac{1}{n}\frac{1}{(2\pi)^n}\int_{0}^\infty\cdots\int_{0}^\infty\de t_1\cdots\de t_n \, \frac{1}{t_1+t_2}\cdots\frac{1}{t_n+t_1}\\
&=\frac{3}{2\pi}\int_{0}^\infty\frac{\de t_1}{t_1}\int_{-\infty}^\infty\de \alpha\sum_{n\text{ even}}\frac{(\frac1 2 \text{sech}(\pi\alpha))^n}{n}\\
&\quad-\frac{1}{2\pi}\int_{0}^\infty\frac{\de t_1}{t_1}\int_{-\infty}^\infty\de \alpha\sum_{n\text{ odd}}\frac{(\frac1 2 \text{sech}(\pi\alpha))^n}{n}\\
&= \left( \frac{3}{2 \pi} \frac{\pi}{36} - \frac{1}{2\pi} \frac{\pi}{6} \right) \int_0^\infty \frac{\de t_1}{t_1} \\
&= -\frac{1}{24}\int_{0}^\infty\frac{\de t_1}{t_1} \\
&= -\frac{1}{8}\int_{0}^\infty\frac{\de x_1}{x_1} \, . \label{eq:log-divergent-result}
\end{align}
\endgroup

The result \eqref{eq:log-divergent-result} is logarithmically divergent, so we need to regularize it.
We cut off the integral near $0$ at $x_1 = \eps_0$, and cut off the
integral near $\infty$ at $x_1 = \eps_\infty^{-1}$. 
Then we get
\begin{equation}
  \log \IP{1}_{\nab_\cW(\tPsi)} = \frac18 \log \eps_0 \eps_{\infty},
\end{equation}
i.e.
\begin{equation}
  \IP{1}_{\nab_\cW(\tPsi)} = (\eps_0 \eps_\infty)^{ \frac18} \, .
\end{equation}
Note that this regularization is a bit different from what we described
in our general scheme above; 
there we cut off all the integrals at a distance $\eps$,
rather than just the final one. This change affects the
normalization of the final result, but not the form of the divergence.
So we conclude that
\begin{equation}
  \IP{1}_{\nab_\cW(\tPsi)} = N (\eps_0 \eps_\infty)^{ \frac18} 
\end{equation}
for some normalization constant $N$, which is determined in principle
but not computed here.
This justifies our claim in \autoref{sec:defining-nab-map}.

\section{Explicit Heisenberg blocks} \label{sec:explicit-blocks}

So far we have been describing a nonabelianization map
\begin{equation}
 \nab_\cW: \Conf(\tC,\Heis) \quad \to \quad \Conf(C,\Vir_{c=1} \otimes \Heis) \, .
\end{equation}

The existence of such a map is already interesting in the abstract. It 
becomes particularly useful if we have a way to make elements in 
$\Conf(\tC,\Heis)$. In this section we describe one such way.
The basic ingredients are meromorphic forms, theta functions and Bergman kernels 
on $\tC$, well known in the literature on free fields on 
Riemann surfaces. Discussions close in spirit to ours are given in e.g.
\cite{VERLINDE1987357,Takhtajan:2001sv,Raina:1989ba}. One novelty in our presentation
is that we emphasize the organizational role of the log-Verlinde loop operators.

Throughout this section we work on a compact Riemann surface $\tC$ of genus $\tg$.
In this section, it is not important that $\tC$ arises as a 
double cover of another surface; we use the notation $\tC$ because
we have the application to nonabelianization ultimately in mind.

\subsection{Log-Verlinde operators on Heisenberg blocks}

We consider loop operators acting on $\Conf(\tC,\Heis)$, defined by
\begin{equation} \label{eq:log-verlinde}
\ell_\gamma = \oint_\gamma \tJ \, .
\end{equation}
What this means is
that for an arbitrary conformal 
block $\tPsi \in \Conf(\tC,\Heis)$, $\ell_{\gamma}(\tPsi)$ is given by
\begin{equation} 
\IP{\cdots}_{\ell_{\gamma}(\tPsi)} = \oint_{\gamma} \IP{\cdots \tJ(p)^z}_{\tPsi} \de z(p) \, .
\end{equation}
We can also represent this in a more condensed notation, writing $\tJ$ for 
$\tJ^z \de z$:
\begin{equation}
\IP{\cdots}_{\ell_{\gamma}(\tPsi)} = \IP{\cdots\oint_{\gamma}\tJ}_{\tPsi} \, .
\end{equation}
To see that this indeed gives a well defined
operator on $\Conf(\tC,\Heis)$ we use the fact that
the OPE \eqref{eq:heisenberg-ope} has no residue term, and thus we can freely deform the 
contour $\gamma$ across insertions of $\tJ$.
We call the $\ell_\gamma$ \ti{log-Verlinde operators}, anticipating a relation to the
Verlinde operators, to be discussed in \autoref{sec:verlinde-operators}.

The log-Verlinde operators associated to intersecting loops do not commute with one another:
instead, as we will now show, they obey
\begin{equation} \label{eq:J-commutator}
 \left[\ell_\gamma, \ell_{\mu}\right] = - 2 \pi \I \IP{\gamma,\mu}
\end{equation}
where $\IP{\cdot,\cdot}$ denotes the intersection pairing.
For simplicity
we draw pictures for the case of $\tC = T^2$ and $\IP{\gamma,\mu} = -1$, but the computation is similar for arbitrary $\tC$ and
$\gamma$, $\mu$. We take $\gamma$ to be a straight line from bottom to top and $\mu$ to be from left to right. Then define
\begin{align} c_1 & \equiv\IP{\cdots}_{\ell_{\mu}(\ell_{\gamma}(\tPsi))}=\IP{\cdots\oint_{\mu}\tJ}_{\ell_{\gamma}(\tPsi)}\\
& = \oint_{w\in\mu}\IP{\cdots \tJ(w)\oint_{\gamma(w)}\tJ}_\tPsi\\
& = \oint_{w\in\mu}\oint_{z\in\gamma(w)}\IP{\cdots \tJ(w)\tJ(z)}_\tPsi\\
&= \quad\begin{tikzpicture}[baseline=-.5ex]
\coordinate (c1) at (0,0);
\coordinate (c1cd) at (0,-0.2);
\coordinate (c1cu) at (0,0.2);
\coordinate (c1cr) at (0.2,0);
\coordinate (v1uu) at (0,0.6);
\coordinate (v1u) at (0,0.4);
\coordinate (v1d) at (0,-0.4);
\coordinate (v1dd) at (0,-0.6);
\coordinate (v1l) at (-0.3,0);
\coordinate (v1ll) at (-0.6,0);
\coordinate (v1rr) at (0.6,0);
\draw [line width=0.3mm,->] (c1cd) to[out=0,in=270 ] (c1cr);
\draw [line width=0.3mm] (c1cr) to[out=90,in=0] (c1cu);
\draw [line width=0.3mm,->] (c1cu) -- (v1u);
\draw [line width=0.3mm] (v1u) -- (v1uu);
\draw [line width=0.3mm] (v1d) -- (c1cd);
\draw [line width=0.3mm,->] (v1dd) -- (v1d);
\draw [line width=0.3mm,->] (v1ll) -- (v1l);
\draw [line width=0.3mm] (v1l) -- (c1);
\coordinate (v1ul) at (-0.6,0.6);
\coordinate (v1dl) at (-0.6,-0.6);
\coordinate (v1ur) at (0.6,0.6);
\coordinate (v1dr) at (0.6,-0.6);
\draw[line width=0.3mm,gray,dashed] (v1dl) to (v1ul);
\draw[line width=0.3mm,gray,dashed] (v1dl) to (v1dr);
\draw[line width=0.3mm,gray,dashed] (v1dr) to (v1ur);
\draw[line width=0.3mm,gray,dashed] (v1ur) to (v1ul);
\node [circle, fill, scale=0.2] at (-0.6,0) {};
\node [circle, fill, scale=0.2] at (c1) {};
\end{tikzpicture}
\quad+\quad
\begin{tikzpicture}[baseline=-.5ex]
\coordinate (c1) at (0,0);
\coordinate (c1cd) at (0,-0.2);
\coordinate (c1cu) at (0,0.2);
\coordinate (c1cl) at (-0.2,0);
\coordinate (v1uu) at (0,0.6);
\coordinate (v1u) at (0,0.4);
\coordinate (v1d) at (0,-0.4);
\coordinate (v1dd) at (0,-0.6);
\coordinate (v1r) at (0.3,0);
\coordinate (v1rr) at (0.6,0);
\coordinate (v1ll) at (-0.6,0);
\draw [line width=0.3mm,->] (c1cd) to[out=180,in=270 ] (c1cl);
\draw [line width=0.3mm] (c1cl) to[out=90,in=180] (c1cu);
\draw [line width=0.3mm,->] (c1cu) -- (v1u);
\draw [line width=0.3mm] (v1u) -- (v1uu);
\draw [line width=0.3mm] (v1d) -- (c1cd);
\draw [line width=0.3mm,->] (v1dd) -- (v1d);
\draw [line width=0.3mm] (c1) -- (v1rr);
\draw [line width=0.3mm,->]  (c1) -- (v1r);
\node [circle, fill, scale=0.2] at (c1) {};
\coordinate (v1ul) at (-0.6,0.6);
\coordinate (v1dl) at (-0.6,-0.6);
\coordinate (v1ur) at (0.6,0.6);
\coordinate (v1dr) at (0.6,-0.6);
\draw[line width=0.3mm,gray,dashed] (v1dl) to (v1ul);
\draw[line width=0.3mm,gray,dashed] (v1dl) to (v1dr);
\draw[line width=0.3mm,gray,dashed] (v1dr) to (v1ur);
\draw[line width=0.3mm,gray,dashed] (v1ur) to (v1ul);
\node [circle, fill, scale=0.2] at (0.6,0) {};
\end{tikzpicture}  \, ,
\end{align}
where $\gamma(w)$ means we regard the contour $\gamma$ as a function of $w \in \mu$, such that $\gamma(w)$ 
is homologous to the original $\gamma$ but deformed to avoid $w$ (thus $\gamma(w)$ necessarily depends
discontinuously on $w$ as indicated above.)
Similarly, define
\begin{align} c_2&\equiv\IP{\cdots}_{\ell_{\gamma}(\ell_{\mu}(\tPsi))}
=\oint_{z\in\gamma}\oint_{w\in\mu(z)}\IP{\cdots \tJ(w)\tJ(z)}_\tPsi \\
&= \quad \begin{tikzpicture}[baseline=-.5ex]
\coordinate (c1) at (0,0);
\coordinate (c1cr) at (0.2,0);
\coordinate (c1cu) at (0,0.2);
\coordinate (c1cl) at (-0.2,0);
\coordinate (v1ll) at (-0.6,0);
\coordinate (v1l) at (-0.4,0);
\coordinate (v1d) at (0,-0.3);
\coordinate (v1dd) at (0,-0.6);
\coordinate (v1uu) at (0,0.6);
\coordinate (v1r) at (0.4,0);
\coordinate (v1rr) at (0.6,0);
\draw [line width=0.3mm] (c1cu) to[out=0,in=90 ] (c1cr);
\draw [line width=0.3mm,->] (c1cl) to[out=90,in=180] (c1cu);
\draw [line width=0.3mm,->] (v1ll) -- (v1l);
\draw [line width=0.3mm] (v1l) -- (c1cl);
\draw [line width=0.3mm] (c1cr) -- (v1r);
\draw [line width=0.3mm,->] (v1r) -- (v1rr);
\draw [line width=0.3mm] (v1d) -- (c1) ;
\draw [line width=0.3mm,->]  (v1dd) -- (v1d);
\coordinate (v1ul) at (-0.6,0.6);
\coordinate (v1dl) at (-0.6,-0.6);
\coordinate (v1ur) at (0.6,0.6);
\coordinate (v1dr) at (0.6,-0.6);
\draw[line width=0.3mm,gray,dashed] (v1dl) to (v1ul);
\draw[line width=0.3mm,gray,dashed] (v1dl) to (v1dr);
\draw[line width=0.3mm,gray,dashed] (v1dr) to (v1ur);
\draw[line width=0.3mm,gray,dashed] (v1ur) to (v1ul);
\node [circle, fill, scale=0.2] at (0,-0.6) {};
\node [circle, fill, scale=0.2] at (c1) {};
\end{tikzpicture}
\quad+\quad
\begin{tikzpicture}[baseline=-.5ex]
\coordinate (c1) at (0,0);
\coordinate (c1cr) at (0.2,0);
\coordinate (c1cd) at (0,-0.2);
\coordinate (c1cl) at (-0.2,0);
\coordinate (v1ll) at (-0.6,0);
\coordinate (v1l) at (-0.4,0);
\coordinate (v1u) at (0,0.3);
\coordinate (v1uu) at (0,0.6);
\coordinate (v1dd) at (0,-0.6);
\coordinate (v1r) at (0.4,0);
\coordinate (v1rr) at (0.6,0);
\draw [line width=0.3mm] (c1cd) to[out=0,in=270 ] (c1cr);
\draw [line width=0.3mm,->] (c1cl) to[out=270,in=180] (c1cd);
\draw [line width=0.3mm,->] (v1ll) -- (v1l);
\draw [line width=0.3mm] (v1l) -- (c1cl);
\draw [line width=0.3mm] (c1cr) -- (v1r);
\draw [line width=0.3mm,->] (v1r) -- (v1rr);
\draw [line width=0.3mm] (v1u) -- (v1uu) ;
\draw [line width=0.3mm,->]  (c1) -- (v1u);
\coordinate (v1ul) at (-0.6,0.6);
\coordinate (v1dl) at (-0.6,-0.6);
\coordinate (v1ur) at (0.6,0.6);
\coordinate (v1dr) at (0.6,-0.6);
\draw[line width=0.3mm,gray,dashed] (v1dl) to (v1ul);
\draw[line width=0.3mm,gray,dashed] (v1dl) to (v1dr);
\draw[line width=0.3mm,gray,dashed] (v1dr) to (v1ur);
\draw[line width=0.3mm,gray,dashed] (v1ur) to (v1ul);
\node [circle, fill, scale=0.2] at (0,0.6) {};
\node [circle, fill, scale=0.2] at (c1) {};
\end{tikzpicture} 
\, .
\end{align}
Then we have
\begin{align}
c_1-c_2
&= \quad\begin{tikzpicture}[baseline=-.5ex]
\coordinate (c1) at (0,0);
\coordinate (c1cd) at (0,-0.2);
\coordinate (c1cu) at (0,0.2);
\coordinate (c1cr) at (0.2,0);
\coordinate (v1uu) at (0,0.6);
\coordinate (v1u) at (0,0.4);
\coordinate (v1d) at (0,-0.4);
\coordinate (v1dd) at (0,-0.6);
\coordinate (v1l) at (-0.3,0);
\coordinate (v1ll) at (-0.6,0);
\coordinate (v1rr) at (0.6,0);
\draw [line width=0.3mm,->] (c1cd) to[out=0,in=270 ] (c1cr);
\draw [line width=0.3mm] (c1cr) to[out=90,in=0] (c1cu);
\draw [line width=0.3mm,->] (c1cu) -- (v1u);
\draw [line width=0.3mm] (v1u) -- (v1uu);
\draw [line width=0.3mm] (v1d) -- (c1cd);
\draw [line width=0.3mm,->] (v1dd) -- (v1d);
\draw [line width=0.3mm,->] (v1ll) -- (v1l);
\draw [line width=0.3mm] (v1l) -- (c1);
\coordinate (v1ul) at (-0.6,0.6);
\coordinate (v1dl) at (-0.6,-0.6);
\coordinate (v1ur) at (0.6,0.6);
\coordinate (v1dr) at (0.6,-0.6);
\draw[line width=0.3mm,gray,dashed] (v1dl) to (v1ul);
\draw[line width=0.3mm,gray,dashed] (v1dl) to (v1dr);
\draw[line width=0.3mm,gray,dashed] (v1dr) to (v1ur);
\draw[line width=0.3mm,gray,dashed] (v1ur) to (v1ul);
\node [circle, fill, scale=0.2] at (-0.6,0) {};
\node [circle, fill, scale=0.2] at (c1) {};
\end{tikzpicture}
\quad+\quad
\begin{tikzpicture}[baseline=-.5ex]
\coordinate (c1) at (0,0);
\coordinate (c1cd) at (0,-0.2);
\coordinate (c1cu) at (0,0.2);
\coordinate (c1cl) at (-0.2,0);
\coordinate (v1uu) at (0,0.6);
\coordinate (v1u) at (0,0.4);
\coordinate (v1d) at (0,-0.4);
\coordinate (v1dd) at (0,-0.6);
\coordinate (v1r) at (0.3,0);
\coordinate (v1rr) at (0.6,0);
\coordinate (v1ll) at (-0.6,0);
\draw [line width=0.3mm,->] (c1cd) to[out=180,in=270 ] (c1cl);
\draw [line width=0.3mm] (c1cl) to[out=90,in=180] (c1cu);
\draw [line width=0.3mm,->] (c1cu) -- (v1u);
\draw [line width=0.3mm] (v1u) -- (v1uu);
\draw [line width=0.3mm] (v1d) -- (c1cd);
\draw [line width=0.3mm,->] (v1dd) -- (v1d);
\draw [line width=0.3mm] (c1) -- (v1rr);
\draw [line width=0.3mm,->]  (c1) -- (v1r);
\node [circle, fill, scale=0.2] at (c1) {};
\coordinate (v1ul) at (-0.6,0.6);
\coordinate (v1dl) at (-0.6,-0.6);
\coordinate (v1ur) at (0.6,0.6);
\coordinate (v1dr) at (0.6,-0.6);
\draw[line width=0.3mm,gray,dashed] (v1dl) to (v1ul);
\draw[line width=0.3mm,gray,dashed] (v1dl) to (v1dr);
\draw[line width=0.3mm,gray,dashed] (v1dr) to (v1ur);
\draw[line width=0.3mm,gray,dashed] (v1ur) to (v1ul);
\node [circle, fill, scale=0.2] at (0.6,0) {};
\end{tikzpicture}
\quad-\quad
\begin{tikzpicture}[baseline=-.5ex]
\coordinate (c1) at (0,0);
\coordinate (c1cr) at (0.2,0);
\coordinate (c1cu) at (0,0.2);
\coordinate (c1cl) at (-0.2,0);
\coordinate (v1ll) at (-0.6,0);
\coordinate (v1l) at (-0.4,0);
\coordinate (v1d) at (0,-0.3);
\coordinate (v1dd) at (0,-0.6);
\coordinate (v1uu) at (0,0.6);
\coordinate (v1r) at (0.4,0);
\coordinate (v1rr) at (0.6,0);
\draw [line width=0.3mm] (c1cu) to[out=0,in=90 ] (c1cr);
\draw [line width=0.3mm,->] (c1cl) to[out=90,in=180] (c1cu);
\draw [line width=0.3mm,->] (v1ll) -- (v1l);
\draw [line width=0.3mm] (v1l) -- (c1cl);
\draw [line width=0.3mm] (c1cr) -- (v1r);
\draw [line width=0.3mm,->] (v1r) -- (v1rr);
\draw [line width=0.3mm] (v1d) -- (c1) ;
\draw [line width=0.3mm,->]  (v1dd) -- (v1d);
\coordinate (v1ul) at (-0.6,0.6);
\coordinate (v1dl) at (-0.6,-0.6);
\coordinate (v1ur) at (0.6,0.6);
\coordinate (v1dr) at (0.6,-0.6);
\draw[line width=0.3mm,gray,dashed] (v1dl) to (v1ul);
\draw[line width=0.3mm,gray,dashed] (v1dl) to (v1dr);
\draw[line width=0.3mm,gray,dashed] (v1dr) to (v1ur);
\draw[line width=0.3mm,gray,dashed] (v1ur) to (v1ul);
\node [circle, fill, scale=0.2] at (0,-0.6) {};
\node [circle, fill, scale=0.2] at (c1) {};
\end{tikzpicture}
\quad-\quad
\begin{tikzpicture}[baseline=-.5ex]
\coordinate (c1) at (0,0);
\coordinate (c1cr) at (0.2,0);
\coordinate (c1cd) at (0,-0.2);
\coordinate (c1cl) at (-0.2,0);
\coordinate (v1ll) at (-0.6,0);
\coordinate (v1l) at (-0.4,0);
\coordinate (v1u) at (0,0.3);
\coordinate (v1uu) at (0,0.6);
\coordinate (v1dd) at (0,-0.6);
\coordinate (v1r) at (0.4,0);
\coordinate (v1rr) at (0.6,0);
\draw [line width=0.3mm] (c1cd) to[out=0,in=270 ] (c1cr);
\draw [line width=0.3mm,->] (c1cl) to[out=270,in=180] (c1cd);
\draw [line width=0.3mm,->] (v1ll) -- (v1l);
\draw [line width=0.3mm] (v1l) -- (c1cl);
\draw [line width=0.3mm] (c1cr) -- (v1r);
\draw [line width=0.3mm,->] (v1r) -- (v1rr);
\draw [line width=0.3mm] (v1u) -- (v1uu) ;
\draw [line width=0.3mm,->]  (c1) -- (v1u);
\coordinate (v1ul) at (-0.6,0.6);
\coordinate (v1dl) at (-0.6,-0.6);
\coordinate (v1ur) at (0.6,0.6);
\coordinate (v1dr) at (0.6,-0.6);
\draw[line width=0.3mm,gray,dashed] (v1dl) to (v1ul);
\draw[line width=0.3mm,gray,dashed] (v1dl) to (v1dr);
\draw[line width=0.3mm,gray,dashed] (v1dr) to (v1ur);
\draw[line width=0.3mm,gray,dashed] (v1ur) to (v1ul);
\node [circle, fill, scale=0.2] at (0,0.6) {};
\node [circle, fill, scale=0.2] at (c1) {};
\end{tikzpicture}\\
&= \quad \begin{tikzpicture}[baseline=-.5ex]
\coordinate (c1) at (0,0);
\coordinate (c1cd) at (0,-0.2);
\coordinate (c1cu) at (0,0.2);
\coordinate (c1cr) at (0.2,0);
\coordinate (c1cl) at (-0.2,0);
\coordinate (v1uu) at (0,0.6);
\coordinate (v1u) at (0,0.4);
\coordinate (v1d) at (0,-0.4);
\coordinate (v1dd) at (0,-0.6);
\coordinate (v1l) at (-0.3,0);
\coordinate (v1ll) at (-0.6,0);
\coordinate (v1rr) at (0.6,0);
\coordinate (ls) at (0,-0.8);
\draw [line width=0.3mm,->] (c1cd) to[out=0,in=270 ] (c1cr);
\draw [line width=0.3mm] (c1cr) to[out=90,in=0] (c1cu);
\draw [line width=0.3mm,->](c1cl) -- (c1);
\coordinate (v1ul) at (-0.6,0.6);
\coordinate (v1dl) at (-0.6,-0.6);
\coordinate (v1ur) at (0.6,0.6);
\coordinate (v1dr) at (0.6,-0.6);
\draw[line width=0.3mm,gray,dashed] (v1dl) to (v1ul);
\draw[line width=0.3mm,gray,dashed] (v1dl) to (v1dr);
\draw[line width=0.3mm,gray,dashed] (v1dr) to (v1ur);
\draw[line width=0.3mm,gray,dashed] (v1ur) to (v1ul);
\node [circle, fill, scale=0.2] at (c1) {};
\end{tikzpicture}
\quad+\quad
\begin{tikzpicture}[baseline=-.5ex]
\coordinate (c1) at (0,0);
\coordinate (c1cd) at (0,-0.2);
\coordinate (c1cu) at (0,0.2);
\coordinate (c1cl) at (-0.2,0);
\coordinate (c1cr) at (0.2,0);
\coordinate (v1uu) at (0,0.6);
\coordinate (v1u) at (0,0.4);
\coordinate (v1d) at (0,-0.4);
\coordinate (v1dd) at (0,-0.6);
\coordinate (v1r) at (0.3,0);
\coordinate (v1rr) at (0.6,0);
\coordinate (v1ll) at (-0.6,0);
\coordinate (ls) at (0,-0.8);
\draw [line width=0.3mm,->] (c1cd) to[out=180,in=270 ] (c1cl);
\draw [line width=0.3mm] (c1cl) to[out=90,in=180] (c1cu);
\draw [line width=0.3mm,->]  (c1) -- (c1cr);
\coordinate (v1ul) at (-0.6,0.6);
\coordinate (v1dl) at (-0.6,-0.6);
\coordinate (v1ur) at (0.6,0.6);
\coordinate (v1dr) at (0.6,-0.6);
\draw[line width=0.3mm,gray,dashed] (v1dl) to (v1ul);
\draw[line width=0.3mm,gray,dashed] (v1dl) to (v1dr);
\draw[line width=0.3mm,gray,dashed] (v1dr) to (v1ur);
\draw[line width=0.3mm,gray,dashed] (v1ur) to (v1ul);
\node [circle, fill, scale=0.2] at (c1) {};
\end{tikzpicture}
\quad-\quad
\begin{tikzpicture}[baseline=-.5ex]
\coordinate (c1) at (0,0);
\coordinate (c1cr) at (0.2,0);
\coordinate (c1cu) at (0,0.2);
\coordinate (c1cl) at (-0.2,0);
\coordinate (v1cd) at (0,-0.2);
\coordinate (v1ll) at (-0.6,0);
\coordinate (v1l) at (-0.4,0);
\coordinate (v1d) at (0,-0.3);
\coordinate (v1dd) at (0,-0.6);
\coordinate (v1r) at (0.4,0);
\coordinate (v1rr) at (0.6,0);
\coordinate (v1ll) at (-0.6,0);
\coordinate (ls) at (0,-0.8);
\draw [line width=0.3mm] (c1cu) to[out=0,in=90 ] (c1cr);
\draw [line width=0.3mm,->] (c1cl) to[out=90,in=180] (c1cu);
\draw [line width=0.3mm,->] (v1cd) -- (c1) ;
\coordinate (v1ul) at (-0.6,0.6);
\coordinate (v1dl) at (-0.6,-0.6);
\coordinate (v1ur) at (0.6,0.6);
\coordinate (v1dr) at (0.6,-0.6);
\draw[line width=0.3mm,gray,dashed] (v1dl) to (v1ul);
\draw[line width=0.3mm,gray,dashed] (v1dl) to (v1dr);
\draw[line width=0.3mm,gray,dashed] (v1dr) to (v1ur);
\draw[line width=0.3mm,gray,dashed] (v1ur) to (v1ul);
\node [circle, fill, scale=0.2] at (c1) {};
\end{tikzpicture}
\quad-\quad
\begin{tikzpicture}[baseline=-.5ex]
\coordinate (c1) at (0,0);
\coordinate (c1cr) at (0.2,0);
\coordinate (c1cd) at (0,-0.2);
\coordinate (c1cu) at (0,0.2);
\coordinate (c1cl) at (-0.2,0);
\coordinate (v1ll) at (-0.6,0);
\coordinate (v1dd) at (0,-0.6);
\coordinate (v1l) at (-0.4,0);
\coordinate (v1u) at (0,0.3);
\coordinate (v1uu) at (0,0.6);
\coordinate (v1r) at (0.4,0);
\coordinate (v1rr) at (0.6,0);
\coordinate (ls) at (0,-0.8);
\draw [line width=0.3mm] (c1cd) to[out=0,in=270 ] (c1cr);
\draw [line width=0.3mm,->] (c1cl) to[out=270,in=180] (c1cd);
\draw [line width=0.3mm,->]  (c1) -- (c1cu);
\coordinate (v1ul) at (-0.6,0.6);
\coordinate (v1dl) at (-0.6,-0.6);
\coordinate (v1ur) at (0.6,0.6);
\coordinate (v1dr) at (0.6,-0.6);
\draw[line width=0.3mm,gray,dashed] (v1dl) to (v1ul);
\draw[line width=0.3mm,gray,dashed] (v1dl) to (v1dr);
\draw[line width=0.3mm,gray,dashed] (v1dr) to (v1ur);
\draw[line width=0.3mm,gray,dashed] (v1ur) to (v1ul);
\node [circle, fill, scale=0.2] at (c1) {};
\end{tikzpicture} \, .
\end{align}
Now we replace the integrand by its most singular part $\frac{1}{(z-w)^2} \IP{\cdots}$ (this is justified since we
can take the circle to be arbitrarily small, which will
kill all less singular terms),
and then use the fact that $\partial_z \partial_w \log(z-w) = \frac{1}{(z-w)^2}$. This gives finally
\begin{equation}
c_1 - c_2 = \left( \frac{\pi \I}{2} + \frac{\pi \I}{2} - \left(- \frac{\pi \I}{2}\right) - \left(- \frac{\pi \I}{2}\right) \right) \IP{\cdots}  \\ = 2\pi\ri \IP{\cdots}
\end{equation}
as desired.

\subsection{Constructing Heisenberg blocks explicitly} \label{sec:heisenberg-explicit-construction}

Fix a choice of $A$ and $B$ cycles on $\tC$, with the intersection condition
$\IP{A_i, B_j} = \delta_{ij}$,
and also fix a vector 
\begin{equation}
a = (a_1, \dots, a_{\tg}) \in \bbC^{\tg} \, .  
\end{equation}
We will construct a block $\tPsi_a \in \Conf(\tC, \Heis)$
determined by these data.
The block $\tPsi_a$ will be a joint eigenvector
of the log-Verlinde operators $\ell_{A_i}$ acting on $\Conf(\tC, \Heis)$, with eigenvalues $a_i$, i.e.
\begin{equation} \label{eq:Psi-A-cycle-property}
\ell_{A_i} \tPsi_a = a_i \tPsi_a \, .  
\end{equation}
In fact, this property
determines $\tPsi_a$ up to scale.
It would be impossible to diagonalize the operators $\ell_\gamma$ on \ti{all} $1$-cycles $\gamma$, 
because of \eqref{eq:J-commutator}. We will determine the overall scale of 
$\tPsi_a$ by the additional conditions
\begin{equation} \label{eq:Psi-B-cycle-property}
\ell_{B_i} \tPsi_a = 2 \pi \I \partial_{a_i} \tPsi_a \, , \qquad \IP{1}_{\tPsi_{a=0}} = 1 \, .
\end{equation}

We need some preliminaries on compact Riemann surfaces. 
Let $(\omega_1, \dots, \omega_{\tg})$ be the basis of holomorphic 1-forms dual to $(A_1, \dots, A_\tg)$, and let
\begin{equation}
 \eta_a = \sum_{i=1}^{\tg} a_i \omega_i \, . 
\end{equation}
Let $B(p,q)$ denote the Bergman kernel on $\tC$, normalized on the $A$ cycles:
this is the unique section of $T^* \tC \boxtimes T^* \tC$ over $\tC \times \tC$
which obeys $B(p,q) = B(q,p)$,
is holomorphic
except for a singularity
\begin{equation}
 B(p,q) = \frac{\de z(p) \boxtimes \de z(q)}{(z(p)-z(q))^2} + \regular 
\end{equation}
along the diagonal, and obeys $\oint_{p \in A_i} B(p,q) = 0$.
Finally let $\tau$ be the period matrix of $\tC$,
$\tau_{ij} = \oint_{B_j} \omega_i$.

For example, say $\tg=1$ and $\tC = \bbC / (\bbZ \oplus \tau \bbZ)$, with
the standard $A$ and $B$ cycles, and the standard coordinate $z \sim z+1 \sim z+\tau$. Then
\begin{equation} \label{eq:explicit-genus-1-data}
 \eta_a = a_1 \, \de z, \qquad B(z,w) = \left( \wp(\tau, z-w) + \frac{\pi^2}{3} E_2(\tau) \right) \de z \boxtimes \de w \, .
\end{equation}

We now give a direct construction of Heisenberg blocks $\tPsi_a$ with the properties \eqref{eq:Psi-A-cycle-property}, \eqref{eq:Psi-B-cycle-property}.
The correlation function
\begin{equation}
  \IP{\tJ(p_1) \cdots \tJ(p_n)}_{\tPsi_a}
\end{equation}
is $\e^{\frac{1}{4 \pi \I} a \cdot \tau a}$ times a
sum of Feynman diagrams with $n$ vertices labeled $p_1$, \dots, $p_n$, 
with all vertices either $0$-valent or $1$-valent;
a $0$-valent vertex gives a factor $\eta_a(p_i)$,
and an edge gives a factor $B(p_i,p_j)$.
\begin{center}
\begin{tikzpicture}   
    \draw[fill=black](0,0) circle (1pt) node [above] {$p_1$};
    \draw[fill=black](0,0) circle (1pt) node [below, yshift=-0.5cm] {$\eta_a(p_1)$};
    \draw[fill=black](2,0) circle (1pt) node [above] {$p_2$};
    \draw[fill=black](3,0) circle (0pt) node [below, yshift=-0.5cm] {$B(p_2,p_3)$};
    \draw[fill=black](4,0) circle (1pt) node [above] {$p_3$};
    \draw (2,0) -- (4,0);
    \draw[fill=black](6,0) circle (1pt) node [above] {$p_4$};
    \draw[fill=black](6,0) circle (1pt) node [below, yshift=-0.5cm] {$\eta_a(p_4)$};
\end{tikzpicture}
\end{center}
So, for example,
\begin{align}
  \IP{1}_{\tPsi_a} &= \e^{\frac{1}{4 \pi \I} a \cdot \tau a} \, ,  \\
  \IP{\tJ(p)}_{\tPsi_a} &= \e^{\frac{1}{4 \pi \I} a \cdot \tau a} \eta_a(p) \, ,  \\
  \IP{\tJ(p) \tJ(q)}_{\tPsi_a} &= \e^{\frac{1}{4 \pi \I} a \cdot \tau a} (\eta_a(p) \eta_a(q) + B(p,q)) \, , \\
  \IP{\tJ(p) \tJ(q) \tJ(r)}_{\tPsi_a} &= \e^{\frac{1}{4 \pi \I} a \cdot \tau a} (\eta_a(p) \eta_a(q) \eta_a(r) + B(p,q) \eta_a(r) + B(p,r) \eta_a(q) + B(q,r) \eta_a(p)) \, .  
\end{align}
(Again here we used a condensed notation, suppressing the local coordinate dependence, which is the same on both sides.)
One can check directly that $\tPsi_a$ has all the claimed properties.\footnote{To check \eqref{eq:Psi-A-cycle-property}, we need to use the fact that $\oint_{p \in A_i} B(p,q) = 0$; to check \eqref{eq:Psi-B-cycle-property}, we need $\oint_{p \in B_i} B(p,q) = 2 \pi \I \omega_i(q)$; see e.g. \cite{Eynard-lectures} for these properties.}

Having defined $\tPsi_a$ we can consider its fermion correlators.
Suppose given a spin structure
$K_\tC^{\frac12}$ and points $p, q \in \tC$ lying in a patch with coordinate $z$.
Then using \eqref{eq:heisenberg-unfusion} we get
\begin{equation} \label{eq:free-fermion-correlator-psia}
  \IP{\tikz[remember picture, baseline]{
  \node[anchor=base, inner sep=0pt] (psiplus1) {{$\psi_+(p)^z$}};
}
\,
\tikz[remember picture, baseline]{
  \node[anchor=base, inner sep=0pt] (psiminus1) {{$\psi_-(q)^z$}};
} }_{\tPsi_a} = \frac{1}{z(p) - z(q)} \exp \left[ \frac{a \cdot \tau a}{4 \pi \I} + \int_q^p \eta_a + \frac12 \int_q^p \int_q^p B(r_1,r_2) - \frac{\de z(r_1) \de z(r_2)}{(z(r_1) - z(r_2))^2} \right] \, .
\end{equation}
\begin{tikzpicture}[overlay, remember picture]
  \coordinate (start1) at ($(psiplus1.south)!0.5!(psiplus1.north)$);
  \coordinate (end1) at ($(psiminus1.south)!0.5!(psiminus1.north)$);
  \coordinate (start1Shifted) at ([yshift=-1.8ex]start1); 
  \coordinate (end1Shifted) at ([yshift=-1.8ex]end1);
  \draw[thick] (start1Shifted) -- (end1Shifted);
  \draw[thick] (start1Shifted) -- ++(0,0.5ex);
  \draw[thick] (end1Shifted) -- ++(0,0.5ex);

  \coordinate (mid1) at ($(start1Shifted)!0.5!(end1Shifted)$);
  \node at ($(mid1)+(0,-1.8ex)$) {$\ell$}; 
\end{tikzpicture}

\subsection{Heisenberg blocks with primaries inserted}

All of the foregoing can be extended to the case when we insert primaries $V_{\alpha_i}(q_i)$ on $\tC$, as 
follows.
We again fix a choice of $A$ and $B$ cycles on $\tC$, now taking care that they do not pass 
through any of the $q_i$, and fix $a = (a_1, \dots, a_{\tg}) \in \bbC^{\tg}$.
We will construct a block $\tPsi_a \in \Conf(\tC, \Heis; V_{\alpha_1}(q_1) \cdots V_{\alpha_k}(q_k))$
determined by these data. As before, $\tPsi_a$ will be engineered to 
obey \eqref{eq:Psi-A-cycle-property}, \eqref{eq:Psi-B-cycle-property}.
To construct $\tPsi_a$, 
let $\eta_a$ be the unique meromorphic 1-form on $\tC$ which has $\oint_{A_i} \eta_a = a_i$
and has poles at the $q_i$ with residues $\alpha_i$.
Then we have
\begin{equation}
 \oint_{B_i} \eta_a = \sum_{j=1}^{\tg} \tau_{ij} a_j + c_i 
\end{equation}
for some constants $c_i \in \bbC$.
The correlation functions 
\begin{equation}
  \IP{\tJ(p_1) \cdots \tJ(p_n) V_{\alpha_1}(q_1) \cdots V_{\alpha_k}(q_k)}_{\tPsi_a}
\end{equation}
are defined by the same rules as above, except that the prefactor is modified to include an additional 
term $\frac{1}{2 \pi \I} c \cdot a$, so e.g.
\begin{align}
  \IP{V_{\alpha_1}(q_1) \cdots V_{\alpha_k}(q_k)}_{\tPsi_a} &= \e^{\frac{1}{4 \pi \I} a \cdot \tau a + \frac{1}{2\pi \I} c \cdot a} \, \\
  \IP{\tJ(p) V_{\alpha_1}(q_1) \cdots V_{\alpha_k}(q_k)}_{\tPsi_a} &= \e^{\frac{1}{4 \pi \I} a \cdot \tau a + \frac{1}{2\pi \I} c \cdot a} \eta_a(p) \, ,
\end{align}
where we recall that $\eta_a$ is now meromorphic rather than holomorphic.

\subsection{Diagonalizing Verlinde operators on Heisenberg blocks} \label{sec:diagonalizing-abelian-verlinde}

We have just constructed a family of conformal blocks $\tPsi_a \in \Conf(\tC, \Heis)$, labeled by $a \in \bbC^{\tg}$,
and characterized up to overall normalization by \eqref{eq:Psi-A-cycle-property}, \eqref{eq:Psi-B-cycle-property}.
By taking linear combinations of the $\tPsi_a$ we now construct another useful family.

As we already remarked, we cannot simultaneously diagonalize the operators $\ell_\gamma$. 
But there is a closely related algebra which we can diagonalize.
Consider the \ti{Verlinde operators} $L_\gamma$ defined by\footnote{The exponential is defined as the sum $\sum_{n=0}^\infty \frac{\ell_\gamma^n}{n!}$, where $\ell_\gamma^n$ in turn is defined by point splitting, using $n$ slightly displaced copies of $\gamma$.}
\begin{equation} \label{eq:abelian-log-verlinde-relation}
  L_\gamma = \exp \ell_\gamma \, .
\end{equation}
(Again the name ``Verlinde'' anticipates \autoref{sec:verlinde-operators} below.)
It follows from \eqref{eq:J-commutator} and the Baker-Campbell-Hausdorff formula 
that these operators obey the twisted torus algebra,
\begin{equation} \label{eq:verlinde-multiplicativity}
  L_\gamma L_\mu = (-1)^{\IP{\gamma,\mu}} L_{\gamma + \mu} \, .
\end{equation}
(In particular, $L_\gamma$ and $L_\mu$ commute with one another.)
This algebra can also be described as the $\GL(1)$ skein algebra $\Sk_{-1}(\tC, \GL(1))$,
or dually as $\cO(\cM(\tC, \GL(1)))$, where $\cM(\tC, \GL(1))$ is the moduli space parameterizing
twisted $\GL(1)$-connections over $\tC$.

We can describe the action
of $L_\gamma$ on the blocks $\tPsi_a$: namely, by
\eqref{eq:Psi-A-cycle-property}, \eqref{eq:Psi-B-cycle-property}  we have
\begin{equation}
  L_{A_i} \tPsi_a = \exp(a_i) \tPsi_a, \qquad L_{B_i} \tPsi_a = \tPsi_{a + 2 \pi \I e_i} \, ,
\end{equation}
and this determines the action of all $L_\gamma$ using \eqref{eq:verlinde-multiplicativity}.

To build a common eigenvector of the $L_\gamma$, fix parameters $\left((x_1, \dots, x_\tg),(y_1, \dots, y_\tg)\right) \in \bbC^{2\tg}$, and define a block $\tPsi_{x,y} \in \Conf(\tC,\Heis)$ by
\begin{equation}\label{eq:diagonalizing-abelian-verlinde}
  \tPsi_{x,y} = \sum_{n \in \bbZ^\tg} \exp\left(-\frac{(x + 2 \pi \I n) \cdot y}{2 \pi \I}\right) \tPsi_{a = x + 2 \pi \I n} \, .
\end{equation}
Then we have
\begin{equation}
  L_{A_i} \tPsi_{x,y} = \exp(x_i) \tPsi_{x,y} \, , \qquad L_{B_i} \tPsi_{x,y} = \exp(y_i) \tPsi_{x,y} \, ,
\end{equation}
so $\tPsi_{x,y}$ indeed diagonalizes all of 
the $L_\gamma$. 
The eigenvalues $(\e^x, \e^y)$ can be understood more invariantly as
specifying a point $X \in \cM(\tC, \GL(1))$.
Using \eqref{eq:diagonalizing-abelian-verlinde} 
we can also describe the action of the log-Verlinde operators on the $\tPsi_{x,y}$:
\begin{equation} \label{eq:log-verlinde-on-psixy}
  \ell_{A_i} \tPsi_{x,y} = - 2 \pi \I \partial_{y_i} \tPsi_{x,y} \, , \qquad \ell_{B_i} \tPsi_{x,y} = (2 \pi \I \partial_{x_i} + y_i) \tPsi_{x,y} \, .
\end{equation}

Computing correlation functions explicitly
in the block $\tPsi_{x,y}$ using \eqref{eq:diagonalizing-abelian-verlinde}, we find:
\begin{itemize}
\item The $0$-point function is a Riemann theta function with characteristics,
\begin{align}
  \IP{1}_{\tPsi_{x,y}} &= \sum_{n \in \bbZ^g} \exp\left(-\frac{(x + 2 \pi \I n) \cdot y}{2 \pi \I}\right) \IP{1}_{\tPsi_{a = x + 2 \pi \I n}} \\
  &= \sum_{n \in \bbZ^g} \exp\left( -\frac{(x + 2 \pi \I n) \cdot y}{2 \pi \I} + \frac{1}{4 \pi \I} (x + 2 \pi \I n) \cdot \tau (x + 2 \pi \I n) \right) \\
\label{0pointtheta}  &= \exp\left(-\frac{x \cdot y}{2 \pi \I} + \frac{x \cdot \tau x}{4 \pi \I} \right) \Theta \left(\tau, u \right) \\
  &= \Theta\left[\frac{x}{2 \pi \I} \bigg\vert \frac{-y}{2 \pi \I}\right] (\tau, 0) \label{eq:theta-with-characteristics}
\end{align}
where $u \in \bbC^\tg$ is
\begin{equation}
  u = \frac{-y+\tau x}{2 \pi \I} \, .
\end{equation}
\item The $1$-point function of $\tJ$ is a derivative of the theta function,
\begin{equation}
  \IP{\tJ(p)}_{\tPsi_{x,y}} = - 2 \pi \I \sum_{i=1}^\tg \omega_i(p) \partial_{y_i} \IP{1}_{\tPsi_{x,y}} \, .
\end{equation}
Higher-point correlation functions of $\tJ$ are higher derivatives of theta functions.

\item Fix $p$, $q$ in a patch with coordinate $z$, with a leash in the patch, and a spin structure $K^{\frac12}$
and a choice of $\sqrt{\de z}$ in the patch. Then
the free fermion $2$-point function is
\Needspace{5\baselineskip}
\begin{multline}
  \IP{
    \tikz[remember picture, baseline]{
      \node[anchor=base, inner sep=0pt] (psiplus_mult) {$\psi_+(p)^z$};
    }
    \,
    \tikz[remember picture, baseline]{
      \node[anchor=base, inner sep=0pt] (psiminus_mult) {$\psi_-(q)^z$};
    }
  }_{\tPsi_{x,y}} = \\
  \exp\left[-\frac{x \cdot y}{2 \pi \I} + \frac{x \cdot \tau x}{4 \pi \I} + x\cdot\int_q^p \omega + \frac{1}{2} \left( \int_q^p \int_q^p B(r_1,r_2) - \frac{\de z(r_1) \de z(r_2)}{(z(r_1) - z(r_2))^2} \right) \right] \frac{\Theta \left(\tau, u + \int_q^p \omega \right)}{z(p) - z(q)} \, .
\end{multline}
\begin{tikzpicture}[overlay, remember picture]
  \coordinate (start) at (psiplus_mult.north);
  \coordinate (end) at (psiminus_mult.north);
  \coordinate (startShifted) at ([yshift=1ex]start);
  \coordinate (endShifted) at ([yshift=1ex]end);
  \draw[thick] (startShifted) -- (endShifted);
  \draw[thick] (startShifted) -- ++(0,-0.5ex);
  \draw[thick] (endShifted) -- ++(0,-0.5ex);
\end{tikzpicture}\noindent\unskip
It follows that the normalized 2-point function is
\begin{equation} \label{eq:normalized-fermion-2-point-diagonal-block}
  \frac{ \IP{
    \tikz[remember picture, baseline]{
      \node[anchor=base, inner sep=0pt] (psiplus) {$\psi_+(p)^z$};
    }
    \,
    \tikz[remember picture, baseline]{
      \node[anchor=base, inner sep=0pt] (psiminus) {$\psi_-(q)^z$};
    }
  }_{\tPsi_{x,y}} }{ \IP{1}_{\tPsi_{x,y}} } = 
  \frac{\exp\left[x\cdot\int_q^p \omega\right] \Theta \left(\tau, u + \int_q^p \omega \right)}{\Theta(\tau,u) \, E(p,q)^z} 
\end{equation}
\begin{tikzpicture}[overlay, remember picture]
  \coordinate (start) at (psiplus.north);
  \coordinate (end) at (psiminus.north);
  \coordinate (startShifted) at ([yshift=1ex]start);
  \coordinate (endShifted) at ([yshift=1ex]end);
  \draw[thick] (startShifted) -- (endShifted);
  \draw[thick] (startShifted) -- ++(0,-0.5ex);
  \draw[thick] (endShifted) -- ++(0,-0.5ex);
\end{tikzpicture}\noindent\unskip
where $E$ denotes the prime form, which in our notation is
\begin{equation}
  E(p,q)^z = (z(p) - z(q)) \, {\exp\left[-\frac12 \left( \int_q^p \int_q^p B(r_1,r_2) - \frac{\de z(r_1) \de z(r_2)}{(z(r_1) - z(r_2))^2} \right) \right]} \, .
\end{equation}
The normalized 2-point function \eqref{eq:normalized-fermion-2-point-diagonal-block} is also known as
the twisted Szeg\"o kernel.

\item The normalized fermion higher-point functions can
also be expressed in terms of this kernel, as follows.
Suppose all $p_i$ and $q_j$ lie in a single coordinate patch with coordinate $z$, and we take all leashes to lie in this patch,
and use a fixed spin structure and a fixed choice of $\sqrt{\de z}$ for all fermion insertions.
Then the normalized $2n$-fermion correlation functions are determinants of matrices of normalized $2$-fermion correlation functions:
\begin{equation} \label{eq:free-fermion-determinant-formula}
    \frac{\left\langle \cdots \prod_{i=1}^n 
    \tikz[remember picture, baseline]{
        \node[anchor=base, inner sep=0pt] (psiplus_num) {$\psi_+(p_i)$};
    }
    \,
    \tikz[remember picture, baseline]{
        \node[anchor=base, inner sep=0pt] (psiminus_num) {$\psi_-(q_i)$};
    }
    \right\rangle_{\tPsi_{x,y}}}{\left\langle \cdots\right\rangle_{\tPsi_{x,y}}} = \det \left( \left[ \frac{\left\langle \cdots 
    \tikz[remember picture, baseline]{
        \node[anchor=base, inner sep=0pt] (psiplus_det) {$\psi_+(p_i)$};
    }
    \,
    \tikz[remember picture, baseline]{
        \node[anchor=base, inner sep=0pt] (psiminus_det) {$\psi_-(q_j)$};
    }
    \right\rangle_{\tPsi_{x,y}}}{\left\langle \cdots\right\rangle_{\tPsi_{x,y}}} \right]_{i,j=1}^{n} \right) \, .
\end{equation}
\begin{tikzpicture}[overlay, remember picture]
    \coordinate (start1) at (psiplus_num.north);
    \coordinate (end1) at (psiminus_num.north);
    \coordinate (start1Shifted) at ([yshift=1ex]start1);
    \coordinate (end1Shifted) at ([yshift=1ex]end1);
    \draw[thick] (start1Shifted) -- (end1Shifted);
    \draw[thick] (start1Shifted) -- ++(0,-0.5ex);
    \draw[thick] (end1Shifted) -- ++(0,-0.5ex);

    \coordinate (start2) at (psiplus_det.north);
    \coordinate (end2) at (psiminus_det.north);
    \coordinate (start2Shifted) at ([yshift=1ex]start2);
    \coordinate (end2Shifted) at ([yshift=1ex]end2);
    \draw[thick] (start2Shifted) -- (end2Shifted);
    \draw[thick] (start2Shifted) -- ++(0,-0.5ex);
    \draw[thick] (end2Shifted) -- ++(0,-0.5ex);
\end{tikzpicture}\noindent\unskip
The formula \eqref{eq:free-fermion-determinant-formula} 
is a close relative of Fay's multisecant identity.
One can prove it using 
the fact that both sides have the same
monodromy around loops on $\tC$, have the same singularities 
when some $p_i \to q_j$ (and no other singularities),
and have zeroes when some $p_i = p_j$
or $q_i = q_j$. This proof is discussed in e.g. \cite{Raina:1989ba}.

\end{itemize}

Changing our choice of $A$ and $B$ cycles by 
an element of $\Sp(2g,\bbZ)$ changes the normalization of $\tPsi_{x,y}$
by a factor, which can be read out from the modular properties of the
Riemann theta function. For instance:
\begin{itemize}
\item taking $A'_i = A_i$ and
$B'_i = B_i + c_{ij} A_j$, where all $c_{ij} \in 2 \bbZ$, gives
$\tau' = \tau + c$ and $y' = y + c x$,
and then
$\tPsi'_{x',y'} = \exp \left(- \frac{x \cdot c x}{4 \pi \I} \right) \tPsi_{x,y}$.
\item 
taking $A'_i = B_i$, $B'_i = -A_i$,
gives $\tau' = - \tau^{-1}$, $x' = y$, $y' = -x$,
and then $\tPsi'_{x',y'} = (\det(-\I \tau))^{\frac12} \exp \left(\frac{x \cdot y}{2 \pi \I}\right) \tPsi_{x,y}$.
\end{itemize}

\subsection{The line bundle of eigenblocks} \label{sec:heisenberg-line-bundle}

As we have just discussed, for each $X \in \cM(\tC,\GL(1))$ we have a corresponding
$1$-dimensional space of Verlinde eigenblocks in $\Conf(\tC,\Heis)$. These eigenspaces make up a line bundle $\tcL$ over $\cM(\tC,\GL(1))$. 

One of the important geometric features of $\tcL$ 
is that it carries a holomorphic connection, 
whose curvature is the standard (Atiyah-Bott-Goldman) holomorphic symplectic form on 
$\cM(\tC, \GL(1))$.
This connection can be built directly from the 
log-Verlinde operators \eqref{eq:log-verlinde}.
Indeed, note that from \eqref{eq:J-commutator} we get
\begin{equation} \label{eq:l-L-commutator}
  [\ell_\gamma, L_{\mu}] = - 2 \pi \I \IP{\gamma,\mu} L_{\mu} \, .
\end{equation}
Thus $\ell_\gamma$ can be used to shift the eigenvalue of $L_\mu$. Said more precisely:
for any $\gamma \in H_1(\tC,\bbZ)$ there is a corresponding vector field $v_\gamma$ on $\cM(\tC, \GL(1))$,
which acts on functions by $v_\gamma(X_\mu) = \IP{\gamma,\mu} X_\mu$.
From \eqref{eq:l-L-commutator} it follows that the operator
\begin{equation}
  \tnabla_\gamma = v_\gamma - \frac{1}{2 \pi \I} \ell_\gamma
\end{equation}
preserves the eigenline bundle $\tcL$. As $\gamma$ varies, the $v_\gamma$ span $T\cM(\tC, \GL(1))$,
and their lifts $\tnabla_\gamma$ give a connection in $\tcL$.
The curvature of this connection is determined by \eqref{eq:J-commutator}:
\begin{equation}
  F(v_\gamma, v_\mu) = \frac{1}{2 \pi \I} \IP{\gamma, \mu} \, .
\end{equation}
This is indeed the Atiyah-Bott form on $\cM(\tC, \GL(1))$.

Here is another viewpoint on this connection. A tangent vector to $\cM(\tC, \GL(1))$ can be represented
by a closed complex 1-form $\beta \in \Omega^{1}(\tC)$. The variation of a Verlinde
eigenblock $\Psi$ in the direction $\beta$ is
\begin{equation} \label{eq:derivative-abelian}
  \IP{ \cdots }_{\tnabla_\beta \tPsi} = \partial_\beta \IP{ \cdots }_{\tPsi} - \IP{ \cdots  \int_\tC \beta \tJ }_{\tPsi} \, .
\end{equation}
In other words, $\tJ$ is the operator which generates an infinitesimal variation of the flat connection,
much as $\tT^\Heis$ generates an infinitesimal variation of the conformal structure.
We recover the previous description of the connection by choosing $\beta$ to be a delta-function supported on a loop in $\tC$.

Our specific construction of the eigenblock
$\tPsi_{x,y}$ by the formula \eqref{eq:diagonalizing-abelian-verlinde} provides a 
local trivialization of the line bundle $\tcL$.
The normalization of $\tPsi_{x,y}$ depends in a quasiperiodic way on $(x,y)$:
\begin{equation} \label{eq:L-construction}
  \tPsi_{x + 2 \pi \I e_i, y} = \tPsi_{x,y}, \qquad \tPsi_{x, y + 2 \pi \I e_i} = \exp\left(-x_i\right) \tPsi_{x,y} \, .
\end{equation}
Moreover, using \eqref{eq:log-verlinde-on-psixy} we see that, 
relative to the local gauge $\tPsi_{x,y}$, the connection
$1$-form is
\begin{equation} \label{eq:connection-local-gauge-explicit}
  A = \frac{1}{2 \pi \I} \sum_{i=1}^\tg y_i \, \de x_i \, .
\end{equation}

\subsection{Variation of moduli} \label{sec:variation-of-moduli}

In this section we briefly discuss how Heisenberg blocks behave under 
variation of the moduli of $\tC$ in the moduli space $\cM_{\tg}$ of genus $\tg$ curves.

First take the 
special case $\tg = 1$. In this case we have $\tC = \bbC / (\bbZ \oplus \tau\bbZ)$
and we can choose the complex projective structure induced by the standard coordinate $z$
on $\bbC$. Then we get a connection on the spaces of conformal blocks
as in \autoref{sec:connections-on-blocks}.
Because the $L_\gamma$ are topological this connection
must preserve the eigenspaces; said otherwise, the connection in the line bundle $\tcL \to \cM(\tC, \GL(1))$
extends to a connection in a line bundle over a larger moduli space, $\tcL \to \cM(\tC, \GL(1)) \times \cM_1$.
We use the notation $\tnabla$ for both connections.

To compute $\tnabla$ it is enough to consider the $0$-point
function. The tangent vector $\partial_\tau$ to $\cM_1$
comes from the Beltrami differential $\mu^z = \frac{1}{\im \tau}$. Then
using \eqref{eq:T-connection-rough} we have
\begin{align} \label{eq:deltataupsi}
  \IP{1}_{\tnabla_\tau \tPsi_{x,y}} &= \partial_\tau \left( \IP{1}_{\tPsi_{x,y}} \right) - \frac{1}{2 \pi \I} \int_\tC \mu(p)^z \IP{T(p)^z}_{\tPsi_{x,y}} \de z \de \overline{z} 
  \\ &=
     \partial_\tau \IP{1}_{\tPsi_{x,y}} - \frac{1}{4 \pi \I} \IP{\nop{\tJ(0)^2}}_{\tPsi_{x,y}}
\end{align}
using translation invariance.
Using the explicit formulas \eqref{eq:explicit-genus-1-data}
and \eqref{eq:diagonalizing-abelian-verlinde},
we obtain
\begin{align}
\IP{\nop{\tJ(0)^2}}_{\tPsi_{x,y}}&=\lim_{p\rightarrow 0}\IP{\tJ(p)\tJ(0)-\frac{1}{p^2}}_{\tPsi_{x,y}}\\
&=\left(-4\pi^2\partial_y^2+ \frac{\pi^2}{3} E_2(\tau)\right)\IP{1}_{\tPsi_{x,y}} \, .
\end{align}
Then \eqref{eq:deltataupsi} reduces to
\begin{align}
  \IP{1}_{\tnabla_\tau \tPsi_{x,y}} &= \left( (\partial_\tau - \pi \I \partial_y^2) + \frac{\pi \I}{12} E_2(\tau) \right) \IP{1}_{\tPsi_{x,y}} \\ 
  &= \frac{\pi \I}{12} E_2(\tau) \IP{1}_{\tPsi_{x,y}} \, ,
\end{align}
so we conclude the connection form in this direction is
\begin{equation} \label{eq:A-explicit-genus1}
  A = \frac{\pi \I}{12} E_2(\tau) \, \de \tau = \de \log \eta(\tau) \, .
\end{equation}
Said otherwise, the renormalized eigenblocks
\begin{equation} 
  \hat\tPsi_{x,y} = \eta(\tau)^{-1} \tPsi_{x,y}
\end{equation}
are covariantly constant under variations of $\tC$.

Now let us discuss the analogous structure for higher genus $\tC$: it is similar to the $\tg = 1$ case, 
only with less explicit formulas.
We choose a local section $\tS$ of the bundle of complex projective
structures over $\cM_{\tg}$; in contrast to the $\tg = 1$ case, we do not have a particularly natural choice here, so we
just leave it general. Having made this choice we get a connection $\tnabla$ on the bundle of conformal blocks over $\cM_{\tg}$, 
as described in \autoref{sec:connections-on-blocks}. Choosing some particular $(x,y)$, this connection 
has $\tnabla \tPsi_{x,y} = A \tPsi_{x,y}$, for some local $1$-form $A$ on $\cM_{\tg}$, the analogue of \eqref{eq:A-explicit-genus1} above.
Contracting this $1$-form with a tangent vector to $\cM_\tg$, i.e. a Beltrami differential $\tmu$ on $\tC$, 
should give us a number; 
a similar computation to the one we made in the $\tg = 1$ case gives this number as
\begin{equation} \label{eq:Mg-connection-form}
  A \cdot \tmu = \frac12 \int_{\tC} \tmu^z \left(\lim_{p \to q} \left( B(p,q)^z - \frac{1}{(z(p)-z(q))^2} \right) \right) \, .
\end{equation}
On the right side, we use coordinates $z$ in the atlas determined by the chosen complex projective structure $\tS$; 
thus $A$ depends on this choice as expected.
On the other hand, $A$ is independent of $(x,y)$, again as expected.\footnote{The connection form $A$ is not modular invariant: it depends on the choice of $A$ and $B$ cycles through the Bergman kernel. This property is to be expected, since $A$ represents the connection $\tnabla$ relative to the trivialization given by the blocks $\tPsi_{x,y}$, which are also not modular invariant. 
(It is already visible in case $\tg = 1$, where
it arises from the inhomogeneous term in the modular transformation of $E_2$.) 
The connection $\tnabla$ itself is modular invariant as it should be: it depends only on $\tS$, not additionally on a choice of $A$ and $B$ cycles.}

Depending on which $\tS$ we choose, this connection over $\cM_\tg$ may be flat or not; 
$\tS$ for which the connection is flat are called \ti{admissible} (see e.g. \cite{Korotkin:2017yxn,bkn} for discussion
of various examples of admissible projective structures).
If $\tS$ is admissible, then there is at least locally a 
function $\eta_\tS$ on $\cM_\tg$ such that the renormalized eigenblocks
\begin{equation} \label{eq:renormalized-eigenblocks-general}
  \hat\tPsi_{x,y,\tS} = \eta_\tS^{-1} \tPsi_{x,y}
\end{equation}
are covariantly constant. Explicitly $\eta_\tS$ can be obtained by integrating the connection form
\eqref{eq:Mg-connection-form}. It is determined only up to an overall constant.
Finally,
using the formula \eqref{eq:diagonalizing-abelian-verlinde}, 
it follows from the covariant constancy of $\hat\tPsi_{x,y,\tS}$ 
that the renormalized blocks
\begin{equation} \label{eq:renormalized-blocks-general}
  \hat\tPsi_{a,\tS} = \eta_\tS^{-1} \tPsi_{a}
\end{equation}
are also covariantly constant.

The normalization factor $\eta_\tS$ is a higher-genus analogue of the Dedekind eta function, and an important
object in its own right, although we cannot say much about it here.
Many variants of this function have been studied in the literature; see for instance the very useful
review \cite{Korotkin:2018cfo} where they are called Bergman tau functions, in the case where $\tS$ is the projective structure 
determined by an abelian differential on $\tC$.

\subsection{Mutations} \label{sec:mutations}

We have been considering the nonabelianization map $\nab_\cW$ associated to one 
spectral network $\cW$ at a time. Loosely speaking, we think of the different
maps $\nab_\cW$ as providing different ``coordinatizations'' of $\Conf(C, \Vir_{c=1} \otimes \Heis)$,
labeling conformal blocks by their simpler counterparts in $\Conf(\tC, \Heis)$.
To get a complete understanding of $\Conf(C, \Vir_{c=1} \otimes \Heis)$ from this point of view, then, we would 
need to understand the change-of-coordinate maps. We have not completely solved this problem, but we
comment a bit here on what we expect.

Here is the most fundamental example. Consider two spectral networks $\cW^\pm$ which differ
by a transformation associated to a $1$-cycle $\gamma$ on $\tC$, 
in the sense of the figure below. (We call this transformation a \ti{flip} of the
spectral network, because it would induce a flip of the corresponding dual triangulation
as discussed in \cite{MR3115984}.)
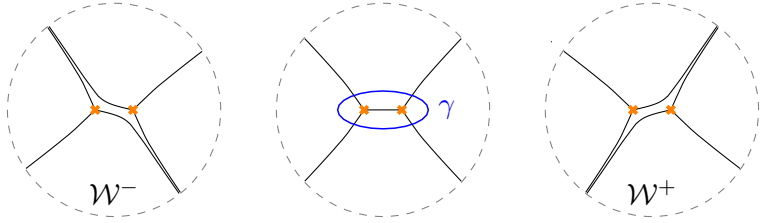
\begin{center}
\begin{tikzpicture}[scale=0.5]
\draw plot [smooth] coordinates {(0.5, -2.37764*10^-9) (0.814824, -0.696241) (1.01312, -1.03903) (1.17609, -1.30296) (1.31807, -1.5252) (1.44562, -1.72065) (1.56245, -1.89703) (1.6709, -2.05899) (1.77256, -2.20953) (1.86858, -2.35076) };
\draw plot [smooth] coordinates {(0.5, 5.19779*10^-10)(0.319611, 0.0400614) (0.206459, 0.0675617) (0.106273, 0.0942645) (0.0119252, 0.122428) (-0.07944, 0.153976) (-0.16864, 0.191232) (-0.254099, 0.236638) (-0.331641, 0.290473) (-0.397895, 0.348871) (-0.453508, 0.407113) (-0.782282, 0.849523) (-0.990707, 1.15393) (-1.15843, 1.39885) (-1.3032, 1.60918) (-1.43262, 1.79628) (-1.5508, 1.96639) (-1.66027, 2.12342) (-1.76274, 2.26995)};
\draw plot [smooth] coordinates {(0.5, 1.85786*10^-9) (1.03915, 0.515344) (1.33331, 0.763144) (1.56823, 0.953903) (1.77025, 1.11463) (1.95041, 1.25605) (2.11465, 1.38376) (2.2666, 1.50106) (2.4087, 1.61013)};
\draw plot [smooth] coordinates {((-0.5, 2.37764*10^-9) (-0.814824, 0.696241) (-1.01312, 1.03903) (-1.17609, 1.30296) (-1.31807, 1.5252) (-1.44562, 1.72065) (-1.56245, 1.89703) (-1.6709, 2.05899) (-1.77256, 2.20953) (-1.86858, 2.35076)};
\draw plot [smooth] coordinates {(-0.5, -5.19779*10^-10) (-0.319611, -0.0400614) (-0.206459, -0.0675617) (-0.106273, -0.0942645) (-0.0119252, -0.122428) (0.07944, -0.153976) (0.16864, -0.191232) (0.254099, -0.236638) (0.331641, -0.290473) (0.397895, -0.348871) (0.453508, -0.407113) (0.782282, -0.849523) (0.990707, -1.15393) (1.15843, -1.39885) (1.3032, -1.60918) (1.43262, -1.79628) (1.5508, -1.96639) (1.66027, -2.12342) (1.76274, -2.26995)};
\draw plot [smooth] coordinates {(-0.5, -1.85786*10^-9)(-1.03915, -0.515344) (-1.33331, -0.763144) (-1.56823, -0.953903) (-1.77025, -1.11463) (-1.95041, -1.25605) (-2.11465, -1.38376) (-2.2666, -1.50106) (-2.4087, -1.61013) (-2.54265, -1.71248)};
\draw [line width=0.2cm, white] (0,0) circle (3cm);
\draw[dashed,gray] (0,0) circle (2.82cm);
\node at (0,-1.6) [below] {$\cW^-$};
\drawbranchpointmarker{-0.5, 0};
\drawbranchpointmarker{0.5, 0};
\end{tikzpicture}
\hspace{0.1cm}
\begin{tikzpicture}[scale=0.5]
\draw plot [smooth] coordinates {(0.5, 0)  (0.93592, -0.615053)  (1.18603, -0.913777) (1.38783, -1.14376)  (1.5622, -1.33749)  (1.71812, -1.50792) (1.86051, -1.66178)  (1.99241, -1.80308)  (2.11587, -1.93445) };
\draw plot [smooth] coordinates {(0.5, 0)  (0.93592, 0.615053)  (1.18603, 0.913777)  (1.38783, 1.14376)  (1.5622, 1.33749)  (1.71812, 1.50792)  (1.86051, 1.66178)  (1.99241, 1.80308)  (2.11587, 1.93445)};
\draw plot [smooth] coordinates {(-0.5, 2.16506*10^-9) (-0.93592, 0.615053) (-1.18603, 0.913777) (-1.38783, 1.14376) (-1.5622, 1.33749) (-1.71812, 1.50792) (-1.86051, 1.66178) (-1.99241, 1.80308) (-2.11587, 1.93445) };
\draw plot [smooth] coordinates {(-0.5, -2.16506*10^-9) (-0.93592, -0.615053) (-1.18603, -0.913777) (-1.38783, -1.14376) (-1.5622, -1.33749) (-1.71812, -1.50792) (-1.86051, -1.66178) (-1.99241, -1.80308) (-2.11587, -1.93445) (-2.23234, -2.05771)};
\draw (-0.5, 0) to (0.5, 0);
\draw [line width=0.2cm, white] (0,0) circle (3cm);
\draw[dashed,gray] (0,0) circle (2.82cm);
\drawbranchpointmarker{-0.5, 0};
\drawbranchpointmarker{0.5, 0};
\draw [domain=0:360,line width=0.2mm,blue] plot ({1.2*cos(\x)}, {0.5*sin(\x)});
\node at (1.2,0) [right] {$\color{blue}\gamma$};
\end{tikzpicture}
\hspace{0.1cm}
\begin{tikzpicture}[scale=0.5]
\draw plot [smooth] coordinates {(0.5, -1.85786*10^-9) (1.03915, -0.515344) (1.33331, -0.763144) (1.56823, -0.953903) (1.77025, -1.11463) (1.95041, -1.25605) (2.11465, -1.38376) (2.2666, -1.50106) (2.4087, -1.61013) (2.54265, -1.71248)};
\draw plot [smooth] coordinates {(0.5, -5.19779*10^-10) (0.319611, -0.0400614) (0.206459, -0.0675617) (0.106273, -0.0942645) (0.0119252, -0.122428) (-0.07944, -0.153976) (-0.16864, -0.191232) (-0.254099, -0.236638) (-0.331641, -0.290473) (-0.397895, -0.348871) (-0.453508, -0.407113) (-0.782282, -0.849523) (-0.990707, -1.15393) (-1.15843, -1.39885) (-1.3032, -1.60918) (-1.43262, -1.79628) (-1.5508, -1.96639) (-1.66027, -2.12342) (-1.76274, -2.26995) (-1.85941, -2.40783) };
\draw plot [smooth] coordinates {(0.5, 2.37764*10^-9) (0.814824, 0.696241) (1.01312, 1.03903) (1.17609, 1.30296) (1.31807, 1.5252) (1.44562, 1.72065) (1.56245, 1.89703) (1.6709, 2.05899) (1.77256, 2.20953) (1.86858, 2.35076) (1.9598, 2.4842)};
\draw plot [smooth] coordinates {(-0.5, 1.85786*10^-9) (-1.03915, 0.515344) (-1.33331, 0.763144) (-1.56823, 0.953903) (-1.77025, 1.11463) (-1.95041, 1.25605) (-2.11465, 1.38376) (-2.2666, 1.50106) (-2.4087, 1.61013) (-2.54265, 1.71248) (-2.66972, 1.80921)};
\draw plot [smooth] coordinates {(-0.5, 5.19779*10^-10)(-0.319611, 0.0400614) (-0.206459, 0.0675617) (-0.106273, 0.0942645) (-0.0119252, 0.122428) (0.07944, 0.153976) (0.16864, 0.191232) (0.254099, 0.236638) (0.331641, 0.290473) (0.397895, 0.348871) (0.453508, 0.407113) (0.782282, 0.849523) (0.990707, 1.15393) (1.15843, 1.39885) (1.3032, 1.60918) (1.43262, 1.79628) (1.5508, 1.96639) (1.66027,  2.12342) (1.76274, 2.26995) (1.85941, 2.40783)};
\draw plot [smooth] coordinates {(-0.5, -2.37764*10^-9) (-0.814824, -0.696241) (-1.01312, -1.03903) (-1.17609, -1.30296) (-1.31807, -1.5252) (-1.44562, -1.72065) (-1.56245, -1.89703) (-1.6709, -2.05899) (-1.77256, -2.20953) (-1.86858, -2.35076) (-1.9598, -2.4842) };
\draw [line width=0.2cm, white] (0,0) circle (3cm);
\draw[dashed,gray] (0,0) circle (2.82cm);
\drawbranchpointmarker{-0.5, 0};
\drawbranchpointmarker{0.5, 0};
\node at (0,-1.6) [below] {$\cW^+$};
\end{tikzpicture}
\end{center}
Now we consider the operator on $\Conf(\tC, \Heis)$ given by
\begin{equation}
 k_\gamma = \sum_{n = 1}^\infty \left( \frac{1}{n^2} - \frac{(-1)^n \ell_\gamma}{n} \right) L_\gamma^n \, .
\end{equation}
When $k_{A_i}$ acts on the blocks $\tPsi_a$, with $\re a_i \le 0$ and
$a_i \notin 2 \pi \I \bbZ$, it gives a convergent expression:
\begin{align}
  k_{A_i} \tPsi_a &= \left( \Li_2(\e^{a_i}) + a_i \log (1 + \e^{a_i}) \right) \tPsi_a \, .
\end{align}
It follows that, when $\re x_i \le 0$ and $x_i \notin 2 \pi \I \bbZ$,
\begin{equation}
  k_{A_i} \tPsi_{x,y} = (\Li_2(\e^{x_i}) - 2 \pi \I \log(1 + \e^{x_i}) \partial_{y_i} ) \tPsi_{x,y} \, .
\end{equation}
The formulas above actually admit analytic continuation in $a$ or $(x,y)$, 
and one might hope that there is a better definition of $k_\gamma$
which would make this continuation manifest. We will not pursue that here; instead we make do with the
domains given above.
Now we propose that if we define the mutation operator $\cK_\gamma$ by
\begin{equation} \label{eq:mutation-explicit}
  \cK_\gamma = \exp \left( \frac{k_\gamma}{2 \pi \I} \right) \, ,
\end{equation}
then $\cK_\gamma$ fits into a diagram
\begin{center}
\begin{tikzcd}
\Conf(\tC, \Heis) \arrow[rd, "\nab_{\mathcal{W^-}}"] \arrow[rr, dashed, "\cK_\gamma"] & & \Conf(\tC, \Heis) \arrow[ld, "\nab_{\mathcal{W^+}}"'] \\
& \Conf(C, \Vir_{c=1} \otimes \Heis) & 
\end{tikzcd}
\end{center}
which commutes up to a constant:
in other words, we have
\begin{equation} \label{eq:dilog-intertwining}
  \nab_{\cW^-} = \xi \nab_{\cW^+} \, \circ \, \cK_\gamma
\end{equation}
for some $\xi \in \bbC^\times$.
We discuss some of the motivation of \eqref{eq:dilog-intertwining} in \autoref{sec:bundle-geometry} below.
Unfortunately, we do not have a proof of \eqref{eq:dilog-intertwining}; we hope to provide one in the future.

\section{Verlinde loop operators} \label{sec:verlinde-operators}

One of the important structures on Virasoro conformal blocks is the action of Verlinde loop operators.
See \cite{MR0954762} for the original definition of these operators in 
rational CFT, \cite{Alday:2009fs,Drukker:2009id} for the extension to general Virasoro blocks,
\cite{Gaiotto:2024tpl} for a more recent treatment.
In this section we review the essential properties of these operators,
and explain in what sense they are compatible with our nonabelianization map for conformal blocks.

\subsection{Definition of Verlinde loop operators on Heisenberg blocks} \label{sec:verlinde-abelian}

We begin with the simpler case of the Heisenberg blocks, where we can understand the Verlinde operators
in a completely explicit way. The Verlinde operators are linear endomorphisms of $\Conf(\tC,\Heis)$,
built from three basic ingredients:

\begin{itemize}
\item We have the unfusion map which creates two nearby fermion
insertions $\psi_+(p) \psi_-(q)$, via the explicit construction given in \autoref{sec:free-fermions}:
\begin{equation}\label{eq:heisenberg-unfusionm}
  \Unfus_{p,q}: \Conf(\tC, \Heis) \to \Conf(\tC, \Heis; \psi_+(p) \psi_{-}(q)) \otimes K_\tC^{\frac12}(p) \otimes K_\tC^{\frac12}(q) \, .
\end{equation}
\item There is also the fusion map, which takes the leading singularity when two fermions collide:
\begin{equation}
  \Fus_{p,q}: \Conf(\tC, \Heis; \psi_+(p) \psi_{-}(q)) \otimes K_\tC^{\frac12}(p) \otimes K_\tC^{\frac12}(q) \to \Conf(\tC, \Heis) \, .
\end{equation}
This map is given explicitly by
\begin{equation}
  \IP{\cdots}_{\Fus_{p,q}(\tPsi)} = \lim_{p \to q} \frac{z(p) - z(q)}{\sqrt{\de z(p)} \sqrt{\de z(q)}} \IP{\cdots \psi_+(p) \psi_-(q)}_\tPsi \, ,
\end{equation}
where on the right side we use the connection on conformal blocks to move the points $p$, $q$.
\item Finally, if $\gamma$ is an oriented loop on $\tC$ based at $p$, 
we have a map
\begin{equation}
 \Hol_{\gamma,q}: \Conf(\tC, \Heis; \psi_+(p) \psi_{-}(q)) \otimes K_\tC^{\frac12}(p) \otimes K_\tC^{\frac12}(q) \circlearrowleft
\end{equation}
which continues the $\psi_+$ insertion around $\gamma$, using the connection on conformal blocks.
\end{itemize}
The Verlinde operator is the composition of these three, modified by a sign:
\begin{equation} \label{eq:L-heis}
  L_\gamma = \tsigma(\gamma) \cdot \Fus_{p,q} \circ \Hol_{\gamma,q} \circ \Unfus_{p,q}
\end{equation}
where $\tsigma: H_1(\tC, \bbZ) \to \{ +1, -1 \} $ is the quadratic refinement associated to
the spin structure $K^{\frac12}_\tC$ \cite{MR588283}. With this sign included, the operator $L_\gamma$
is independent of the choice of spin structure, and they obey the relation\footnote{To check this, we use the fact that $\Unfus_{p,q} \circ \Fus_{p,q}$ is the identity operator on $\Conf(\tC, \Heis; \psi_+(p) \psi_{-}(q)) \otimes K_\tC^{\frac12}(p) \otimes K_\tC^{\frac12}(q)$.}
\begin{equation}
  L_\gamma L_\mu = (-1)^{\IP{\gamma,\mu}} L_{\gamma+\mu} \, .
\end{equation}
By direct computation using \eqref{eq:heisenberg-unfusion} 
one can check that \eqref{eq:L-heis} agrees with the
concrete formula \eqref{eq:abelian-log-verlinde-relation} which we used above.

\subsection{Definition of Verlinde loop operators on Virasoro-Heisenberg blocks} \label{sec:verlinde-nonabelian}

Next let us discuss the more difficult case of Verlinde operators acting on Virasoro-Heisenberg blocks.
To construct these, we need to generalize the three ingredients above:
\begin{itemize}
\item We need an unfusion map which creates two nearby degenerate-field
insertions $\chi_{\frac12}(p) \chi_{-\frac12}(q)$,
\begin{equation}
  \Unfus_{p,q}: \Conf(C, \Vir_{c=1} \otimes \Heis) \to \Conf(C, \Vir_{c=1} \otimes \Heis; \chi_{\frac12}(p) \chi_{-\frac12}(q)) \otimes K_C^{\frac12}(p) \otimes K_C^{\frac12}(q) \, .
\end{equation}
In the works \cite{Alday:2009fs,Drukker:2009id,Gaiotto:2024tpl}, unfusion is constructed using 
the factorization property of conformal blocks. It is not clear to us whether this property should be 
expected to hold for arbitrary elements of $\Conf(C, \Vir_{c=1} \otimes \Heis)$ (see e.g. \cite{Teschner:2008qh} for related
discussion). We will proceed pragmatically as follows. We are only interested in the specific conformal blocks that
lie in the image of nonabelianization maps.
So, suppose we fix a spectral network $\cW$, such that
the conformal block $\Psi$ which we consider arises as $\Psi = \nab_\cW(\tPsi)$.
In this case, we can leverage the unfusion map which we already have on Heisenberg blocks, defining
\begin{equation} \label{eq:def-unfus-gl2}
  \Unfus_{p,q}(\Psi) = \nab_\cW\left( \Unfus_{p^{(1)},q^{(1)}} (\tPsi) + \Unfus_{p^{(2)},q^{(2)}} (\tPsi) \right) \, .
\end{equation}
Then there is one point we need to check: suppose that $\Psi = \nab_\cW(\tPsi)$ and also
$\Psi = \nab_{\cW'}(\tPsi')$. Then, does $\Unfus_{p,q}$ depend on whether we use 
$\cW$ or $\cW'$ in \eqref{eq:def-unfus-gl2}?
Fortunately the answer is no,
because $\Unfus_{p,q}$ is defined by operator insertions away from the 
spectral network, which thus commute
with the mutation operator $\cK_\gamma$ we discussed in \autoref{sec:mutations}.

\item There is also the fusion map, which takes the leading singularity when two denegerate fields collide:
\begin{equation}
  \Fus_{p,q}: \Conf(C, \Vir_{c=1} \otimes \Heis; \chi_{\frac12}(p) \chi_{-\frac12}(q)) \otimes K_C^{\frac12}(p) \otimes K_C^{\frac12}(q) \to \Conf(C, \Vir_{c=1} \otimes \Heis) \, .
\end{equation}
This map is given explicitly by
\begin{equation}
  \IP{\cdots}_{\Fus_{p,q}(\Psi)} = \lim_{p \to q} \frac{z(p) - z(q)}{\sqrt{\de z(p)} \sqrt{\de z(q)}} \IP{\cdots \chi_{\frac12}(p) \chi_{-\frac12}(q)}_\Psi \, .
\end{equation}
This is parallel to the Heisenberg case.
\item Finally, if $\wp$ is an oriented loop on $C$ based at $p$, 
we have a map
\begin{equation}
 \Hol_{\wp,q}: \Conf(C, \Vir_{c=1} \otimes \Heis ; \chi_{\frac12}(p) \chi_{-\frac12}(q)) \otimes K_C^{\frac12}(p) \otimes K_C^{\frac12}(q) \circlearrowleft
\end{equation}
which continues the $\chi_{\frac12}$ insertion around $\wp$, using the connection on conformal blocks.
This is again parallel to the Heisenberg case.
\end{itemize}
The Verlinde operator is the composition
\begin{equation}
  L_\wp = \sigma(\wp) \cdot \Fus_{p,q} \circ \Hol_{\wp,q} \circ \Unfus_{p,q} \, ,
\end{equation}
where $\sigma$ is the quadratic refinement associated to the chosen spin structure $K_C^{\frac12}$.
As we have explained, $L_\wp$ may not be defined on the whole of $\Conf(C, \Vir_{c=1} \otimes \Heis)$, but it is 
defined at least on those conformal blocks which are in the image of $\nab_\cW$, and this is all that we will use.

More generally, instead of a loop $\wp$ on $C$, we could consider a web on $C$, with oriented legs carrying various labels corresponding to different possible degenerate insertions, and 3-leg junctions corresponding to possible fusions. This kind of web again determines a Verlinde operator, as described in \cite{Coman:2015lna}, by composition of elementary pieces corresponding to the legs and vertices of the web.

\subsection{Verlinde operators for \texorpdfstring{$c \neq 1$}{c != 1}}

For orientation, we briefly comment about the case of more general $c$.
Then there is a similar construction of Verlinde operators acting on $\Conf(C, \Vir_{c} \otimes \Heis)$. These Verlinde operators depend on a choice of a parameter $b \in \bbC$, obeying
\begin{equation}
c = 1 + 6 Q^2, \qquad Q = b + b^{-1} \, .
\end{equation}
For fixed $Q$ there are two solutions $b$, giving 
two distinct Verlinde operators $L^b_\wp$. These operators generate two
skein algebras $\Sk_\fq(C, \GL(2))$, with
$\fq = \e^{\pi \I b^{2}}$, as discussed e.g. in \cite{Alday:2009fs,Drukker:2009id,Bullimore:2013xsa,Coman:2015lna}.

We can also give an analogous construction of Verlinde operators acting on Heisenberg blocks:
just define $L_\gamma^b = \exp \I b \ell_\gamma$.
From \eqref{eq:J-commutator} we see that they obey the relations $L_\gamma^b L_\mu^b = \fq L_{\mu_\gamma}^b$,
which define the skein algebra $\Sk_\fq(\tC, \GL(1))$.

In this paper we are only interested in the case $c = 1$; 
then $b = \I$ and $b = -\I$ give
the same Verlinde operators up to reversal of orientation of
the loops, so there is no loss of generality in
considering only $b = -\I$. The corresponding skein algebras have $\fq = -1$, 
and in particular they are commutative. This commutativity is important for 
our purposes: it 
means that we can contemplate simultaneous eigenblocks of the full
algebras of Verlinde operators.

\subsection{Abelianization and Verlinde operators} \label{sec:abelianization-and-verlinde}

We have just discussed two kinds of Verlinde operators: 
the $L_\gamma$ acting on $\Conf(\tC,\Heis)$, and the $L_\wp$ acting on $\Conf(C,\Vir_{c=1} \otimes \Heis)$.
These two types of operators are connected through the nonabelianization maps
$\nab_\cW$.

To explain this we first recall that the spectral network $\cW$ determines a map between the algebras
of Verlinde operators \cite{Gaiotto:2010be,MR3115984,Neitzke:2020jik},
\begin{equation}
  \nab_\cW^\Sk :  \Sk_\fq(C,\GL(2)) \to \Sk_\fq(\tC,\GL(1)) \, .
\end{equation}
Its specialization to $\fq = -1$ is
equivalent to the nonabelianization map for twisted flat connections from \cite{MR3115984,Hollands:2013qza},
\begin{equation}
  \nab_\cW^\flat :  \cM(\tC,\GL(1)) \to \cM(C,\GL(2)) \, .
\end{equation}
For spectral networks of a suitable type, this map is
in turn equivalent to a spectral coordinate system on a dense subset of 
$\cM(C,\GL(2))$ (e.g. Fock-Goncharov coordinates or complex Fenchel-Nielsen coordinates)
\cite{MR3115984,Hollands:2013qza}.

Now how is this related to conformal blocks?
The nonabelianization map for conformal blocks
intertwines the two actions of Verlinde operators: 
given a Verlinde operator $L \in \Sk_{-1}(C, \GL(2))$,
we have the commuting diagram
\begin{center}
\begin{tikzcd}
\Conf(\tC, \Heis) \arrow[r, "\nab_{\mathcal{W}}^{\Sk}(L)"] \arrow[d, "\nab_{\mathcal{W}}"'] & \Conf(\tC, \Heis) \arrow[d, "\nab_{\mathcal{W}}"] \\
\Conf(C, \Vir_{c=1} \otimes \Heis) \arrow[r, "L"] & \Conf(C, \Vir_{c=1} \otimes \Heis)
\end{tikzcd}
\end{center}
i.e.
\begin{equation} \label{eq:verlinde-intertwining}
  L \cdot \nab_\cW(\tPsi) = \nab_\cW\left( \nab^\Sk_\cW(L) \cdot \tPsi \right) \, .
\end{equation}
Indeed, this is a shadow of a stronger statement:
each ingredient in the construction of Verlinde operators --- unfusion, parallel transport, and
fusion --- separately intertwines with nonabelianization. The spectral network $\cW$ plays no role in the unfusion and fusion steps,
which occur in a small neighborhood of some generic point of $C$, and intertwine with nonabelianization essentially by definition. 
The most interesting and nontrivial part is the statement that nonabelianization intertwines with parallel transport: more precisely, parallel transport around 
a loop $\wp$ on $C$ intertwines with transport around a corresponding combination of loops $\nab_\cW^\Sk(\wp)$ on the cover.
Fortunately we have already discussed this, in \autoref{sec:degenerate-primaries} above.

\subsection{Verlinde eigenblocks}

As we have explained, the Verlinde operators acting on
Virasoro-Heisenberg blocks
generate the commutative algebra
$\Sk_{-1}(C, \GL(2)) = \cO(\cM(C, \GL(2)))$, and so
it makes sense to seek conformal 
blocks $\Psi \in \Conf(C, \Vir_{c=1} \otimes \Heis)$ which are
simultaneous eigenvectors of these operators. 

In the Heisenberg case we showed in \autoref{sec:diagonalizing-abelian-verlinde}
that each joint eigenvalue $X \in \cM(\tC,\GL(1))$ 
has a corresponding $1$-dimensional eigenspace in $\Conf(\tC, \Heis)$.
In the nonabelian case,
a joint eigenvalue of the Verlinde operators is a
point $\lambda$ of $\Spec \Sk_{-1}(C, \GL(2))$, also known as the twisted character 
variety $\cM(C, \GL(2))$.
In this case we have not proven that the eigenspaces are one-dimensional, but we 
can give a construction of an eigenblock for each generic $\lambda$.
Indeed, suppose we have an abelian block $\tPsi$ which 
is a simultaneous eigenblock of the
Verlinde operators $L_\gamma$, with eigenvalue $X \in \cM(\tC, \GL(1))$. 
Then \eqref{eq:verlinde-intertwining} says that $\nab_\cW(\tPsi)$ is a simultaneous eigenblock
of the Verlinde operators $L_\wp$, with eigenvalue $\lambda = \nab^\flat_\cW(X) \in \cM(C, \GL(2))$.
We conjecture that this recipe gives all the eigenblocks
for generic $\lambda$.

\subsection{Eigenblocks and connections} \label{sec:eigenblocks-and-connections}

Since the eigenvalues of the Verlinde operators parameterize (twisted) flat $\GL(2)$-connections over $C$, it is natural to wonder: how, given a particular eigenblock, do we see its corresponding flat connection? One answer is that we can realize it via the parallel transport of degenerate fields, as we now explain.

Suppose given a conformal block $\Psi \in \Conf(C, \Vir_{c=1} \otimes \Heis)$ which is 
realized as $\nab_\cW(\tPsi) = \Psi$. 
Choose a spin structure on $C$,
and consider the block $\Unfus_{p,q}(\Psi)$ 
with two degenerate insertions $\chi_{\frac12}(p) \chi_{-\frac12}(q)$,
valued in $K^{\frac12}(p) \otimes K^{\frac12}(q)$. 
We can use the connection on such blocks (\autoref{sec:connections-on-blocks}) to continue the $p$ variable along arbitrary paths in $C \setminus \{ q \}$.
For general $\Psi$, this continuation need not close on any finite-dimensional space. However, when $\tPsi$ (and hence $\Psi$) is a Verlinde eigenblock, 
the continuation of $\Unfus_{p,q}(\Psi)$ does close on a rank $2$ bundle over $C \setminus \{ q \}$ 
with connection,
\begin{equation}
\cE^+(\Psi,q) \subset \Conf(C, \Vir_{c=1} \otimes \Heis; \chi_{\frac12}(\cdot) \chi_{-\frac12}(q)) \otimes K^{\frac12} \otimes K^{\frac12}(q) \, .
\end{equation}
Indeed, we can describe $\cE^+(\Psi,q)$ concretely: if $p$, $q$ are not on
the spectral network $\cW$, then
\begin{equation}
  \cE^+(\Psi,q)_p = \nab_\cW\left( \Span \left( \Unfus_{p^{(1)},q^{(j)}} (\tPsi), \Unfus_{p^{(2)},q^{(j)}} (\tPsi) \right) \right) \, ,  
\end{equation}
for either $j=1$ or $j=2$. It follows that the connection in $\cE^+(\Psi,q)$ has trivial monodromy around $q$ (this boils down to the fact that 
the free fermion blocks on $\tC$ have trivial monodromy when one fermion goes around another).
Moreover, the connection in $\cE^+(\Psi,q)$
is in the class $\lambda \in \cM(C, \GL(2))$.
Similarly, if we continue $q$ holding $p$ fixed we get a bundle $\cE^-(\Psi,p)$ with 
connection over $C \setminus \{ p \}$, in the class $\lambda^*$ (dual to $\lambda$).

Finally, continuing both $p$ and $q$ gives a connection in a rank $4$ bundle $\cE(\Psi)$
over $(C \times C) \setminus \Delta$.
Given $p,q,i,j$, we get a block 
\begin{equation}
b^{(i,j)}(\Psi,p,q) = \nab_\cW\left(\Unfus_{p^{(i)},q^{(j)}}(\tPsi)\right) \in \cE(\Psi)_{(p,q)}   \, .
\end{equation}
This block depends
on a leash connecting $p^{(i)}$ to $q^{(j)}$; changing the leash changes the block by a scalar factor. In any case,
the blocks $b^{(i,j)}(\Psi,p,q)$ for $i,j = 1,2$ span the $4$-dimensional vector space $\cE(\Psi)_{(p,q)}$.

It is interesting to take the limit $q \to p$: we define
\begin{equation} \label{eq:def-Xi}
  \IP{\cdots \Xi^{(i,j)}(p)^z}_\Psi = \lim_{q \to p}  \IP{\cdots \left( \chi_\frac12(p)^z \chi_{-\frac12}(q)^z - \frac{\delta_{ij}}{z(p)-z(q)} \right) }_{b^{(i,j)}(\Psi,p,q)} \, .
\end{equation}
The insertion $\Xi^{(i,j)}(p)$ still depends on the choice of a leash from $p^{(i)}$ to $p^{(j)}$.
More invariantly, we can organize the $\Xi^{(i,j)}(p)$ into an operator $\Xi(p)$ which is valued in $\End(V_\lambda(p))$,
where $V_\lambda$ denotes a $\GL(2)$-bundle with connection in the class $\lambda$.

From \eqref{eq:def-Xi} we can see directly the dictionary
\begin{equation} \label{eq:xi-diagonal-dictionary}
\Xi^{(i,i)}(p) \rightsquigarrow \tJ(p^{(i)}) \, ,
\end{equation}
and in particular the trace of $\Xi$
maps to the Heisenberg generator $J$ on $C$. For the off-diagonal parts the dictionary is
\begin{equation}
\Xi^{(i,j)}(p) \rightsquigarrow \psiplus(p^{(i)}) \psiminus(p^{(j)}) \, .  
\end{equation}

\subsection{The line bundle of eigenblocks} \label{sec:bundle-geometry}

The Verlinde eigenblocks
in $\Conf(C,\Vir_{c=1} \otimes \Heis)$ make up a sheaf $\cL$ over
$\cM(C,\GL(2))$, which we conjecture is generically a line bundle.
In this language, we can see nonabelianization of conformal blocks 
as a lift of $\nab^\flat_\cW$ to the line bundles of eigenblocks:
\begin{center}
\begin{tikzcd}
\tcL \arrow[r, "\nab_\cW"] \arrow[d] & \cL \arrow[d] \\
\cM(\tC, \GL(1)) \arrow[r, "\nab^\flat_\cW"] & \cM(C, \GL(2))
\end{tikzcd}
\end{center}
Since $\tcL$ has concrete local trivializations by the eigenblocks $\tPsi_{x,y}$, this map in particular gives
local trivializations of $\cL$ by the eigenblocks $\Psi^\cW_{x,y} = \nab_\cW(\tPsi_{x,y})$.

If we think of $\cL$ as an abstract line bundle for a moment, forgetting its origin as a space of
eigenblocks, then this basic setup has appeared in various places before: in particular it is in 
\cite{MR2233852,Alexandrov:2011ac,Neitzke:2011za,Coman:2020qgf,Bertola:2019nvr,Freed:2022yae}.
In \cite{MR2233852,Alexandrov:2011ac,Neitzke:2011za,Bertola:2019nvr} the line bundle is treated mainly 
as an abstract geometric object. In \cite{Freed:2022yae} it arises from classical complex 
Chern-Simons theory. The reference \cite{Coman:2020qgf} is closest to our current point of view: 
there, as here, $\cL$ is the line bundle of Verlinde eigenblocks.

To give a complete description of $\cL$, it is not enough to know that it has local trivializations $\Psi^\cW_{x,y}$:
we also need to know how the local trivializations depend on $\cW$.
On general grounds the answer must take the form
\begin{equation} \label{eq:L-transition-map}
  \Psi^\cW_{x,y} = \alpha^{\cW,\cW'} \Psi^{\cW'}_{x',y'}
\end{equation}
where $(x,y)$ and $(x',y')$ differ by a symplectomorphism. In particular, if $\cW$ and $\cW'$ are related by a flip 
as in \autoref{sec:mutations}, then this symplectomorphism takes the form\footnote{The formula \eqref{eq:log-mutation}
is related to the mutation law for the ``cluster $X$ coordinates'' on $\cM(C, \PSL(2))$, in the sense of \cite{MR2233852};
see e.g. \cite{Gaiotto:2009hg,IwakiNakanishi2014ExactWKB} for an account of the precise relation.}
\begin{equation} \label{eq:log-mutation}
x' = x, \qquad y' = y - \log(1 + \e^{x_i}) \, ,
\end{equation}
The question of finding $\alpha^{\cW,\cW'}(x,y)$ was also
addressed in \cite{MR2233852,Alexandrov:2011ac,Neitzke:2011za,Coman:2020qgf,Bertola:2019nvr,Freed:2022yae}, from various points of view; see also the related works \cite{Lisovyy:2016qig,Iwaki:2023cek,Marino:2024yme,Iorgov:2013uoa,Its:2014lga,Lisovyy:2018mnj}
where essentially the same object appears.
In all of these references it turns out that $\alpha^{\cW,\cW'}(x,y)$ is a relative of the dilogarithm function,
of the form
\begin{equation} \label{eq:transition-dilog}
 \alpha^{\cW,\cW'}(x,y) = \xi \exp \left( \frac{1}{2 \pi \I} \Li_2(\pm \e^{x_i}) \right) \, .
\end{equation}
This kind of formula for $\alpha^{\cW,\cW'}(x,y)$ would follow from our conjectural description of the mutation operator $\cK_\gamma$ in \autoref{sec:mutations},
and indeed this is one of the main motivations for that conjecture.

Another important geometric feature of $\cL$ is a holomorphic connection $\nabla$, whose curvature is the Atiyah-Bott
symplectic form on $\cM(C,\GL(2))$. This connection is a nonabelian analogue of the connection $\tnabla$
in  $\tcL \to \cM(C, \GL(1))$ which we explained in \autoref{sec:heisenberg-line-bundle},
and indeed it can be constructed by pulling $\tnabla$ through $\nab_\cW$.
For this one needs to know that the transition map \eqref{eq:L-transition-map} is compatible with $\tnabla$;
this would follow from the formula \eqref{eq:transition-dilog} (indeed this is enough to 
\ti{determine} that formula.)

It would be desirable to understand the origin of the connection $\nabla$ in $\cL$ more directly,
in the language of conformal blocks. This is trickier than for $\tnabla$,
because now we do not have log-Verlinde operators available.
Still, we can make a tentative proposal, as follows.
We consider an eigenblock $\Psi$ with eigenvalue $\lambda \in \cM(C, \GL(2))$.
A tangent vector to $\cM(C, \GL(2))$ at $\lambda$ is a covariantly closed 1-form
$\beta \in \Omega^{1}(C, \End V_\lambda)$.
Then we generalize \eqref{eq:derivative-abelian} to
\begin{equation} \label{eq:derivative-nonabelian}
  \IP{ \cdots }_{\nabla_\beta \Psi} = \partial_\beta \IP{ \cdots }_{\Psi} - \IP{ \cdots  \int_C \tr \left(\beta \Xi \right) }_{\Psi}
\end{equation}
where $\Xi$ denotes the nonabelian current valued in $\End V_\lambda$
which we constructed in \eqref{eq:def-Xi}.

\section{Expectations in examples}

For any spectral network $\cW$ subordinate to a double cover $\tC$, 
and any choice of $A$ and $B$ cycles on $\tC$,
we have defined a family of conformal blocks $\nab_\cW(\tPsi_a) \in \Conf(C, \Vir_{c=1} \otimes \Heis)$.
In this section we make some proposals for how these blocks should be related to 
formulas and conjectures already in the literature.

\subsection{Triangulations and Goncharov-Shen blocks} \label{sec:triangulations}

First, suppose we mark points $p_1, \dots, p_n$ of $C$,
with $n \ge 1$, and fix an
ideal triangulation $T$ of $C$, with vertices at the $p_i$.
Also fix parameters $\beta_1, \dots, \beta_n \in \bbC$.
Finally fix a covering $\pi: \tC \to C$ and spectral network $\cW_T$ which induces the triangulation $T$ 
as described in \cite{Gaiotto:2009hg,MR3115984}.
The covering $\tC$ then has genus $\tg = 4g-3+n$ and $2n$ punctures.
Now suppose we choose $A$ and $B$ cycles on $\tC$.
Then we get a family of conformal blocks
parameterized by $a \in \bbC^\tg$,
\begin{equation}
 \Psi^T_a = \nab_{\cW_T}(\tPsi_a) \in \Conf\left(C, \Vir_{c=1} \otimes \Heis ; \chi_{\beta_1}(p_1) \cdots \chi_{\beta_n}(p_n) \right) \, .
\end{equation}
We proposed in \autoref{sec:mutations} that 
changing the triangulation changes
the $\Psi^T_a$ by a certain intertwining operator built from the dilogarithm function.
Changing the $A$ and $B$ cycles by an action of 
$\Sp(2\tg,\bbZ)$
changes the $\Psi^T_a$ by a (generalized) Fourier transform
as we discussed in \autoref{sec:explicit-blocks}.

Now we recall a proposal of Goncharov-Shen \cite{Goncharov2019}.
Roughly, this proposal states that an ideal triangulation $T$ 
should determine Virasoro conformal blocks
\begin{equation}
 \Xi^T_{b} \in \Conf(C, \Vir_{c} ; W_{h_1}(p_1) \cdots W_{h_n}(p_n) ) \, ,
\end{equation}
depending on $b \in \bbC^{3g-3+n}$ and some discrete polarization data.
When $T$ undergoes a flip 
the blocks $\Xi^T_{b}$ should transform by intertwiners involving 
the Faddeev quantum dilogarithm, which reduces to the ordinary dilogarithm when $c=1$.
For $c=1$ these properties are very similar to those we expect for the 
$\Psi^T_a$ which we constructed above. By a modification 
of our construction (projecting out the $\Heis$ part) one should be
able to construct the desired $\Xi^T_{b}$ on the nose,
and thus establish the proposal of Goncharov-Shen.
It would be very desirable to carry this out.

\subsection{Pants decompositions and Liouville blocks} \label{sec:pants-decompositions}

Next, 
suppose $C$ is the Riemann sphere with nondegenerate operator insertions $W_{\beta_1^2}(p_1)$, \dots, $W_{\beta_n^2}(p_n)$. 
Choose a decomposition of $C$ into pairs of pants,
represented by a ``comb'' diagram 
like the one below:
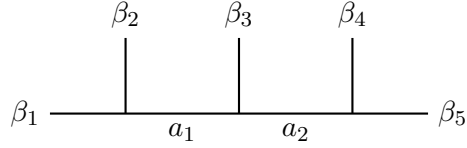
\begin{center}
\begin{tikzpicture}
  \draw[thick] (0,0) -- (5,0);
  
  \draw[thick] (1,0) -- (1,1) node[above] {$\beta_2$};
  \draw[thick] (2.5,0) -- (2.5,1) node[above] {$\beta_3$};
  \draw[thick] (4,0) -- (4,1) node[above] {$\beta_4$};

  \node at (1.75,0) [below] {$a_1$};
  \node at (3.25,0) [below] {$a_2$};

  \node at (0,0) [left] {$\beta_1$};
  \node at (5,0) [right] {$\beta_5$};
\end{tikzpicture}
\end{center}
Each vertex corresponds to
one of the pairs-of-pants, 
and the legs labeled $a_i$ correspond to
the $n-3$ internal tubes. We also fix an additional decoration, namely a tripod drawn on each
pair-of-pants, with one leg ending on each boundary component. Let $P$ denote the datum of the
pants decomposition together with the decoration. Then:

\begin{itemize}
\item There is a conformal block
\begin{equation}
\Psi^\Liouville_P(a) \in \Conf(C, \Vir_{c}; W_{\beta_1^2}(p_1) \cdots W_{\beta_{n}^2}(p_{n}))
\end{equation}
determined by these data; see \cite{Teschner:2008qh} for an account of its construction.
These blocks are the ones which are used in Liouville theory on the sphere.

The blocks $\Psi^\Liouville_P(a)$ also
appear in the AGT correspondence \cite{Alday:2009aq}, which identifies the vacuum correlator 
$\IP{1}_{\Psi^\Liouville_P(a)}$
with the Nekrasov partition function of a 
linear quiver $\cN=2$ theory determined by the pants decomposition $P$,
with $n-3$ $\SU(2)$ gauge nodes and two flavor nodes:\footnote{Here on the CFT side we mean the full vacuum correlator, without factorizing it into three-point functions times other factors; likewise, on the gauge theory side we mean the full Nekrasov partition function, including the 1-loop factors.}
\begin{equation}\label{eq:agtZ}
  \IP{1}_{\Psi^\Liouville_P(a)} = Z_{\Nek}(\eps_1 = - \eps_2 = 1, m = \beta, a) \, .
\end{equation}

\item On the other hand, 
the pants decomposition $P$ can be induced by a spectral network
$\cW_P$ of ``Fenchel-Nielsen'' type \cite{Hollands:2013qza},\footnote{$\cW_P$ restricts on each pair of pants 
to ``molecule I'' of \cite{Hollands:2013qza}.} with associated covering $\pi: \tC \to C$.
The cover $\tC$ has genus $\tg = n-3$ and $2n$ punctures,
and natural cycles $A_i$, $B_i$ lying over each internal pant-leg $\wp_i$. (To determine the $B$ cycles we need to use the decoration.)
Given these cycles, we have the Heisenberg blocks described in \autoref{sec:heisenberg-explicit-construction},
\begin{equation}
\tPsi_{a} \in \Conf(\tC, \Heis; V_{\alpha_1}(p_1^{(1)}) V_{-\alpha_1}(p_1^{(2)}) \cdots V_{\alpha_n}(p_n^{(1)}) V_{-\alpha_n}(p_n^{(2)}) ) \, .  
\end{equation}

Then, let $\tS$ denote the projective structure on $\tC$ induced by the
standard coordinate on the base $C$. 
We define a normalized block $\hat\tPsi_{a,\tS} = \eta_\tS^{-1} \tPsi_{a}$ as in \eqref{eq:renormalized-blocks-general},
engineered to be parallel for the connection $\tnabla^\tS$.

Now we want to apply nonabelianization to $\hat\tPsi_{a,\tS}$.\footnote{Because $\tS$ is singular at the branch points, 
the connection $\nabla^\tS$ lives not over $\cM_{\tg,2n}$ but over a moduli space 
parameterizing surfaces equipped with a choice of local coordinate at each branch point, 
and so the normalized block $\hat\tPsi_{a,\tS}$ depends on this choice of local coordinate.
Happily, this dependence cancels with the coordinate dependence in $\nab_{\cW_P}$,
so that $\nab_{\cW_P}(\hat\tPsi_{a,\tS})$ is independent of the local coordinate.}
We will encounter one more new subtlety: because $\cW_P$ involves double walls, our definition of $\nab_{\cW_P}$ involves coincident insertions of $\psi_+$ and $\psi_-$. We adopt the ``symmetric'' convention that all ill-defined integrals are to be defined by principal value.

Ultimately, applying nonabelianization will give a block
\begin{equation}
  \nab_{\cW_P}(\hat\tPsi_{a,\tS}) \in \Conf(C, \Vir_{c=1} ; W_{\beta_1^2}(p_1) \cdots W_{\beta_{n}^2}(p_{n})) \, .
\end{equation}
(We have suppressed the Heisenberg part, which is trivial since $C$ has genus zero.)

\end{itemize}

So far we held the punctures $p_i$ and the covering $\pi: \tC \to C$ fixed.
Now let us consider the whole construction in a family, allowing the $p_i$ to vary,
with $\pi: \tC \to C$ varying through ``even'' variations as discussed in \autoref{sec:compatibility}.
Then we propose that the two blocks we have considered agree, up to an overall
normalization factor which is independent of all continuous parameters:
\begin{equation} \label{eq:nab-multiple-np}
  \nab_{\cW_P}(\hat\tPsi_{a,\tS}) = N \Psi^\Liouville_P(a) \, .
\end{equation}
It would be very desirable to verify \eqref{eq:nab-multiple-np}
directly. It would be sufficient to do this in 
the degeneration limit where $C$ splits into
thrice-punctured spheres.\footnote{One might wonder why there is not a
relative normalization factor,
depending on a point of $\cM_{0,n}$; the reason is that 
the blocks on both sides of \eqref{eq:nab-multiple-np} 
are parallel for the connection in the bundle of Virasoro blocks 
over $\cM_{0,n}$.}

For $C$ of higher genus we expect the same type of relation, 
but there will be some extra complications. First, we expect that
the construction of the blocks $\Psi_P^\Liouville(a)$ in this case involves a choice of complex projective structure $S$ on $C$, and to fix the normalizations correctly we should use the corresponding $\tS = \pi^* S$ on $\tC$. Second, for $C$ of higher genus 
we will need more care in separating out the Heisenberg from the Virasoro parts.

\section{Nonabelianization and \texorpdfstring{$\tau$}{tau}-functions} \label{sec:tau-functions}

Finally we discuss how our picture of the Virasoro blocks relates to $\tau$-functions in the sense of integrable systems.

\subsection{Painlev\'e \texorpdfstring{$\tau$}{tau}-functions and the Kyiv formula}

We begin with a motivating special case.
It is known that certain $c=1$ Virasoro conformal blocks correspond 
to $\tau$-functions of the Painlev\'e equations, via the
celebrated \ti{Kyiv formula}.
There are various versions of this statement, 
see e.g. \cite{Gamayun:2012ma,Gamayun:2013auu,MR3322384,Bonelli:2019boe,Jeong:2020uxz}.
Each says that a certain linear combination of $0$-point functions in Liouville conformal blocks 
gives a Painlev\'e $\tau$-function:
\begin{equation}
\btau_P = \sum_{n \in \bbZ} \exp\left(-\frac{(x + 2 \pi \I n) y}{2 \pi \I}\right) \IP{1}_{\Psi^\Liouville_P(a = x+2\pi\ri n)} \, .\label{eq:taukyiv}
\end{equation}
For instance, in the original example of \cite{Gamayun:2012ma}, $C$ is the sphere with four primary field
insertions, and the Painlev\'e time is the cross-ratio of their positions.
Then the parameters $(\e^x,\e^y)$ are labels which parameterize the space of solutions of the Painlev\'e equation.

Now note that \eqref{eq:taukyiv} resembles the formula \eqref{eq:diagonalizing-abelian-verlinde} which we used to define eigenblocks of the 
Verlinde operators acting on $\Conf(\tC,\Heis)$. 
We can rewrite \eqref{eq:taukyiv} to make this resemblance
more transparent.
First, substituting in \eqref{eq:nab-multiple-np} we have
\begin{equation}
\btau_P = N^{-1} \sum_{n \in \bbZ} \exp\left(-\frac{(x + 2 \pi \I n) y}{2 \pi \I}\right) \IP{1}_{\nab_{\cW_P}(\hat\tPsi_{a = x+2\pi\ri n}, \tS)} \, .
\end{equation}
Then, using \eqref{eq:diagonalizing-abelian-verlinde}
and the fact that $\nab_{\cW_P}$ is a linear map, we can rewrite this in 
the simpler form
\begin{equation} \label{eq:tau-reinterpreted}
  \btau_P = N^{-1} \IP{1}_{\nab_{\cW_P}(\hat\tPsi_{x,y,\tS})} \, .
\end{equation}

The block appearing on the right side, $\nab_{\cW_P}(\hat\tPsi_{x,y,\tS})$,
has a conceptual meaning: it is an eigenblock
of the Verlinde operators acting on $\Conf(C,\Vir_{c=1} \otimes \Heis)$.
Thus we have arrived at the statement of \cite{MR3322384} 
that the $0$-point function 
in a Verlinde eigenblock is a $\tau$-function.
More precisely, we do not use an \ti{arbitrary} Verlinde eigenblock,
but rather the specific block $\nab_{\cW_P}(\hat\tPsi_{x,y,\tS})$; we  
will put this choice in a more general context below.

\subsection{Other spectral networks}

So far what we have done is just to reinterpret the Kyiv formula as
\eqref{eq:tau-reinterpreted}. Now let us discuss some natural generalizations.

First we remark that there was nothing special about the spectral network $\cW_P$; for any spectral network $\cW$ we could similarly define
\begin{equation} \label{eq:tau-eigenblock-general}
  \btau_\cW = N^{-1} \IP{1}_{\nab_{\cW}(\hat\tPsi_{x,y,\tS})} \, .
\end{equation}
The function $\btau_\cW$ differs from $\btau_P$
by a function of $(x,y)$, depending on the discrete 
choice of $\cW$, but not depending 
on the Painlev\'e times, i.e. the moduli of $C$.
Reversing the steps above we arrive at a Kyiv-type formula for this function:
\begin{equation} \label{eq:kyiv-strong}
\btau_\cW = \sum_{n \in \bbZ} \exp\left(-\frac{(x + 2 \pi \I n) y}{2 \pi \I}\right)\IP{1}_{\nab_{\cW}(\hat\tPsi_{a = x + 2 \pi \I n,\tS})}.
\end{equation}
The summands $\IP{1}_{\nab_{\cW}(\hat\tPsi_{a,\tS})}$ appearing on the right side are 
analogues of the Nekrasov partition function, but not necessarily linked to a weak-coupling limit;
thus one might view \eqref{eq:kyiv-strong} as a strong-coupling analogue of the Kyiv formula.
Formulas of this kind have been written in e.g. 
\cite{Its:2014lga, Coman:2020qgf,Gavrylenko:2020gjb, Bonelli:2016qwg,Bonelli:2019boe,MR4115013};
it would be interesting to see whether \eqref{eq:kyiv-strong}
reproduces them.

\subsection{Other surfaces}

We could also consider more general $C$.
The notion of isomonodromy $\tau$-function is well 
understood only in some specific examples,
where $C$ has genus $0$ or $1$. It is not completely clear whether there is a notion of 
isomonodromy $\tau$-function more generally, e.g. for a surface of genus $g$, with or 
without primary field insertions.
The formula \eqref{eq:tau-eigenblock-general} does
make sense for general $C$, so we can use it as a provisional definition
of $\tau$-function more generally.\footnote{When $C$ varies, we require that $\pi: \tC \to C$ varies only by an 
even variation, as we did in \autoref{sec:pants-decompositions} above.}

We can also formulate this definition in a more intrinsic way, without mentioning nonabelianization 
directly.
The key idea, again, is that the $\tau$-function is the $0$-point function in
a Verlinde eigenblock, $\btau = \IP{1}_\Psi$.
The eigenblock property by itself is not enough to determine $\btau$,
because any scalar multiple of an eigenblock is still an eigenblock.
We get more constraints by requiring $\Psi$ to behave well with respect to
the (twisted) connection on the bundle of eigenblocks 
over $\cM_g \times \cM(C, \GL(2))$. Namely, we fix a spectral network $\cW$,
inducing a local coordinate system $(x,y)$ on $\cM(C, \GL(2))$ through the map $\nab_\cW^\flat$.
Then we require
\begin{equation}
  \nabla \Psi = \frac{1}{2\pi\I} \left(\sum_{i=1}^\tg y_i \, \de x_i\right) \Psi \, ,
\end{equation}
i.e. $\Psi$ is parallel in the $\cM_g$ directions and its derivative in the $\cM(C, \GL(2))$ directions
is in a simple fiducial form. (It would be impossible for $\Psi$ to be parallel in all directions,
since the connection $\nabla$ has curvature.)
By \eqref{eq:connection-local-gauge-explicit}, this property is enough to fix $\Psi = \nab_\cW(\hat \tPsi_{x,y})$ up to an overall constant, and thus it determines the function $\btau = \IP{1}_\Psi$
up to an overall constant.

This method of fixing the normalization of $\tau$-functions by requiring them to obey
differential equations with respect to all parameters has appeared before, e.g.
\cite{MR2579465,MR4328054}.

\subsection{A Fredholm determinant representation}

In this section we give a more explicit description of $\btau_\cW$,
in terms of Fredholm determinants.

We begin from the definition \eqref{eq:tau-eigenblock-general},
and observe that the correlation function appearing there
can be viewed as a Fredholm determinant, in the following sense.
Expand out the definition \eqref{eq:nab-0-point}:
\begin{equation}
    \IP{1}_{\nab_\cW(\tPsi)} = \IP{E_\ren(\cW)}_{\tPsi} = \lim_{\eps \to 0} \eps^{-\frac{k}{8}} \IP{\sum_{n=0}^\infty \frac{1}{n! (2 \pi \I)^n} \left( \int_{\cW_\eps}\,
    \tikz[remember picture, baseline]{
  \node[anchor=base, inner sep=0pt] (psiplus1) {{$\psiplus(q^{(+)})^{z^{(+)}}$}};
}
\,
\tikz[remember picture, baseline]{
  \node[anchor=base, inner sep=0pt] (psiminus1) {{$\psiminus(q^{(-)})^{z^{(-)}}$}};
}
\, \de z(q) \right)^n }_\tPsi \, .
\end{equation}
\begin{tikzpicture}[overlay, remember picture]
  \coordinate (start1) at ($(psiplus1.south)!0.5!(psiplus1.north)$);
  \coordinate (end1) at ($(psiminus1.south)!0.5!(psiminus1.north)$);
  \coordinate (start1Shifted) at ([yshift=-1.8ex]start1); 
  \coordinate (end1Shifted) at ([yshift=-1.8ex]end1);
  \draw[thick] (start1Shifted) -- (end1Shifted);
  \draw[thick] (start1Shifted) -- ++(0,0.5ex);
  \draw[thick] (end1Shifted) -- ++(0,0.5ex);

  \coordinate (mid1) at ($(start1Shifted)!0.5!(end1Shifted)$);
  \node at ($(mid1)+(0,-1.8ex)$) {$\ell_{\mathcal{G_{\epsilon}}(q)}$}; 
\end{tikzpicture}
The $n$-th term in this sum is an integrated correlation function of $2n$ fermions,
\begin{equation}
  \frac{1}{n!} \prod_{i=1}^n \int_{\cW_\eps} \frac{\de z(q_i)}{2 \pi \I} \IP{ \prod_{i=1}^n\,
      \tikz[remember picture, baseline]{
  \node[anchor=base, inner sep=0pt] (psiplus1) {{$\psiplus(q_i^{(+)})^{z^{(+)}}$}};
}
\,
\tikz[remember picture, baseline]{
  \node[anchor=base, inner sep=0pt] (psiminus1) {{$\psiminus(q_i^{(-)})^{z^{(-)}}$}};
}
} _\tPsi\, .
\end{equation}
\begin{tikzpicture}[overlay, remember picture]
  \coordinate (start1) at ($(psiplus1.south)!0.5!(psiplus1.north)$);
  \coordinate (end1) at ($(psiminus1.south)!0.5!(psiminus1.north)$);
  \coordinate (start1Shifted) at ([yshift=-1.8ex]start1); 
  \coordinate (end1Shifted) at ([yshift=-1.8ex]end1);
  \draw[thick] (start1Shifted) -- (end1Shifted);
  \draw[thick] (start1Shifted) -- ++(0,0.5ex);
  \draw[thick] (end1Shifted) -- ++(0,0.5ex);

  \coordinate (mid1) at ($(start1Shifted)!0.5!(end1Shifted)$);
  \node at ($(mid1)+(0,-1.8ex)$) {$\ell_{\mathcal{G_{\epsilon}}(q_i)}$}; 
\end{tikzpicture}

Using \eqref{eq:free-fermion-determinant-formula}, when $\tPsi$ is a Verlinde eigenblock, we can express these correlation functions as determinants of two-point functions, giving
\Needspace{4\baselineskip}
\begin{equation}
  \frac{\IP{1}_{\nab_{\cW}(\tPsi)}}{\IP{1}_\tPsi} = \lim_{\eps \to 0} \eps^{-\frac{k}{8}} \sum_{n=0}^\infty \frac{1}{n!} \prod_{i=1}^n \int_{\cW_\eps} \frac{\de z(q_i)}{2 \pi \I} \det \left( \left[ \frac{\IP{
  \tikz[remember picture, baseline]{
  \node[anchor=base, inner sep=0pt] (psiplus1) {{$\psiplus(q_i^{(+)})^{z^{(+)}} $}};
}
\,
\tikz[remember picture, baseline]{
  \node[anchor=base, inner sep=0pt] (psiminus1) {{$\psiminus(q_j^{(-)})^{z^{(-)}}$}};
}
}_\tPsi}{\IP{1}_\tPsi} \right]_{i,j=1}^{n} \right) \, .
\end{equation}
\begin{tikzpicture}[overlay, remember picture]
  \coordinate (start) at (psiplus1.north);
  \coordinate (end) at (psiminus1.north);
  \coordinate (startShifted) at ([yshift=1ex]start);
  \coordinate (endShifted) at ([yshift=1ex]end);
  \draw[thick] (startShifted) -- (endShifted);
  \draw[thick] (startShifted) -- ++(0,-0.5ex);
  \draw[thick] (endShifted) -- ++(0,-0.5ex);
  \coordinate (mid) at ($(startShifted)!0.5!(endShifted)$);
  \node at ($(mid)+(0,1.5ex)$) {};
\end{tikzpicture}
This expression has another interpretation, as a regularization of a Fredholm determinant in the sense of \cite{MR1554993}:
\begin{equation} \label{eq:regularized-determinant-0-point}
  \frac{\IP{1}_{\nab_{\cW}(\tPsi)}}{\IP{1}_\tPsi} = \det_\reg (1 + \cI) = \lim_{\eps \to 0} \eps^{-\frac{k}{8}} \det(1 + \cI_\eps)
\end{equation}
where $\cI_\eps$ is an integral operator, acting on the space of $K_C^{\frac12}$-valued functions on ${\cW_\eps}$, given by convolution 
\begin{equation}
  (\cI_\eps f) (q) = \int_{\cW_\eps} \cK(p,q) f(p) 
\end{equation}
with the $K_C^{\frac12} \boxtimes K_C^{\frac12}$-valued kernel
\begin{equation} \label{eq:integral-kernel}
  \cK(p,q) = \frac{1}{2\pi \I} \frac{\IP{
  \tikz[remember picture, baseline]{
  \node[anchor=base, inner sep=0pt] (psiplus1) {{$\psiplus(p^{(+)})$}};
}
\,
\tikz[remember picture, baseline]{
  \node[anchor=base, inner sep=0pt] (psiminus1) {{$\psiminus(q^{(-)})$}};
}
}_\tPsi}{\IP{1}_\tPsi} \, .
\end{equation}
\begin{tikzpicture}[overlay, remember picture]
  \coordinate (start) at (psiplus1.north);
  \coordinate (end) at (psiminus1.north);
  \coordinate (startShifted) at ([yshift=1ex]start);
  \coordinate (endShifted) at ([yshift=1ex]end);
  \draw[thick] (startShifted) -- (endShifted);
  \draw[thick] (startShifted) -- ++(0,-0.5ex);
  \draw[thick] (endShifted) -- ++(0,-0.5ex);
  \coordinate (mid) at ($(startShifted)!0.5!(endShifted)$);
  \node at ($(mid)+(0,1.5ex)$) {};
\end{tikzpicture}
(We emphasize that $\cK(p,q)$ has no singularity at $p=q$, because the $\psiplus$ and $\psiminus$ insertions are
taken on different sheets of $\tC$.)

Now we apply this in the case $\tPsi = \hat\tPsi_{x,y,\tS} = \eta_{\tS}^{-1} \tPsi_{x,y}$.
Then we have $\IP{1}_\tPsi = \eta_{\tS}^{-1} \Theta\left[\frac{x}{2 \pi \I} \big\vert \frac{-y}{2 \pi \I}\right] (\tau, 0)$. As discussed above, 
we take the complex projective structure $\tS = \pi^* S$; also as above, we consider variations of $\pi: \tC \to C$ which are
even, in the sense of \autoref{sec:compatibility}.
Then we arrive at our final result for 
$\btau_\cW$:
\begin{equation} \label{eq:tau-formula}
  \btau_\cW = \frac{\Theta\left[\frac{x}{2 \pi \I} \big\vert \frac{-y}{2 \pi \I}\right] (\tau, 0)}{N \eta_{\pi^* S}} \times \det_{\reg}(1 + \cI_{x,y}) \, ,
\end{equation}
where:
\begin{itemize}
\item $\cI_{x,y}$ denotes an integral operator acting on sections 
of $K^\frac12_C$ over $\cW$,
whose kernel is \eqref{eq:integral-kernel}, explicitly given by the 
twisted Szeg\"o kernel \eqref{eq:normalized-fermion-2-point-diagonal-block}.
\item $\det_\reg$ is the regularization of the Fredholm 
determinant defined in \eqref{eq:regularized-determinant-0-point}.
\item $\eta_{\pi^* S}$ is the function on $\cM_\tg$ discussed in \autoref{sec:variation-of-moduli}, here evaluated on a family of
curves $\tC$ obtained by lifting variations of $C$ to even variations of $(C, \tC, \pi)$. This function is determined only up to an overall multiplicative constant, and depends on the choice of complex projective structure $S$ on $C$.
\item $\Theta$ denotes the theta function with characteristics, defined in \eqref{eq:theta-with-characteristics}, using the period matrix $\tau$ of $\tC$.
\item $N$ is an arbitrary complex constant, independent of continuous parameters ($x$, $y$ and the complex structure modulus 
of $C$).
We could have absorbed $N$ in the ambiguity of $\eta_{\pi^* S}$, but keep it in
for maximal consistency with the earlier equations.
\end{itemize}

One feature of \eqref{eq:tau-formula} deserves special comment.\footnote{We thank the referee for raising this point.}
The zeroes of $\btau_\cW$ have a meaning. For example, in situations when $\btau_\cW$ is an isomonodromy tau function,
zeroes of $\btau_\cW$ arise at loci where the isomonodromic variation becomes singular, i.e. the 
monodromy data $(x,y)$ cannot be realized by the most generic sort of connection on $C$.
In \eqref{eq:tau-formula}
there are two possible sources of such a zero: either the theta function or $\det_\reg(1 + \cI_{x,y})$
could vanish. When the theta function vanishes, though, the kernel $\cK(p,q)$ also becomes singular, so 
$\det_\reg(1 + \cI_{x,y})$ is not well defined. Our expectation is that this factor 
develops a pole which cancels the zero of the theta function, so that the combined $\btau_\cW$ is regular
and nonvanishing at this locus. On the other hand, the zeroes of $\det_\reg(1 + \cI_{x,y})$ should give rise
to actual zeroes of $\btau_\cW$. It would be interesting to verify these expectations directly.

\section*{Declarations}
\subsection*{Conflict of interest statement}
The authors have no competing interests to declare that are relevant to the content of this article.
\subsection*{Data availability statement}
The study didn’t involve any data.

\appendix

\section{Heisenberg and Virasoro conformal blocks}
\label{app:conformal-blocks}

By \emph{conformal block} we will mean a system of correlation functions obeying chiral Ward identities.
This approach is taken e.g. in \cite{MR1338609,MR0869564,MR2082709,MR3971924}.
It has the advantage that it involves no arbitrary choices such as pants decompositions: 
the space of conformal blocks is a canonically defined vector space,
depending only on the data of a vertex algebra and a Riemann surface, plus
the specification of primary fields inserted at punctures (if any).

Although conformal blocks can be defined for any vertex algebra, in this paper we will only
use a few specific vertex algebras, and so we give the definition directly for those.

\subsection{Heisenberg blocks} \label{sec:heisenberg-blocks}

Suppose given a Riemann surface $C$.

\subsubsection{The definition}

We are going to define a complex vector space 
$\Conf(C,\Heis)$, the space of Heisenberg conformal blocks on $C$.

An element $\Psi \in \Conf(C,\Heis)$ means a system of correlation
functions, as follows. For every $n$, and any collection of patches 
$U_i$ on $C$ with local coordinate systems $z_i: U_i \hookrightarrow \bbC$,
we have a function
\begin{equation} \label{eq:heis-correlation-function}
  \IP{J( p_1 )^{z_1} \cdots J( p_n )^{z_n}}_\Psi : U_1 \times U_2 \times \cdots \times U_n \to \bbC \, .
\end{equation}

This collection of functions has the following properties:
\begin{enumerate}
  \item Each $\IP{J( p_1 ) \cdots J( p_n )}_\Psi$ is meromorphic in the $p_i$, with singularities only
  when some $p_i = p_j$.
  \item The collection is invariant under the symmetric group $S_n$, so e.g. 
  \begin{equation}
  \IP{J(p_1)^{z_1} J(p_2)^{z_2} \cdots}_\Psi = \IP{J(p_2)^{z_2} J(p_1)^{z_1} \cdots }_\Psi \, .  
  \end{equation}
    Here and below, $\cdots$ denotes an arbitrary product of insertions $J(p_i)^{z_i}$, with
  the same product on both sides of the equation. 
  \item If $U_1 = U_2 = U$ and $z_1 = z_2 = z$ (i.e. we use a single common coordinate system around
  $p_1$ and $p_2$), then the singularity of the $n$-point function as $p_1 \to p_2$ is determined
  by the $(n-2)$-point function, as
  \begin{equation} \label{eq:pole-condition-heisenberg-blocks}
    \IP{J(p_1)^z J(p_2)^z \cdots }_\Psi \ = \  \frac{1}{(z(p_1) - z(p_2))^2} \IP{\cdots}_\Psi + \regular.
  \end{equation}
More informally,
 \eqref{eq:pole-condition-heisenberg-blocks} says that the OPE relation \eqref{eq:heisenberg-ope} ``holds in correlation functions.''
  \item If $z$ and $z'$ are two local coordinate systems around $p$, then the correlation functions are related
  by
  \begin{equation} \label{eq:coordinate-transformation-heisenberg-blocks}
    \IP{J(p)^{z'} \cdots}_\Psi = \left(\frac{\de z(p)}{\de z'(p)}\right) \IP{J(p)^{z} \cdots}_\Psi \, .
  \end{equation}
\end{enumerate}
The condition \eqref{eq:coordinate-transformation-heisenberg-blocks} says
that the holomorphic multi-1-form
\begin{equation}
\IP{J(p_1)^{z_1} \cdots J(p_n)^{z_n}}_\Psi \, \de z_1(p_1) \boxtimes \cdots \boxtimes \de z_n(p_n)
\end{equation}
is well defined, independent of the choices of local coordinate systems.

Note that all our conditions are linear over the complex numbers, 
so $\Conf(C,\Heis)$ is indeed a vector space, with the rule
\begin{equation}
  \IP{\cdots}_{a \Psi + b \Psi'} = a \IP{\cdots}_\Psi + b \IP{\cdots}_{\Psi'} \, .
\end{equation}

\subsubsection{Including primaries}

Now fix points $q_1, \dots, q_k \in C$
and weights $\alpha_1, \dots, \alpha_k \in \bbC$.
Then we also define a vector space $\Conf(C, \Heis; V_{\alpha_1}(q_1) \cdots V_{\alpha_k}(q_k))$,
the space of Heisenberg conformal blocks on $C$ with primary fields $V_{\alpha_i}(q_i)$ inserted.

The definition is just as above, with the following changes.
We now denote the correlation functions by the notation
\begin{equation}
 \IP{J(p_1)^{z_1} \cdots J(p_n)^{z_n} V_{\alpha_1}(q_1) \cdots V_{\alpha_k}(q_k)}_\Psi : U_1 \times U_2 \times \cdots \times U_n \to \bbC \, .
\end{equation}
(We emphasize that for now they are functions only of the $p_i$, not of the $q_j$; the $q_j$ are held fixed throughout the definition
of $\Conf(C, \Heis; V_{\alpha_1}(q_1) \cdots V_{\alpha_k}(q_k))$.)
These correlation functions now have poles at $p_i = q_j$ (as well as at $p_i = p_j$ as before).
The singularity of the $n$-point function as $p_1 \to q_j$ is determined by the $(n-1)$-point function, as
\begin{equation} \label{eq:pole-condition-heisenberg-blocks-primary}
    \IP{J(p_1)^z \cdots V_{\alpha_j}(q_j) \cdots}_\Psi \ = \  \frac{\alpha_j}{z(p_1) - z(q_j)} \IP{\cdots V_{\alpha_j}(q_j) \cdots}_\Psi + \regular.  
\end{equation}
More informally,
 \eqref{eq:pole-condition-heisenberg-blocks-primary} says that the OPE relation \eqref{eq:heisenberg-primary-ope} ``holds in correlation functions.''

\subsubsection{Examples}

The simplest case is $C = \bbC\bbP^1$, with no primary fields inserted. In this case
$\Conf(C,\Heis)$ is $1$-dimensional, so we can fix a block $\Psi$ by
choosing the zero-point function: we choose
\begin{equation}
\IP{1}_\Psi = 1 \, .  
\end{equation}
To write the other correlation functions concretely,
we use the standard inhomogeneous coordinate $z$ around
every $p_i$ (assuming no $p_i = \infty$). Then
\begin{equation}
  \IP{J(p)^z}_\Psi = 0, \qquad \IP{J(p_1)^z J(p_2)^z}_\Psi = \frac{1}{(z(p_1) - z(p_2))^2} \, ,
\end{equation}
and more generally the $2n$-point function is a sum over the $\frac{(2n)!}{2^n n!}$ 
ways of grouping the $2n$ insertions into unordered 
pairs $(p_i,p_j)$, with each term the product of
$n$ factors $\frac{1}{(z(p_i) - z(p_j))^2}$, and the $(2n+1)$-point function vanishes.

When $C$ is a compact surface of genus $g > 0$, $\Conf(C,\Heis)$ is infinite-dimensional.
We discuss explicit representations of Heisenberg blocks in \autoref{sec:explicit-blocks}. 
In much of this paper, though, we need not concern ourselves
with the explicit form of the Heisenberg conformal blocks; we just use
the formal properties listed above.

\subsection{Virasoro blocks} \label{sec:virasoro-blocks}

Fix a constant $c \in \bbC$.
As in \autoref{sec:heisenberg-blocks} above,
given a Riemann surface $C$ we have a complex vector space 
$\Conf(C,\Vir_c)$, the space of Virasoro conformal blocks on $C$.
The definition is completely parallel to that in \autoref{sec:heisenberg-blocks}, but for two modifications.
First, the pole of $\IP{T(p) T(q) \cdots}_\Psi$ on the diagonal is
determined by the OPE relation \eqref{eq:virasoro-ope} rather than \eqref{eq:heisenberg-ope}.
Second, the transformation law under changes of coordinates is determined by
\eqref{eq:virasoro-coordinate-change} rather than \eqref{eq:heisenberg-coordinate-change}.

We can also define a variant with primary field insertions, 
$\Conf(C,\Vir_c;W_{h_1}(q_1) \cdots W_{h_k}(q_k))$.
Again the definition is parallel to that in \autoref{sec:heisenberg-blocks}, 
now with the pole of $\IP{T(p) \cdots W_h(q) \cdots}$ 
constrained by the OPE relation \eqref{eq:virasoro-primary-ope}.

As for $\Heis$, the space $\Conf(C,\Vir_c)$ is $1$-dimensional in the case $C = \bbC \bbP^1$, and 
infinite-dimensional if $C$ is a compact surface of genus $g > 0$.

\subsection{Virasoro-Heisenberg blocks}

We will also need to consider a decoupled combination of the two
notions above: a block $\Psi \in \Conf(C, \Vir_c \otimes \Heis)$
means a system of correlation functions
\begin{equation} \label{eq:vir-heis-correlation-function}
  \IP{T(p_1) \cdots T(p_n) J(q_1) \cdots J(q_m)}_\Psi
\end{equation}
where the dependence on the $p_i$ is as in
\autoref{sec:virasoro-blocks},
the dependence on the $q_i$ is as in 
\autoref{sec:heisenberg-blocks}, and
there are no singularities at $p_i = q_j$.

Note that there is a map
$\Conf(C, \Vir_c \otimes \Heis) \to \Conf(C, \Vir_c)$
obtained by considering the correlation functions of $T$ alone.
Likewise there is a map
$\Conf(C, \Vir_c \otimes \Heis) \to \Conf(C, \Heis)$.

\section{Fermionization} \label{app:unfusion-heisenberg}

In this section, we verify the properties of 
free-fermion insertions in Heisenberg blocks
which we claimed in \autoref{sec:free-fermions}.
Throughout this section we work in a fixed contractible patch
with local coordinate $z$, and we frequently simplify our notation, writing $r_i$ for $z(r_i)$, $J(r_i)$ for $J(r_i)^{z}$ and $\psi_{\pm}(p)$ for $\psi_{\pm}(p)^z$. 
We also sometimes write $\int_q^p$ for $\int_\ell$,
when $\ell$ is the leash running from $q$ to $p$ 
in our patch.

\subsection{The normal-ordered exponential}

We recall the definition \eqref{eq:heisenberg-unfusion}:
\Needspace{4\baselineskip}
\begin{equation}\label{eq:app-heisenberg-unfusion}
  \tikz[remember picture, baseline]{
    \node[anchor=base, inner sep=0pt] (psiplus) {$\psi_+(p)^z$};
  }
  \,
  \tikz[remember picture, baseline]{
    \node[anchor=base, inner sep=0pt] (psiminus) {$\psi_-(q)^z$};
  }
  =
  \frac{1}{z(p)-z(q)} \ \nop{\exp \int_\ell J \, } 
\end{equation}
\begin{tikzpicture}[overlay, remember picture]
  \coordinate (start) at (psiplus.north);
  \coordinate (end) at (psiminus.north);
  \coordinate (startShifted) at ([yshift=1ex]start);
  \coordinate (endShifted) at ([yshift=1ex]end);
  \draw[thick] (startShifted) -- (endShifted);
  \draw[thick] (startShifted) -- ++(0,-0.5ex);
  \draw[thick] (endShifted) -- ++(0,-0.5ex);
  \coordinate (mid) at ($(startShifted)!0.5!(endShifted)$);
  \node at ($(mid)+(0,1.5ex)$) {$\ell$};
\end{tikzpicture}\noindent\unskip
The normal-ordered exponential here is defined by
\begin{equation}
\nop{\exp \int_q^p J \, }=\sum_{n=0}^\infty T_n(p,q) \, ,
\end{equation}
where
\begin{equation}
T_n(p,q) = \frac{\prod_{j=1}^n\int_q^p\de r_j  \nop{\prod_{k=1}^nJ(r_k)}}{n!}\, ,
\end{equation}
and $\nop{\prod_{k=1}^n J(r_{k})}$ denotes a sum of Feynman diagrams with $n$ vertices labeled $r_1, \dots, r_n$, with all vertices either 0-valent or 1-valent; a $0$-valent vertex gives a factor $J(r_{i})$, and an edge gives a factor $-\frac{1}{(r_i-r_j)^2}$.
\begin{center}
\begin{tikzpicture}
    \draw[fill=black](0,0) circle (1pt) node [above] {$r_1$};
    \draw[fill=black](0,0) circle (1pt) node [below, yshift=-0.5cm] {$J(r_1)$};
    \draw[fill=black](2,0) circle (1pt) node [above] {$r_2$};
    \draw[fill=black](3,0) circle (0pt) node [below, yshift=-0.5cm] {$-\frac{1}{(r_2-r_3)^2}$};
    \draw[fill=black](4,0) circle (1pt) node [above] {$r_3$};
    \draw (2,0) -- (4,0);
\end{tikzpicture}
\end{center}
Each diagram with two $0$-valent vertices $r_i$, $r_j$
has a corresponding diagram with those two vertices connected. It follows that all singularities in correlation functions 
as $r_i \to r_j$ are cancelled, for any pair $i$, $j$.
Thus $\nop{\prod_{k=1}^n J(r_{k})}$ is a well defined operator for all points $(r_1, \dots, r_n)$ in the domain of integration.

\subsection{OPE between \texorpdfstring{$J$}{J} and \texorpdfstring{$\psi_\pm$}{psi+-}}

The OPE \eqref{eq:heisenberg-primary-ope}, applied to the insertion $V_1 = \psi_+$, requires that as $p' \to p$
\begin{equation}
  J(p')^z\psi_+(p)= \frac{\psi_+(p)}{z(p') - z(p)} + \regular .
\end{equation}
To verify that this OPE is indeed satisfied by 
our definition \eqref{eq:app-heisenberg-unfusion}, we need to show that
as $p' \to p$
\begin{equation}
J(p')^z\frac{\nop{\exp \int_q^p J \, }}{z(p)-z(q)}=\frac{1}{z(p')-z(p)}\frac{\nop{\exp \int_q^p J \, }}{z(p)-z(q)}+\regular \, .
\end{equation}
We will show that in fact
\begin{equation} \label{eq:interm-1}
J(p')^z \, T_n(p,q) = \frac{T_{n-1}(p,q)}{z(p')-z(p)} + \regular \, .
\end{equation}
Then it will follow that 
\begin{align}
J(p')\frac{\nop{\exp \int_q^p J \, }}{p-q}&=\frac{1}{p'-p}\frac{\sum_{n=1}^\infty T_{n-1}(p,q)}{p-q}+\regular\\
&=\frac{1}{p'-p}\frac{\nop{\exp \int_q^p J \, }}{p-q}+\regular,
\end{align}
as expected. 

To prove \eqref{eq:interm-1}, we note that as $p' \to r_1$ 
the integrand on the left side 
has a singularity proportional to $\frac{1}{(p' - r_1)^2}$,
with coefficient $\frac{\prod_{j=2}^{n} \int_q^p\de r_j  \nop{\prod_{k=1}^nJ(r_k)}}{n!}$.
(To see this, note that each Feynman diagram where $r_1$ is $0$-valent contributes to the
singularity, with coefficient given by the same diagram with $r_1$ deleted.)
After integration over $r_1$, 
using
\begin{equation}\label{removeregnop}
\int_q^p\frac{1}{(r-p')^2}\rd r=\frac{1}{p'-p}-\frac{1}{p'-q},
\end{equation}
this contributes as $p' \to p$ a singular term
\begin{equation} \label{eq:J-fermion-singular-term}
\frac{1}{p' - p} \frac{\prod_{j=2}^{n} \int_q^p\de r_j  \nop{\prod_{k=1}^nJ(r_k)}}{n!}
\end{equation}
There are similar singular terms as $p' \to r_i$ for any $i = 1, \dots, n$; after integration over $r_i$ and 
relabeling of the remaining variables, they all 
contribute the same 
term \eqref{eq:J-fermion-singular-term}.
Thus altogether we get as $p' \to p$ the singular term
\begin{equation}
\frac{n}{p' - p} \frac{\prod_{j=2}^{n} \int_q^p\de r_j  \nop{\prod_{k=2}^nJ(r_k)}}{n!} = \frac{T_{n-1}(p,q)}{p'-p}
\end{equation}
as desired.

Similarly, we can prove \eqref{eq:app-heisenberg-unfusion} obeys as $p' \to q$
\begin{equation}
J(p')\frac{\nop{\exp \int_q^p J \, }}{p-q}=-\frac{1}{p'-q}\frac{\nop{\exp \int_q^p J \, }}{p-q}+\regular,
\end{equation}
as required by the OPE \eqref{eq:heisenberg-primary-ope}
applied to $V_{-1} = \psi_-$,
\begin{equation}
  J(p')\psi_-(q)= -\frac{\psi_-(q)}{p' - q} + \regular .
\end{equation}

\subsection{Covariant constancy}

Next we verify that \eqref{eq:app-heisenberg-unfusion} satisfies the covariant-constancy equation \eqref{eq:fermion-cc-condition} for variations of $p$.
This amounts to checking that 
\begin{equation}\label{covcons}
\partial_{p}\left(\frac{1}{p-q}\nop{\exp\int_q^pJ}\right)=\nop{J(p) \left(\frac{\nop{\exp \int_q^p J \, }\,}{p-q}\right) } \,\, .
\end{equation}

Concretely we have
\begin{equation}
\nop{\,J(p)\ \left(  \nop{\exp \int_q^p J \, } \right) \,\,}=\sum_n\mathfrak{J}_{n}(p,q)
\end{equation}
where
\begin{equation}
  \fJ_n(p,q) = \lim_{p' \to p} \left( J(p') T_n(p,q) - \frac{T_{n-1}(p,q)}{p'-p} \right) \, ,
\end{equation}
which we showed above is well defined.
Below we will check directly that
\begin{align}
\label{appen:Tn}\partial_{p}{T_n}(p,q) &=\mathfrak{J}_{n-1}(p,q)+\frac{T_{n-2}(p,q)}{p-q} \, .
\end{align}
It follows that
\begin{equation}
\partial_p \left( \frac{\sum_{n=0}^\infty T_n}{p-q} \right) = \frac{\sum_{n=1}^\infty \mathfrak{J}_{n-1}}{p-q}+\frac{\sum_{n=2}^\infty T_{n-2}}{(p-q)^2}-\frac{\sum_{n=0}^\infty T_{n}}{(p-q)^2} = \frac{\sum_{n=0}^\infty \mathfrak{J}_n}{p-q},
\end{equation}
which is the desired \eqref{covcons}.

It only remains to establish \eqref{appen:Tn}. 
We check it directly for $n = 1,2$:
\begin{equation}
\partial_{p} {T_1}(p,q) = \partial_{p}\int_q^p \de r \, J(r) = J(p) = \fJ_0(p,q) \, ,
\end{equation}
\begin{align}
\partial_{p}{T_2}(p,q) & = \frac12 \partial_{p} \int_q^p \int_q^p \de r_1 \de r_2 \left(J(r_1) J(r_2) - \frac{1}{(r_1-r_2)^2} \right)\\ 
&= \int_q^p \de r\left( J(p)J(r)- \frac{1}{(r-p)^2} \right)\\
&= \lim_{p' \to p} \left( J(p') T_1(p,q) - \frac{1}{p' - p}  + \frac{1}{p-q} \right) \\
&= \fJ_1(p,q) + \frac{T_0(p,q)}{p-q}
\end{align}
where we used \eqref{removeregnop} again.
More generally, we can write
\begin{align}
  \partial_p T_n(p,q) &= \partial_p \left( \frac{\prod_{j=1}^{n} \int_q^p\de r_j  \nop{\prod_{k=1}^{n-1} J(r_k)}}{n!} \right) \\
  &= \frac{\prod_{j=1}^{n-1} \int_q^p\de r_j \, \nop{J(p) \prod_{k=1}^{n-1} J(r_k)}}{(n-1)!} \\
  &= \lim_{p' \to p} \frac{\prod_{j=1}^{n-1} \int_q^p\de r_j \, \nop{J(p') \prod_{k=1}^{n-1} J(r_k)}}{(n-1)!}  \, .
\end{align}
Now, under the limit sign, 
we split the sum over Feynman diagrams into two pieces.
The diagrams where $p'$ is a $0$-valent vertex give
$J(p') T_{n-1}(p,q)$. The diagrams where $p'$ is connected
to another vertex give 
\begin{align}
 \frac{1}{(n-2)!} \prod_{j=1}^{n-1} \int_q^p\de r_j \, \frac{\nop{\prod_{k=2}^{n-1} J(r_k)}}{(p' - r_1)^2} &= T_{n-2}(p,q) \int_q^p \de r_1 \frac{1}{(p'-r_1)^2} \\
 &= \frac{T_{n-2}(p,q)}{p'-p} - \frac{T_{n-2}(p,q)}{p'-q} \, .
\end{align}
Combining the two types of diagram 
gives 
\begin{align} \label{eq:Tn-equiv}
\partial_{p}{T_n}(p,q) &=
\lim_{p' \to p} \left( J(p') T_{n-1}(p,q) - \frac{T_{n-2}(p,q)}{p'-p} + \frac{T_{n-2}(p,q)}{p-q} \right) \, ,
\end{align}
which is the desired \eqref{appen:Tn}.

\subsection{OPE between \texorpdfstring{$\psiplus$}{psi+} and \texorpdfstring{$\psiminus$}{psi-}}

Finally we verify that \eqref{eq:app-heisenberg-unfusion} 
satisfies the OPE relation \eqref{eq:fermion-OPE-general}.

Define $T_n^{(i)} = T_n(p_i,q_i)$ and
$T_n^{(1+2)} = T_n(p_1,q_2)$.
We are interested in the behavior of products
$T_n^{(1)} T_m^{(2)}$ as $p_2 \to q_1$.
We first compute for low $n$, $m$:
\begin{equation}
T^{\lo}_0 T^{\lt}_0 = 1 = T_0^{(1+2)} \, ,
\end{equation}
\begin{align}
T^{\lo}_0 T^{\lt}_1+T^{\lo}_1 T^{\lt}_0&=  \int_{q_1}^{p_1} \de r_1J(r_1)+\int_{q_2}^{p_2} \de r_1 J(r_1)\\
& \to \int_{q_2}^{p_1} \de r_1 J(r_1) \\
&= T_1^{(1+2)} \, ,
\end{align}
and more interestingly
\begin{align}
T^{\lo}_0 T^{\lt}_2 + T^{\lo}_2 T^{\lt}_0 
&=  \frac{1}{2} \int_{q_1}^{p_1} \int_{q_1}^{p_1} \de r_1 \de r_2 \nop{J(r_1) J(r_2)} + \frac{1}{2} \int_{q_2}^{p_2} \int_{q_2}^{p_2} \de r_1 \de r_2 \nop{J(r_1) J(r_2)} \, ,
\end{align}
\begin{align}
T^{\lo}_1T^{\lt}_1
&= \int_{q_1}^{p_1} \de r_1 J(r_1) \int_{q_2}^{p_2}\de r_2 J(r_2) \\
&= \int_{q_1}^{p_1}\int_{q_2}^{p_2} \de r_1 \de r_2 \left(\nop{J(r_1) J(r_2) }+ \frac{1}{(r_1-r_2)^2} \right) \\
&= \int_{q_1}^{p_1}\int_{q_2}^{p_2} \de r_1 \de r_2 \, \nop{J(r_1) J(r_2) } + \log\left(r_1-r_2\right)\bigg|_{q_1}^{p_1}\bigg|_{q_2}^{p_2} \, ,
\end{align}
so combining these and taking $p_2 \to q_1$ we get
\begin{equation}
  T^{\lo}_0 T^{\lt}_2 + T^{\lo}_1 T^{\lt}_1 + T^{\lo}_2 T^{\lt}_0 - S \, \to \, T_2^{(1+2)} \, ,
\end{equation}
where we defined
\begin{equation}\label{Sfactor}
S=\log\frac{\left(p_1-p_2\right)\left(q_1-q_2\right)}{\left(q_1-p_2\right)\left(p_1-q_2\right)} \, .
\end{equation}

More generally, let us consider the sum
\begin{equation}
 \sum_{m=0}^{\infty} \sum_{n=0}^\infty \e^{-S} T_m^\lo T_n^\lt \, .
\end{equation}
This can be expressed as a sum over Feynman diagrams
with the same Feynman rules as before, with 
vertices of two colors, integrated over the
two integration contours $\ell_1$, $\ell_2$.
The factor $\e^{-S}$ accounts for edges
connecting vertices of different colors.
Since the Feynman rules are independent of the
colors, we can rewrite them in terms of vertices of a single color,
now with each vertex integrated 
over the combined contour $\ell_1 + \ell_2$.
Said otherwise, we have
\begin{equation}
\sum_{m=0}^{\infty} \sum_{n=0}^\infty \e^{-S} T_m^\lo T_n^\lt =  \sum_{l=0}^\infty \frac{1}{l!} \prod_{i=1}^l \int_{\ell_1+\ell_2} \de r_i \nop{\prod_{j=1}^l J(r_j)} \, .
\end{equation}
As $p_2 \to q_1$, $\ell_1 + \ell_2$ becomes
a single contour running from $q_2$ to $p_1$, which gives
\begin{equation}
\sum_{m=0}^{\infty} \sum_{n=0}^\infty \e^{-S} T_m^\lo T_n^\lt \to \sum_l T_l^{(1+2)} \, .
\end{equation}
Finally we conclude that
\begin{align} 
  \tikz[remember picture, baseline]{
    \node[anchor=base, inner sep=0pt] (psi3plus1) {$\psi_+(p_1)$};
  }
  \,
  \tikz[remember picture, baseline]{
    \node[anchor=base, inner sep=0pt] (psi3minus1) {$\psi_-(q_1)$};
  }
  \,\,
  \tikz[remember picture, baseline]{
    \node[anchor=base, inner sep=0pt] (psi3plus2) {$\psi_+(p_2)$};
  }
  \,
  \tikz[remember picture, baseline]{
    \node[anchor=base, inner sep=0pt] (psi3minus2) {$\psi_-(q_2)$};
  }
  =\frac{\sum_{m=0}^\infty\sum_{n=0}^\infty T^{\lo}_m T^{\lt}_n}{(p_1-q_1)(p_2-q_2)} \to \frac{\sum_{l=0}^\infty T^{{(1+2)}}_l}{(q_1-p_2)(p_1-q_2)}=
  - \frac{
    \tikz[remember picture, baseline]{
      \node[anchor=base, inner sep=0pt] (psi3plus1rhs) {$\psi_+(p_1)$};
    }
    \,
    \tikz[remember picture, baseline]{
      \node[anchor=base, inner sep=0pt] (psi3minus2rhs) {$\psi_-(q_2)$};
    }
  }{p_2 - q_1}
\end{align}
\begin{tikzpicture}[overlay, remember picture]
  \coordinate (start31) at ($(psi3plus1.south)!0.5!(psi3plus1.north)$);
  \coordinate (end31) at ($(psi3minus1.south)!0.5!(psi3minus1.north)$);
  \coordinate (start31Shifted) at ([yshift=-2ex]start31);
  \coordinate (end31Shifted) at ([yshift=-2ex]end31);
  \draw[thick] (start31Shifted) -- (end31Shifted);
  \draw[thick] (start31Shifted) -- ++(0,0.5ex);
  \draw[thick] (end31Shifted) -- ++(0,0.5ex);
  \coordinate (mid31) at ($(start31Shifted)!0.5!(end31Shifted)$);
  \node at ($(mid31)+(0,-1.5ex)$) {$\ell_1$};

  \coordinate (start32) at ($(psi3plus2.south)!0.5!(psi3plus2.north)$);
  \coordinate (end32) at ($(psi3minus2.south)!0.5!(psi3minus2.north)$);
  \coordinate (start32Shifted) at ([yshift=-2ex]start32);
  \coordinate (end32Shifted) at ([yshift=-2ex]end32);
  \draw[thick] (start32Shifted) -- (end32Shifted);
  \draw[thick] (start32Shifted) -- ++(0,0.5ex);
  \draw[thick] (end32Shifted) -- ++(0,0.5ex);
  \coordinate (mid32) at ($(start32Shifted)!0.5!(end32Shifted)$);
  \node at ($(mid32)+(0,-1.5ex)$) {$\ell_2$};

  \coordinate (start3rhs) at ($(psi3plus1rhs.north)!0.5!(psi3plus1rhs.south)$);
  \coordinate (end3rhs) at ($(psi3minus2rhs.north)!0.5!(psi3minus2rhs.south)$);
  \coordinate (start3rhsShifted) at ([yshift=2ex]start3rhs);
  \coordinate (end3rhsShifted) at ([yshift=2ex]end3rhs);
  \draw[thick] (start3rhsShifted) -- (end3rhsShifted);
  \draw[thick] (start3rhsShifted) -- ++(0,-0.5ex);
  \draw[thick] (end3rhsShifted) -- ++(0,-0.5ex);
  \coordinate (mid3rhs) at ($(start3rhsShifted)!0.5!(end3rhsShifted)$);
  \node at ($(mid3rhs)+(0,1.5ex)$) {$\ell_1 + \ell_2$};
\end{tikzpicture}
\noindent
which is the desired \eqref{eq:fermion-OPE-general}.

\section{Abelianization map for irregular singularities}\label{appen:irreg}

Similar to the insertion of primary fields which correspond to regular singularities \eqref{eq:W-dictionary} or \eqref{eq:chi-dictionary}, we also provide the abelianization map for the insertions corresponding to irregular singularities \cite{Gaiotto:2009ma}. Roughly speaking, for the Virasoro algebra, such a state is created by a series expansion in some parameter of $C$ with the coefficients given by a highest weight vector and its descendants. In this appendix, we will focus on a particular example corresponding to the pure $SU(2)$ gauge theory by the AGT correspondence.

In this example, there are two irregular singularities at $0$ and $\infty$ of the same type. They can be described as degree 3 poles of a quadratic differential. We briefly review the construction in \cite{Gaiotto:2009ma} for the Virasoro part. The state inserted at such a puncture is denoted by $|\Delta,\Lambda^2\rangle$ which satisfies

\begin{equation}\label{VirirL1}L_1|\Delta,\Lambda^2\rangle=\Lambda^2|\Delta,\Lambda^2\rangle,\end{equation}
and
\begin{equation}\label{VirirL2}L_2|\Delta,\Lambda^2\rangle=0.\end{equation}
By the Virasoro algebra, this further determines
\begin{equation}\label{VirirLn}L_{n>2}|\Delta,\Lambda^2\rangle=0.\end{equation}

We propose that this state should be mapped to 
\begin{equation}
\label{nonabir}|\Delta,\Lambda^2\rangle \rightsquigarrow\E^{2\Lambda \tJ_{-1}}|0\rangle,
\end{equation}
where
\begin{equation}
\tJ(z)=\sum\limits_{n\in\mathbb{Z}}\frac{\tJ_n}{ z^{n+1}}
\end{equation}
with $z$ the coordinate on the cover and the modes $\tJ_n$ obeying
\begin{equation}
[\tJ_m,\tJ_n]=m\delta_{m+n,0}.
\end{equation}
And the state $|0\rangle$ satisfies
\begin{equation}
\tJ_{n>0} |0\rangle = 0.
\end{equation}

We now check that \eqref{nonabir} indeed satisfies \eqref{VirirL1}, \eqref{VirirL2}, \eqref{VirirLn}. We will use the global coordinate $w$ on $C$ and the global coordinate $z$ on $\tC$ as defined in \autoref{sec:branch-point-singularities}. $z$ also serves as a local coordinate on $C$, and it is related to $w$ by $w=z^2$. Using coordinate $w$,
\begin{equation}
L_1=\frac{1}{2\pi\ri}\oint_{w=0}T(p)^w w(p)^2\rd w(p)=\frac{1}{2\pi\ri}\oint_{w=0}(T^{\tot}(p)^w-T^{\Heis}(p)^w) w(p)^2\rd w(p).
\end{equation}
We are going to deal with the two parts separately. First notice since $-z^{(2)}=z^{(1)}=z$, $\tJ(p^{(2)})^{z^{(2)}}=-\tJ(p^{(2)})^{z^{(1)}}$. Thus we have
\begin{equation}\label{curmap}
  T^\tot(p)^z \rightsquigarrow \frac12 \nop{(\tJ(p^{(1)})^{z^{(1)}})^2 + (\tJ(p^{(2)})^{z^{(1)}})^2} \, .
\end{equation}

Parallel to \autoref{app:unfusion-heisenberg}, we simplify our notations, for example, by replacing $T^{\tot}(p)^w$ by $T^{\tot}(w)$ and $w(p)$ by $w$. Under the change of coordinate
\begin{align}\label{L1tot}
L_1^{\tot}\equiv\frac{1}{2\pi\ri}\oint_{w=0}T^{\tot}(w) w^2\rd w&=\frac{1}{2\pi\ri}\int_{\wp}\frac{z^3}{2} \left(T^{\tot}(z) +  \frac14 \frac{1}{z^2} \right)\rd z\\
&=\frac{1}{2\pi\ri}\int_{\wp}\frac{z^3}{2} \left(\frac12 \nop{(\tJ(z))^2 + (\tJ(-z))^2} +  \frac14 \frac{1}{z^2} \right)\rd z,
\end{align}
where $\wp$ is a path which is a half circle around $z=0$. Using
\begin{align}
\int_{\wp}{z^3}\nop{ (\tJ(-z))^2}  \rd z=\int_{-\wp}{-z^3}\nop{ (\tJ(-(-z)))^2}  \rd (-z)=\int_{-\wp}{z^3}\nop{ (\tJ(z))^2}  \rd z
\end{align}
and
\begin{equation}
\int_\wp\frac18 z\rd z=\int_{-\wp}\frac18 z\rd z,
\end{equation}
\eqref{L1tot} can be rewritten as a loop integral,
\begin{align}
\frac{1}{2\pi\ri}\oint_{w=0}T^{\tot}(w) w^2\rd w=\frac{1}{2\pi\ri}\oint_{z=0}\frac{z^3}{4} \nop{(\tJ(z))^2} \rd z.
\end{align}
In terms of the mode expansion,
\begin{equation}
\nop{(\tJ(z))^2}=\sum_{n,m\in\mathbb{Z}}\frac{\nop{\tJ_n\tJ_m}}{z^{n+m+2}}=\frac{\tJ_1\tJ_1}{z^4},
\end{equation}
where we have used that $\tJ_{n>1}$ annihilates the state. So as a summary of the calculation for the $T^{\tot}$ part, we get
\begin{align}
\frac{1}{2\pi\ri}\oint_{w=0}T^{\tot}(w) w^2\rd w=\frac{\tJ_1\tJ_1}{4}.
\end{align}
Now let's check how it acts on our proposed state \eqref{nonabir}. By using
\begin{equation}
[\tJ_1,\E^{2 \Lambda \tJ_{-1}}]=\sum_n[\tJ_1,\frac{(2 \Lambda)^n}{n!}\tJ_{-1}^n]=\sum_n\frac{(2 \Lambda)^n\tJ_{-1}^{n-1}}{(n-1)!}=2 \Lambda\E^{2 \Lambda \tJ_{-1}},
\end{equation}
we get
\begin{equation}
L_1^{\tot}|\Delta,\Lambda^2\rangle \rightsquigarrow \frac{\tJ_1\tJ_1}{4}\E^{2\Lambda \tJ_{-1}}|0\rangle=\Lambda^2\E^{2\Lambda \tJ_{-1}}|0\rangle.
\end{equation}

Next, we look at the other part involving $T^{\Heis}$. For our purpose, it is easier to first express $L_1^{\Heis}$ in terms of the modes of $J$ as
\begin{equation}
L_1^{\Heis}\equiv\oint_{w=0} T^{\Heis}(w)w^2\rd w=\oint_{w=0} \frac12 \sum_{n,m\in\mathbb{Z}}\frac{\nop{J_nJ_m}}{w^{n+m+2}} w^2\rd w=J_0J_1.
\end{equation}
Again using $-z^{(2)}=z^{(1)}=z$, 
\begin{equation}
J(z) \rightsquigarrow \frac{1}{\sqrt2} \left( \tJ(z) -\tJ(-z)\right) .
\end{equation}

Analogously, for arbitrary $i>0$,
\begin{align}
J_i
&=\int_{\wp} \frac{1}{\sqrt2} \left( \tJ(z) - \tJ(-z) \right) z^{2i}\rd z=\oint_{z=0} \frac{1}{\sqrt2} \tJ(z)  z^{2i}\rd z=\frac{1}{\sqrt2}\tJ_{2i}.
\end{align}
Thus
\begin{equation}
L_{1}^{\Heis}|\Delta,\Lambda^2\rangle \rightsquigarrow 0\ \E^{2\Lambda J_{-1}}|0\rangle=0.
\end{equation}

All together, we get
\begin{equation}
\Lambda^2|\Delta,\Lambda^2\rangle=L_1|\Delta,\Lambda^2\rangle=(L_1^{\tot}-L_1^{\Heis})|\Delta,\Lambda^2\rangle \rightsquigarrow \Lambda^2\E^{2\Lambda J_{-1}}|0\rangle
\end{equation}
as expected.

Following the same technique, we can check
\begin{equation}
0=L_{n>1}|\Delta,\Lambda^2\rangle\rightsquigarrow 0
\end{equation}
is also satisfied.

\bibliography{ab-paper}

\end{document}